\documentclass[10pt]{article}
\pdfoutput=1
\usepackage[english]{babel}
 \usepackage{latexsym,amsmath,amssymb,amsbsy,amstext,amscd,amsfonts}
 \usepackage{graphics,graphicx}
 \usepackage{color}
 \usepackage{tabularx}
 \usepackage{multirow}
\usepackage{booktabs}
\usepackage{epstopdf}
\usepackage{natbib}

\date{\today}

\title{Particle velocity based universal algorithm for numerical simulation of hydraulic fractures}

\author{Michal Wrobel$^{(1,2)}$, Gennady Mishuris$^{(1)}$,
\\
{\it $^{(1)}$\!Department of Mathematics,
Aberystwyth University, }
\\ {\it Ceredigion SY23 3BZ, Wales U.K.,}
\\{\it $^{(2)}$\!EnginSoft TRENTO,  }
\\ {\it Via della Stazione 27, fraz. Mattarello, 38123 TRENTO, Italy}
}

\begin{document}

\maketitle

\begin{abstract}

We develop a new effective mathematical formulation and resulting universal computational algorithm capable of tackling various HF models in the framework of a unified approach. The scheme is not limited to any particular elasticity operator or crack propagation  regime. Its basic assumptions are: i) proper choice of independent and dependent variables (with the direct utilization of a new one - the reduced particle velocity), ii) tracing the fracture front by use of the Stefan condition (speed equation), which can be integrated in closed form and provides an explicit relation between the crack propagation speed and the coefficients in the asymptotic expansion of the crack opening, iii) proper regularization techniques, iv) improved temporal approximation, v) modular algorithm architecture.  The application of the new dependent variable, the reduced particle velocity, instead of the usual fluid flow rate, facilitates the computation of the crack propagation speed from the local relation based on the speed equation. This way, we avoid numerical evaluation of the undetermined limit of the product of fracture aperture and pressure gradient at the crack tip (or alternatively the limit resulting from ratio of the fluid flow rate and the crack opening), which always poses a considerable computational challenge. As a result, the position of the crack front is accurately determined from an explicit formula derived from the speed equation. This approach leads to a robust numerical scheme. Its performance is demonstrated using classical examples of 1D models for hydraulic fracturing: PKN and KGD models under various fracture propagation regimes. Solution accuracy is verified against dedicated analytical benchmarks and  other solutions available in the literature. The scheme can be directly extended to more general 2D and 3D cases.
\end{abstract}

\section{Introduction}

Hydraulic fracture is a process of a crack propagating in a solid material, as a result of pressurized liquid injection. It can be observed in many natural phenomena, like magma driven dykes \citep{Rubin}, sub-glacial drainage of water \citep{Rice} and others \citep{Board,Moschovidis,Pine}. Recently it has been used for reservoir stimulation in the oil and gas industry to maximize hydrocarbon extraction. Although this technology can be backdated to the 1930s \citep{Grebe}, it has been in the last twenty years that hydrofracturing has become commonplace.

Mathematical modeling of this multiphysics process is a challenging task. The main difficulties are: (a) strong non-linearities related to the interaction between the solid and fluid phases, (b) singularities in the physical fields near the fracture front, (c) moving boundaries, (d) degeneration of the governing equations in the near-tip region, (e) pronounced multiscale effects. The complexity of the problem necessitates various simplifications dated back to works of \cite{Sneddon_Elliot,Harrison,Howard,Hubbert,Crittedon}. These studies together  with the later works led to the formulation of the basic 1D models of hydraulic fractures: i) the PKN model \citep{Perkins_Kern,Nordgren}, ii) the KGD model \citep{Khristianovic,Geertsma}, iii) the radial or penny shaped model \citep{Sneddon_1946}. However, in the 1980s the need for more advanced and accurate modeling emerged. The so called pseudo 3D models (P3D) appeared \citep{Warpinski}. They approximate the behaviour of the planar 3D fractures, including those in the stratified reservoirs, with minimal computational costs. Another attempt at advancing the mathematical modeling was the introduction of the planar 3D models (PL3D) \citep{Clifton,Vandamme,Advani}, in which case the crack footprint and the internal fluid flow are described by the 2D mesh of cells and combined with the full 3D elasticity equations, allowing one to determine the fracture aperture as a function of the fluid pressure. In recent years, there have also been attempts to develop full 3D models utilizing various numerical techniques, e.g. finite element method \citep{Lecampion_2009,Hunsweck_2012,Wangen,Chen_2013}, boundary element method (or combination of the two \citep{Carter_Wawrzynek,Yamamoto_2004}), the discrete element methods \citep{Damjanac_2013} or other techniques \citep{Kresse_2013,lavrov_2014}. A broad review of the topic can be found in \cite{Adachi-et-Al-2007} where it is has been shown that, in spite of substantial progress made, there is still a demand for further improvements in efficiency and credibility of computations to tackle multiscale effects, complex geometries and properties of the rock and fracturing fluids, and to possibly perform the computations in real time.

Alongside the development of mathematical models, fundamental research, aimed at identifying the basic solution features related to the underlying physics of the process, has been carried out.
Special attention has been paid to the near-tip behaviour of the solution. The early works introducing the correct tip asymptotics can be backdated to the 1980s (\cite{Spence&Sharp} - for the KGD model, \cite{Kemp} - for the PKN model). More comprehensive studies on this problem were presented in \cite{Desroches_1994} for the zero toughness impermeable case, in \cite{Lenoach_1995} for the zero toughness leak-off dominated variant. In \cite{Carbonell_1999}, the near-tip process in an impermeable elastic medium for the plane strain conditions was modeled with the account of a lag between the fracture tip and the fluid front. \cite{Savitski_2002} have proposed asymptotic solutions in the case of a penny shaped fracture driven by the Newtonian fluid for both small and large toughness values. The analysis for plain strain and penny shaped fractures propagating in the toughness dominated regime in permeable rock was delivered in \cite{Bunger_2005}, giving the early and large time asymptotes. Results pertaining to the plain strain fracture driven by the shear-thinning fluids can be found in \cite{garagash_large_toughenss}.

Simultaneously, recognition of the importance of the near-tip behaviour of solutions has led to classification of the basic fracture propagation modes \citep{Detournay_2004,Gar_scaling,Gar_Det_Ad}. They have been categorized in the parametric space which is encompassed by four limiting physical regimes: i) leak-off dominated, ii) storage dominated, iii) toughness dominated, iv) viscosity dominated.
The hydraulic fracture is considered to evolve in time between these specific modes depending on the injection rate, the rock and fracturing fluid properties. A number of semi-analytical and numerical solutions have been constructed for such asymptotic regimes. In particular, the case of zero toughness impermeable rock was analyzed in \cite{Adachi_Detournay,Savitski_2002}, small toughness zero-leak-off variant in \cite{Garagash_small_toughness}, large toughness impermeable in \cite{Garagash_shut_in}, and finite toughness permeable in \cite{MKP}.

All these efforts have underlined the importance of the multiscale character of the problem. It is now well understood that the coupling between non-linear, non-local and history dependent physical fields results in a complex solution structure, where relative importance of the mentioned processes depends on temporal and spatial scales. It has been proved that the global behaviour of a fluid driven fracture is controlled by the near-tip region, and this has consequences for the computational implementation.
Furthermore, in the hydraulic fracture problem, the nature of the moving boundary results in degeneration of the governing equations and the boundary conditions at the crack tip, which makes
tracing the fracture front an extremely difficult task \citep{Peirce_2014}.
All these factors clearly indicate the challenge in understanding the solution structure (especially the tip asymptotics) and its appropriate application in the computational schemes. In the recent studies by \cite{Lecampion_Brisbane} it has been shown that the algorithms which use the appropriate multi-scale hydraulic fracture asymptote in the near tip region provide much better results than those which do not apply it. Moreover, when accounting for the proper tip asymptotics, very good results can be obtained even for coarse meshing.
The analysis given in \cite{Linkov_3,M_W_L,solver_calkowy,Kusmierczyk} proves that proper mathematical formulation of the problem of hydraulic fracture facilitates the correct introduction of the basic asymptotic features of the solution to the numerical algorithm. This in turn results in an appreciable improvement to the accuracy and efficiency of computations.

We propose a new unified approach that yields universal numerical algorithm capable of tackling various HF models.
Its basic assumptions are: i) proper choice of independent and dependent variables (including a new one - the reduced particle velocity), ii) tracing the fracture front by use of the Stefan condition (speed equation) which can be integrated in a closed form to give an explicit relation between the crack propagation speed and the leading coefficients of the crack opening asymptotics, iii) proper regularization techniques, iv) improved numerical approximation of the temporal derivative of the solution, v) modular algorithm architecture.

The application of the new dependent variable, the reduced particle velocity, instead of the usual fluid flow rate, facilitates the computation of the crack propagation speed from the local relation based on the speed equation. This way we avoid numerical evaluation of the undetermined limit of the product of fracture aperture and pressure gradient at the crack tip (or alternatively the limit resulting from division of the fluid flow rate by the crack opening), which poses a considerable computational challenge. As a result the position of the crack front can be accurately determined explicitly. With regards to the numerical modeling of hydraulic fractures, this condition was originally introduced by Kemp \citep{Kemp} but was later abandoned. Recently, it has been rediscovered by Linkov (\cite{Linkov_1,Linkov_2}). The tip asymptotics is utilized in the numerical scheme together with appropriate regularization techniques. Some elements of the employed scheme have been presented in \cite{Kusmierczyk, M_W_L,solver_calkowy}, where a broad discussion on the advantages of application of proper dependent variables and regularization techniques can be found.
One of the key points of the developed universal algorithm is utilization of the explicit relation between the crack propagation speed and the leading terms of the crack opening asymptotic expansion in the form \eqref{v_0_univ} -- \eqref{LC_KGD_toughness}. For the PKN model it was found and described in \cite{Kusmierczyk,solver_calkowy}, while for the KGD formulation respective results have been recently reported at a number of conferences.

The above developments lead to a robust and efficient numerical scheme. Its performance is demonstrated against classical 1D models: PKN and KGD ones. The solution accuracy is verified against analytical benchmarks and solutions available in the literature. Most of the ideas developed here, can be directly extended to more general 2D and 3D cases.

The paper is organized as follows. In Section 2 a general mathematical description of the problem is given. Moreover the basic idea of fracture front tracing is explained, also the motivation for and advantages of the applied approach are presented. We introduce here a dimensionless formulation of the problem, which is henceforth used. Section 3 contains a detailed characterization of the solution tip asymptotics, and its link to the mechanism of crack tip tracing. A universal mathematical description is proposed for different elasticity operators and crack propagation regimes. A complete definition of the mechanism of crack front tracing based on the speed equation (Stefan condition) is given. In section 4, a new dependent variable, the reduced particle velocity, is introduced. The governing equations are reformulated in terms of the new variable. Section 5 describes the reduction of the problem to a self-similar version for two different time-dependent functions: the power function and the exponential one. Then the self-similar variant of the universal algorithm is presented. The performance of the algorithm is verified against dedicated analytical benchmarks (detailed description in Appendix) as well as the reference solutions available in the literature. Fully analytical benchmarks for the KGD variant of the problem are introduced here for the first time. In Section 6, the idea of a universal solver is adapted to the transient case where the improved approximation of the temporal derivative is one of the key elements. Extensive accuracy analysis is given. Final discussions and conclusions are presented in Section 7.

\section{Problem formulation}

\subsection{Governing equations for 1-D model of hydraulic fracture}

Let us consider a rectilinear crack  $-l<x<l$  fully filled by Newtonian fluid injected at midpoint ($x=0$) at a given rate $q_0(t)$. As a result, the crack front ($x=\pm l$) moves and the crack length, $l=l(t)$, is a function of time.
Below we present a classic set of the governing equations for the 1D formulation of the problem, which can be found in  e.g. \cite{Economides,Adachi_Detournay,Kovalyshen}  for various hydrofracturing models.
As usual, due to symmetry of the problem, we analyse only one of the symmetrical parts of the crack $x\in [0,l(t)]$.

The continuity equation has the form:
\begin{equation}
\label{continuos_1} \frac{\partial w}{\partial t}+\frac{\partial
q}{\partial x}+q_l=0,\quad t > 0,\quad 0< x < l(t),
\end{equation}
The fluid flow inside the fracture is described by the Poiseuille equation which, in case of Newtonian fluid, is described as:
\begin{equation}\label{Poiseulle_1}
q=-\frac{1}{M}w^3\frac{\partial p}{\partial x},\quad t > 0,\quad 0< x < l(t),
\end{equation}

In the above equations $w=w(t,x)$ stands for the crack opening, $q=q(t,x)$ is the
fluid flow rate,   $p=p(t,x)$ ($p=p_{f}-\sigma_0$, $\sigma_0$ -
confining stress) refers to the net fluid pressure. Constant $M$
is computed as $M=12{\mu}$,
where $\mu$ denotes the dynamic viscosity. Function $q_l=q_l(t,x)$, in the right-hand
side of the continuity equation \eqref{continuos_1},  is the volumetric rate of fluid loss due to  the rock
formation in the direction perpendicular to the crack surfaces per unit
length of the fracture. Usually it is assumed to be given (local formulation) \citep{Nordgren,Mikhailov,Kusmierczyk}.
More accurate analysis involves a nonlocal formulation where the mass transfer
in the entire external domain should be taken into account \citep{Kovalyshen_PhD}. We will comment on the possible
behaviour of $q_l$ later on.

These equations are to be supplemented by the relation describing the deformation of rock under applied hydraulic pressure.
Thus, the net pressure in the fracture is given by the relationship:
\begin{equation}
\label{elastic_0} p(x)={\cal A}w(x),\quad 0< x < l(t),
\end{equation}
where operator ${\cal A}$ refers to the chosen model of elasticity (which assumes the predefined fracture geometry).
We consider two most popular linear models (local and nonlocal, respectively):
\begin{itemize}
\item{the PKN model \citep{Nordgren}}
\begin{equation}
\label{elastic_PKN} {\cal A}_1w=k_1w,
\end{equation}
\item{the KGD model \citep{Sneddon}}
\begin{equation}
\label{elastic_KGD}
{\cal A}_2w=\frac{k_2}{2\pi}\int_{0}^{l(t)}\frac{\partial w(t,s)}{\partial s}\frac{s}{x^2-s^2}ds.
\end{equation}
\end{itemize}
In the PKN model, constant $k_1$ was given by \citet{Nordgren} for an
elliptical crack of height $h$, while $E$ and $\nu$ are the elasticity modulus and Poisson's ratio.
Constant $k_2$ in the KGD model follows, for example, from \cite{Sneddon,Mushelishvili}:
\begin{equation}
\label{constants_k}
k_1=\frac{2}{\pi h}\frac{E}{1-\nu},\quad
k_2=\frac{E}{1-\nu^2}.
\end{equation}
The multiplier $k_1$ may depend on $x$ and/or $w$ as well: $k_1=k_1(x,w)$,
constituting the so-called pseudo 3D model (P3D) \citep{Warpinski,Brisbane}. This case can also be considered in the framework of the presented approach.

The inverse operators for \eqref{elastic_PKN} -- \eqref{elastic_KGD} are:
\begin{equation}
\label{inverse_PKN}
{\cal A}_1^{-1} p= \frac{1}{k_1}p,
\end{equation}
\begin{equation}
\label{inverse_KGD}
{\cal A}_2^{-1} p=\frac{4}{\pi k_2}\int_0^{l(t)} p(t,s) \ln \left|\frac{\sqrt{l^2(t)-x^2}+\sqrt{l^2(t)-s^2}}{\sqrt{l^2(t)-x^2}-\sqrt{l^2(t)-s^2}}\right|ds.
\end{equation}
Note that the original Cauchy type singular integral in the elasticity equation (\ref{elastic_KGD}) is defined (compare \cite{Sneddon}) over the entire crack, $ -l(t)<x< l(t)$, and that
representation (\ref{elastic_KGD}) is valid only under the assumption
\begin{equation}
\label{BC_3} \frac{\partial w}{\partial x}(t,0)=0.
\end{equation}
On the other hand, the form \eqref{inverse_KGD} of the integral operator ${\cal A}_2^{-1}$  guarantees this property if the net pressure is smooth enough (differentiable) near the zero point and
the following condition holds:
\begin{equation}
\label{cond_net_pressure}
\int_0^{l(t)} \frac{p(t,s)ds}{\sqrt{l^2(t)-s^2}}<\infty,
\end{equation}
that is easily checked by differentiation.
This condition has a clear physical sense: the Stress Intensity Factor (SIF) defined by the integral in (\ref{cond_net_pressure})
is finite.
Thus, when equation (\ref{inverse_KGD}) instead of (\ref{elastic_KGD}) is utilized, (\ref{BC_3}) is satisfied automatically.
Moreover, using condition (\ref{cond_net_pressure}) one can also prove that
\begin{equation}
\label{cond_dw}
\frac{\partial w}{\partial x}(t,0)= O\left(x\log x\right),\quad x\to0.
\end{equation}

The foregoing equations should be supplemented by the initial and boundary conditions.
Thus, the influx boundary condition and two boundary conditions at the crack tip are:
\begin{equation}
\label{BC_1} q(t,0)=q_0(t),
\end{equation}
\begin{equation}
\label{BC_2} w(t,l(t))=0, \quad q(t,l(t))=0.
\end{equation}

\noindent
We use in this paper the non-zero initial conditions:
\begin{equation}
\label{IC}
l(0)=l_\diamond,\quad
 w(0,x)=w_\diamond(x),\quad x\in(0,l_\diamond).
\end{equation}
Usually, the uniform initial conditions are suggested instead (e.g \cite{Nordgren}):
\begin{equation}
\label{IC_PKN}
w(x,0)=0, \quad l(0)=0.
\end{equation}
However,  for time dependent problems, it is quite common to replace them by the condition (\ref{IC}), where
small values of $l_\diamond=l_{sf}(t_0)$ and $w_\diamond(x)=w_{sf}(x,t_0)$ for $x\le l_\diamond$ are taken from the corresponding
self-similar solution which neglects the leak-off to the formation (early time asymptote). Note that the condition (\ref{IC}) describes then a preexisting hydraulic fracture.


In the case when fracture evolution is analyzed in the framework of Linear Elastic Fracture Mechanics (LEFM),
that is for the the toughness driven regime for the KGD model, the following propagation condition is imposed:
\begin{equation}
\label{K_I_criterion}
K_{I}(t)=K_{IC},
\end{equation}
where $K_{IC}$ is the material toughness \citep{Rice_1968} while $K_{I}$ is the already mentioned SIF computed by the following formula
\citep{Mushelishvili} (compare \eqref{cond_net_pressure}):
\begin{equation}
\label{K_I}
K_{I}= 2\sqrt{\frac{l(t)}{\pi}} \int_0^{l(t)} \frac{p(t,s)ds}{\sqrt{l^2(t)-s^2}},
\end{equation}
Note that the tip asymptote for the crack opening is defined in this case as:
\begin{equation}
\label{K_1_1}
w(t,x)\sim\frac{8(1-\nu^2)}{\sqrt{2\pi}}\frac{K_{I}}{E}\sqrt{l(t)-x},\quad x\to l(t).
\end{equation}

\noindent Finally, the global fluid balance equation takes the form:
\begin{equation}
\label{global_balance}
\int_0^{l(t)}[w(t,x)-w_\diamond(x)]dx-\int_{0}^tq_0(t)dt+\int_0^{l(t)}\int_{0}^tq_l(t,x)dtdx=0.
\end{equation}
Which is usually used to determine the crack length (see e.g. \cite{Adachi-et-Al-2007}).

The above set of equations and conditions constitute the classical 1-D model of hydraulic fracture \citep{Adachi-et-Al-2007,Linkov_3}.

In our analysis we will utilize another dependent variable, the average fluid velocity through the fracture cross-sections (called also a particle velocity), $v$, defined as follows:
\begin{equation}
\label{v_def}
v(t,x)=\frac{q(t,x)}{w(t,x)}=-\frac{1}{M}w^2\frac{\partial p}{\partial x}, \quad 0< x < l(t).
\end{equation}
This variable has been frequently mentioned in publications (see for example \cite{Garagash_shut_in,Gar_Det_Ad}), but has not been used directly
in computational algorithms.  In \cite{Warpinski,Linkov_3} it was suggested to incorporate it as a dependent variable instead of the net pressure or the
fluid flux in order to improve the algorithm performance. Our analysis follows this suggestion.

Throughout this paper we  assume that there are no flow stagnation or inversion points along the fracture, which means that $v(t,x)$ should be finite and positive.
\begin{equation}
\label{v_def_1}
0< v(t,x) < \infty, \quad t>0, \quad 0< x < l(t).
\end{equation}
As a result,  the spacial derivative of the net pressure is negative along the entire fracture
\begin{equation}
\label{_der}
\frac{\partial p}{\partial x}<0, \quad t>0, \quad 0< x < l(t).
\end{equation}
Taking into account the fact that the crack opening vanishes at the crack tip (compare (\ref{BC_2})$_1$), equation \eqref{v_def} yields
\begin{equation}
\label{lim_der}
\frac{\partial p}{\partial x}\to -\infty, \quad x \to l(t).
\end{equation}

\subsection{Description of the crack front movement}

Tracing the fracture front evolution is one of the major challenges in the problem of hydraulic fracture. In the recent paper by   \cite{Det_Peirce_2014}, various approaches to this task have been discussed. In the numerical scheme proposed in sections 5-6 below,  we use a strategy for finding the fracture tip different from that advocated in \cite{Det_Peirce_2014}. Note that the above-mentioned paper briefly discusses our approach but this discussion may be misleading. For this reason and in order to eliminate any confusion, we clarify the key points of our approach.

We emphasize that the standard methods of simulating hydrofracturing mostly employ the crack opening, $w(t,{\bf x})$, and the fluid flow rate, ${\bf q}(t,{\bf x})$ as the dependent variables. This allows one to directly account for respective boundary conditions at the fracture tip (compare (\ref{BC_2})), and has other benefits discussed in \cite{Adachi-et-Al-2007}. However, when determining the crack tip position, this approach causes serious difficulties, as comprehensively analysed in \cite{Det_Peirce_2014}, where in particular, the authors pointed out the major computational problem of evaluation of the fluid front velocity (${\bf x =x}_*\in { \partial \Omega}$) from the equation
\begin{equation}
\label{v_def_general}
{\bf v}_*(t,{\bf x}_*)=\lim_{{\bf x}\to {\bf x}_*}\frac{{\bf q}(t,{\bf x})}{w(t,{\bf x})},
\end{equation}
where both dependent variables vanish at the crack tip.

Our approach utilizes different pair of dependent variables: i) the standard one - the crack opening $w(t,{\bf x})$, and ii) the particle velocity (the average through a channel cross section fluid flow velocity), ${\bf v}(t,{\bf x})$, instead of the fluid flux, ${\bf q}(t,{\bf x})$). The next crucial assumption used in our paper is a condition that
\\[3mm]
$\bullet$ \hspace{5mm} \emph{\bf the particle velocity has finite value at the crack front}.
\\[3mm]
This assumption is satisfied for most hydraulic fracture models, with some reservations in the cases of severe leak-off regimes (e.g. fluid driven regime for the KGD model combined with a Carter law - see e.g. \cite{MKP}). Note that the assumption has a clear physical motivation as all the basic equations were derived neglecting the inertia effects (no acceleration terms are present in them). Obviously, the finite velocity of the fracture tip is consistent with experiments  \citep{Rubin_1983,BdP_2006,GRB_2009,Bunger_2013}.

The first consequence of the chosen set of dependent variables is that, when the first of the conditions (\ref{BC_2}) is satisfied, the second one is fulfilled automatically. Moreover, the difficulties mentioned in \cite{Det_Peirce_2014}, as well as those related to computing of the crack propagation speed from relation \eqref{v_def_general} are fully eliminated.

Naturally, this new approach requires reformulation of all the governing equations in terms of the new dependent variables  $w(t,{\bf x})$ and ${\bf v}(t,{\bf x})$. However, one needs to find:
\\[3mm]
$\circ$ \hspace{3mm} \emph{the relationship between the crack propagation speed, ${\bf v}_{cr}$, and the finite fluid front velocity, ${\bf v}_*$.}
\\[3mm]
This question is not specific to our approach. It arises in any formulation of the hydraulic fracture problem (which is not always, however, clearly highlighted). As stated in \cite{Det_Peirce_2014}, the case where the fluid front coincides with the fracture tip is much more challenging from the computational point of view. In problems with moving boundaries such a condition is usually called \emph{the Stefan condition} \citep{Stefan,Lin,Det_Peirce_2014}. In mathematical modeling of hydraulic fractures it was probably Kemp \citep{Kemp} who first explicitly used this condition.  \emph{The Stefan condition} is usually employed in the analysis of hydrofracturing in its implicit form as a compatibility condition (e.g. \cite{Garagash_small_toughness,garagash_large_toughenss}) and has been defined in an explicit way for the steady state problems \citep{Gar_Det_Ad,Det_Gar_2003}.

Kemp's condition has recently been rediscovered by Linkov \citep{Linkov_1} and called by him \textit{the speed equation}.
In our analysis, we shall use both names, the {\it Stefan condition} and the {\it speed equation}, interchangeably. The recalled condition has the following form in the 1D formulation:

\begin{equation}
\label{SE}
\frac{dl}{dt}=v_*(t)\equiv v(t,l(t)),
\end{equation}
where the left-hand side of (\ref{SE}) is the speed of the fracture tip, while the right-hand side is the fluid front velocity.
Equation (\ref{SE}) is valid under the assumption that the fluid front coincides with the fracture tip.
This implies that there is no lag between them.
Also the invasive zone ahead of the crack (the area ahead of the fracture tip penetrated by fluid)  and, in some cases, the Carter leak-off should be excluded from consideration.
In those three aforementioned cases, the speed equation (\ref{SE}) can be still utilised by supplementing the right hand-side with an additional term which takes into account the respective phenomenon (lag, invasive zone, severe leak-off).

As an additional argument to justify the speed equation (\ref{SE}), one can examine a possible behaviour of the leak-off function $q_l$ near the crack tip.
Note that $q_l$ in (\ref{continuos_1}) describes  the rate of fluid flow from the fracture into the surrounding formation, and generally is not known in advance. To describe this phenomenon, one needs to formulate a coupled problem linking the processes within the fracture to those in the rock formation which depends on various external conditions (geometry, porosity, permeability, fluid saturation and others). This nonlocal formulation complicates the problem enormously.
The usual way to overcome this difficulty is to treat the function $q_l=q_l(t,x,w,p)$ as given one (which anyway can be dependent on the processes within the fracture by the known relation through the crack opening, the net pressure and the properties of the adherent rock). Only in the case of impermeable rock the problem of hydraulic fracture simplifies in this respect and one can set $q_l=0$.

In the \emph{local empirical} formulation, the leak-off function is assumed to be solution dependent relation. We assume the following behaviour of $q_l$ near the crack tip:
\begin{equation}
\label{q_l} q_l(t,x)\sim Q(t)\big(l(t)-x\big)^{\eta},\quad \mbox{as} \quad x
\rightarrow l(t),
\end{equation}
where $\eta\ge -1/2$. Note that $\eta=-1/2$ corresponds to the classic empirical Carter law \cite{Carter}.

It is well known (see for example \cite{Adachi_Detournay,Garagash_small_toughness,MKP,Kovalyshen,Gar_Det_Ad,Kusmierczyk}) that the leading term of the crack opening asymptotic expansion in the near-tip region can be expressed as:
\begin{equation}
w(t,x)\sim w_0(t)(l(t)-x)^{\alpha_0},\quad \mbox{as} \quad x
\rightarrow l(t),
\label{as_opening}
\end{equation}
and it does not depend on the value of $\eta$ ($w_0(t)$ is to be found as an element of the solution). Here, the constant $\alpha_0$ depends only on the particular elasticity operator for the problem and the crack propagation regime \citep{MKP,Kusmierczyk} (see Table \ref{T1}).
However, further terms of the asymptotics of the crack opening and other dependent variables depend essentially on the behaviour of the leak-off function near the crack tip and play a crucial role
in the analysis as they determine the smoothness of the particle velocity function in the vicinity of the fracture front.

Let us then discuss the possible behavior of the function $q_l$, taking into account one of the basic assumptions used when deriving the lubrication equation. Namely, it is assumed that the fluid flow inside a thin channel is predominantly directed along the channel
walls. Indeed, the assumption used when deriving the lubrication approximation is that there is no pressure gradient normal to wall. As a result,  the flow in this direction vanishes (e.g \citet{Zimmerman_1991}). This, in turn, implies that
\begin{equation}
\label{leak_off_assumption_0}
q_l(t,x)\ll q'_x(t,x), \quad t > 0,\quad 0< x < l(t),
\end{equation}
at any point along the crack surfaces including the fracture tip. On the other hand, the fluid flux near the crack tip behaves similarly to the that of the crack opening (since the particle velocity is finite at the crack tip). As a result, the following condition should be accepted:
\begin{equation}
\label{leak_off_assumption}
1+\eta\ge\alpha_0.
\end{equation}
Throughout this paper we assume that the condition (\ref{leak_off_assumption}) holds true in its stronger version:
\begin{equation}
\label{leak_off_assumption_strong}
1+\eta>\alpha_0.
\end{equation}
This condition is equivalent to the speed equation in the form (\ref{SE}), as it
follows immediately from the continuity equation (\ref{continuos_1}).

The assumption (\ref{leak_off_assumption_strong}) fails when one considers the toughness driven KGD model ($\alpha_0=1/2$)  with the classical Carter law ($\eta=-1/2$). In this particular case,  an additional term $-2Q(t)w^{-1}_0(t)$ should be introduced in the right-hand side of (\ref{SE}) to guarantee its validity as one can easily check by substituting (\ref{as_opening}) and (\ref{q_l}) into (\ref{continuos_1}) for $q=wv$.
Here $Q(t)$ is a known function or a functional defined on the solution $Q(t)=F[w,v](t)$.
However, in the case of the fluid driven KGD model ($\eta=-1/2$, $\alpha_0=2/3$), failure to comply with condition \eqref{leak_off_assumption} results  in an infinite value of the crack propagation speed (in this case, many authors tend to accept the validity of the Carter law only at some distance from the fracture tip - see e.g. \cite{MKP}).

The authors believe that the empirical Carter law loses its physical sense at the crack tip. For example, we showed in  \cite{Kusmierczyk} that for the PKN model a perturbation of the law in a very small region near the fracture front (at
a distance less than $10^{-4}$ of the crack length) produces a change in the fracture length, which amounts to a few percent. In other words, a change in the law over a distance less than $1$ mm would result in the deviation of the crack length greater than $1$ m, which is unrealistic.

In the following we restrict ourselves to the case (\ref{q_l}) with assumption (\ref{leak_off_assumption_strong}), and consequently with the speed equation in the form (\ref{SE}). The latter will be used throughout this paper to trace the fracture front. On the other hand, the {\it speed equation} serves as the boundary condition at the crack tip for the dependent variable, which is now the particle velocity, $v$, not the fluid flux as in the standard approach. The advantages of such an approach have already been shown in \cite{Linkov_1,Linkov_2,Linkov_3,M_W_L,Kusmierczyk}.

\subsection{Problem normalization}
\label{basic}

Let us normalize the problem by introducing the following dimensionless
variables:
\[
\tilde x=\frac{x}{l(t)}, \quad \tilde t = \frac{t}{t_n},\quad \tilde w(\tilde t,\tilde x)=\frac{w(t,x)}{l_*}, \quad L(\tilde
t)=\frac{l(t)}{l_*},\quad\tilde q_l(\tilde t,\tilde x)= \frac{t_n}{l_*} q_l( t, x),
\]
\begin{equation}\label{norm_V}
 \quad
\tilde q(\tilde t,\tilde x)=
\frac{t_n}{l_*^2}q(t,x),\quad \tilde p(\tilde t, \tilde x)=\frac{t_n}{M}p\,(t,x),\quad \tilde v(\tilde t,\tilde x)=\frac{t_n}{l_*}v(t,x),
\quad t_n=\frac{M}{k_m},
\end{equation}
where $\tilde x\in[0,1]$, and parameter $m$ takes value either 1 or 2 for the PKN and KGD model, respectively.
The value of $m-1$ coincides with the order of homogeneity of the operator ${\cal A}_m$. Parameter $l_*$ is to be chosen as convenient.
Note that the normalization \eqref{norm_V} is not attributed to any particular influx regime, elasticity operator or asymptotic behaviour of the solution.

In the normalized variables, the continuity equation \eqref{continuos_1} takes the form:
\begin{equation}
\label{cont_norm}
\frac{\partial \tilde w}{\partial \tilde t}-\frac{L'}{L}\tilde x \frac{\partial \tilde w}{\partial \tilde x}+\frac{1}{L}\frac{\partial (\tilde w \tilde v)}{\partial \tilde x}+\tilde q_l=0,
\end{equation}
where the fluid flow rate was replaced by the product of the crack opening and the particle velocity according to \eqref{v_def}.

The normalized particle velocity yields:
\begin{equation}
\label{v_norm}
\tilde v=\frac{\tilde q}{\tilde w}=-\frac{1}{L}\tilde w^2 \frac{\partial \tilde p}{\partial \tilde x},
\end{equation}
and the speed equation \eqref{SE} transforms now to:
\begin{equation}
\label{SE_n}
\frac{d L}{d\tilde t}=-\frac{1}{L(\tilde t)}\left[\tilde w^2 \frac{\partial \tilde A \tilde w}{\partial \tilde x}\right]_{\tilde x=1}.
\end{equation}
The normalized elasticity equation \eqref{elastic_0} takes the form:
\begin{equation}
\label{A_norm_PKN_0}
{\tilde p=}{\cal \tilde A} \tilde w,
\end{equation}
where an identity operator,
\begin{equation}
\label{A_norm_PKN}
{\cal \tilde A}_1 ={\cal \tilde A}_1^{-1}={\cal I},
\end{equation}
corresponds to the PKN model, while the integral operators for the KGD model are
\begin{equation}
\label{A_norm_KGD}
{\cal \tilde A}_2 \tilde w= -\frac{1}{2\pi L(\tilde t)} \int_0^1\frac{\partial \tilde w(\tilde t,\eta)}{\partial \eta} \frac{\eta}{\eta^2-\tilde x^2}d\eta,
\end{equation}
and
\begin{equation}
\label{inv_KGD_n}
\tilde w={\cal \tilde A}_2^{-1} \tilde p=\frac{4}{\pi}L(\tilde t)\int_0^1 \tilde p(t,\eta)\ln\left|\frac{\sqrt{1-\tilde x^2}+\sqrt{1-\eta^2}}{\sqrt{1-\tilde x^2}-\sqrt{1-\eta^2}}\right|d\eta,
\end{equation}
for equations \eqref{elastic_KGD} and \eqref{inverse_KGD}, respectively.
In our computations we shall use an alternative form of \eqref{inv_KGD_n} obtained using integration by parts:
\begin{equation}
\label{inv_KGD_n_1}
\tilde w= -\frac{4}{\pi}L(\tilde t)\int_0^1 \frac{\partial \tilde p(t,\eta)}{\partial \eta}K(\eta,x)d\eta +4\sqrt{\frac{L(\tilde t)}{\pi}}\tilde K_I\sqrt{1-\tilde x^2},
\end{equation}
where the kernel $K(\eta,x)$ is:
\begin{equation}
\label{kernel_KGD_inv}
K(\eta,x)=(\eta-x)\ln\ \left| \frac{\sqrt{1-x^2}+\sqrt{1-\eta^2}}{\sqrt{1-x^2}-\sqrt{1-\eta^2}} \right|+x \ln \left( \frac{1+\eta x+\sqrt{1-x^2}\sqrt{1-\eta^2}}{1+\eta x - \sqrt{1-x^2}\sqrt{1-\eta^2}}\right),
\end{equation}
and $\tilde K_{I}={K_{I}(1-\nu^2)}/({E\sqrt{l_*}})$ stands for the dimensionless toughness. Consequently,
the asymptotic estimate \eqref{K_1_1} can be now rewritten in the form:
\begin{equation}
\label{K_I_n}
\tilde w (\tilde t, \tilde x)=\frac{8}{\sqrt{2\pi}}\tilde K_{I} \sqrt{L(\tilde t)}\sqrt{1-\tilde x},\quad \tilde x \to 1.
\end{equation}
From definition \eqref{K_I} one can determine the dimensionless toughness as:
\begin{equation}
\label{K_I_ndef}
\tilde K_{I}(t)=2\sqrt{ \frac{L(\tilde t)}{\pi}} \int_0^{1}\frac{\tilde p(\tilde t,s)ds}{\sqrt{1-s^2}},
\end{equation}

\noindent
The boundary conditions \eqref{BC_1} -- \eqref{BC_2} are converted to:
\begin{equation}
\label{q_0n}
\tilde w(\tilde t,0)\tilde v(\tilde t,0)=\tilde q_0(\tilde t),\quad \tilde w (\tilde t,1)=0.
\end{equation}

\noindent
The initial conditions are now:
\begin{equation}
\label{IC_n}
L(0)=\frac{l_{\diamond}}{l_*}, \quad \tilde w(0,\tilde x)=\tilde w_*(\tilde x)\equiv\frac{1}{l_*}w_\diamond(l_\diamond\tilde x), \quad \tilde x \in [0,1].
\end{equation}

\noindent
The transformation of the global fluid balance equation \eqref{global_balance} gives:
\begin{equation}
\label{global_balance_n}
\int_0^1\left[L(\tilde t)\tilde w(\tilde t, \tilde x)-L(0)w_*(\tilde x)\right]d \tilde x-\int_0^{\tilde t}\tilde q_0(\tilde t)d\tilde t+\int_0^{\tilde t}L(\tilde t)\int_0^1 \tilde q_l(\tilde t,\tilde x)d \tilde x d\tilde t=0,
\end{equation}
Note that relation \eqref{global_balance_n} is valid under the assumption that the process of hydraulic fracturing is monotonous ($L'(t)>0$).

\vspace{3mm}

\noindent
{\sc Remark 1.} Note that the normalized boundary conditions (compare  \eqref{BC_2}$_2$, \eqref{BC_3})
\begin{equation}
\label{BC_3a}
\tilde q(\tilde t,1)=0, \quad \frac{\partial \tilde w}{\partial \tilde x}(\tilde t,0)=0
\end{equation}
are satisfied automatically and no longer need to be enforced. However, when appropriate, they can be implemented in the code.

For simplicity, from now on, we omit the "$\sim$" symbol for all normalized variables and parameters, and consider
respective dimensionless values only.

\section{Tip asymptotics and the crack propagation speed}

In \cite{Kusmierczyk,solver_calkowy} it was demonstrated that, in order to utilize the speed equation (\ref{SE_n}) efficiently, one needs to properly employ the tip asymptotics.
This is the key point of the proposed method: such an approach makes it possible to evaluate the crack propagation speed without the technical difficulties discussed in \cite{Det_Peirce_2014} and uncertainties related to dividing two infinitesimally small values near the crack tip (the fluid flow rate and the crack opening).
We shall show that there is a unique relationship between the crack length
and the multipliers of one (or two) leading term(s) of the solution's tip asymptotics.
This relationship has a universal form and can be used regardless of the elasticity operator or crack propagation regime.
The analysis presented below is nothing but a direct extension (for the KGD model) of the approach introduced and verified in \cite{Kusmierczyk,solver_calkowy}.

It has been proved that the crack aperture in the vicinity of the fracture tip can be expressed as (see \cite{Gar_Det_Ad,Kusmierczyk} and the references therein,
where most of the information can be found for at least the two leading terms):
\begin{equation}
\label{w_roz}
w(t,x)=w_0(t)(1-x^m)^{\alpha_0}+w_1(t)(1-x^m)^{\alpha_1}+O\Big((1-x^m)^{\alpha_2}\ln^\kappa(1-x^m)\Big),\quad x \to 1,
\end{equation}
where powers $\alpha_i$ are given in Table~\ref{T1} for the respective hydraulic fracture models. In the case of non-local elasticity, it is more convenient to base the asymptotic representation on terms $(1-x^2)$. For the PKN and the fluid driven KGD models $\kappa=0$, while in general for the toughness driven variant of KGD: $\kappa=1$. The powers, starting from the second one (the third one in toughness driven KGD model), were taken for the case with leak-off vanishing at the crack tip at least as fast as for the crack opening: $\eta \geq \alpha_0$ (compare \eqref{q_l}). However, the algorithm proposed in the paper is applicable to any other permissible leak-off regime - see (\ref{leak_off_assumption}) and the discussions there. In such a case the table is to be modified, e.g respective data for the PKN model can be found in \cite{Kusmierczyk}.

\noindent
The asymptotic behaviour (\ref{w_roz}) guarantees that the condition (\ref{v_def_1}) is satisfied near the crack tip.
As a consequence, the asymptotics of particle velocity yields:
\begin{equation}
\label{v_rozw}
v(t,x)=v_0(t)+v_1(t)(1-x^m)^{\beta_1}+O\Big((1-x^m)^{\beta_2}\Big ),\quad x \to 1.
\end{equation}
The values of $\beta_i$ ($0<\beta_1<\beta_2$) are collected in Table~\ref{T1} for various models in the case of impermeable rock.
Note that
$v_0(t)=v(t,1)$, is bounded and equal to the crack propagation speed:
\begin{equation}
\label{v_0}
 v_0(t)=\frac{dL}{dt}=-\frac{1}{L}\left[w^2\frac{\partial p}{\partial x}\right]_{x=1} < \infty.
 \end{equation}

\begin{table}[h]
\vspace{3mm}
\begin{center}
\begin{tabular}{|c|c|c|c|c|c|c|c|c|c|c|}
  \hline
    HF model & $m$ & $\kappa$ & $\alpha_0 $ &$\alpha_1$& $\alpha_2$ &$\beta_1$& $\beta_2$ &$\zeta_0$ &$\zeta_1$ &$\zeta_2$\\
  \hline\hline
  PKN & 1 & 0 &1/3 &4/3 &7/3 &1 &2 &1 &2 &3 \\[0.5mm]
  \hline
  KGD (fluid driven)&2 &0 &2/3 &5/3& 8/3& 1&2&1 &2&3 \\[0.5mm]
  \hline
  KGD (toughness driven) & 2 & 1  & 1/2 &1&3/2&1&3/2&1&3/2&2 \\[0.5mm]
  \hline
\end{tabular}
\end{center}
\vspace{-2mm}

\caption{The values of  basic constants involved in the asymptotic expansions of $w$, $v$ and $\phi$ for the zero leak-off case.}
\label{T1}
\end{table}


It can be checked that, for all considered elasticity equations and fracture propagation regimes,
the limiting value of the right-hand side of equation (\ref{v_0}) is defined by no more than the first two terms of the
asymptotic expansion for the crack opening (\ref{w_roz}). Indeed,
let us adopt the following symbolic representation:
\begin{equation}
\label{C_gen}
\lim_{x\to1}w^{2}\frac{\partial}{\partial
 x}{\cal A}w=-C_{{\cal A}}\frac{{\cal L}(w)}{L^{m-1}}<0,
\end{equation}
where ${\cal L}(w)$ is a functional on the fracture aperture $w$,
related to the form of the elasticity operator, while $C_{{\cal A}}$ is a known constant.
In other words, the formula
\begin{equation}
\label{v_0_univ}
v_0=\frac{1}{L^m} C_{{\cal A}}{\cal L}(w),\quad
\end{equation}
is a universal one (valid for all elasticity formulations) and
constitutes a relation between the crack propagation speed and the multiplier(s) of the leading term(s) of the crack opening tip asymptotics.

The values of the constants $C_{{\cal A}}$ and the functionals ${\cal L}(w)$ for respective models are:
\begin{itemize}
\item{PKN model}
\begin{equation}
\label{LC_PKN}
C_{{\cal A}}=\frac{1}{3},\quad {\cal L}(w)=w_0^3,
\end{equation}
\item{KGD model - fluid driven regime}
\begin{equation}
\label{LC_KGD_fluid}
C_{{\cal A}}=\frac{2}{9\sqrt{3}} ,\quad {\cal L}(w)=w_0^3,
\end{equation}
\item{KGD model - toughness driven regime}
\begin{equation}
\label{LC_KGD_toughness}
C_{{\cal A}}=\frac{1}{\pi},\quad {\cal L}(w)=w_0^2w_1.
\end{equation}
\end{itemize}

\noindent
Note that, when using the asymptotic expansion \eqref{w_roz} based  on arguments $(1-x)$ instead of $(1-x^2)$, one needs to modify relations  \eqref{LC_KGD_fluid}-\eqref{LC_KGD_toughness}.

Taking into account that $v_0(t)=L'(t)$, equation (\ref{v_0_univ}) can be directly integrated to compute the crack length:
\begin{equation}
\label{L_int_PKN}
L(t)=\left[L^{m+1}(0)+(m+1)C_{\cal A}\int_0^t {\cal L}(w) d\tau \right]^\frac{1}{m+1}.
\end{equation}
This universal formula in turn, sets a nonlocal relation between the crack length and the leading term(s) of $w(t,x)$ asymptotic expansion \eqref{w_roz}.

For the toughness driven regime of KGD model the following condition is satisfied:
\begin{equation}
\label{w_0_KI}
w_0(t)=\frac{4}{\sqrt{\pi}}K_I\sqrt{L(t)}.
\end{equation}
As a result, a slightly different approach to define the crack length at any time, $t$, can be applied. Namely,
by combining \eqref{w_0_KI} with \eqref{C_gen}, \eqref{LC_KGD_toughness} and \eqref{v_0} one obtains an alternative formula:
\begin{equation}
\label{L_int_KGD}
L(t)=\left[L^2(0)+ C_{I}K_I^2\int_0^t w_1(\tau) d\tau \right]^\frac{1}{2},\quad  C_{I}=\frac{32}{\pi^2}.
\end{equation}

Moreover, the formula \eqref{w_0_KI} itself can be directly used to determine the fracture length:
\begin{equation}
\label{L_tough_w0}
\quad L(t)= \frac{\pi}{16K^2_I}w^2_0(t).
\end{equation}
Thus, for the toughness driven KGD model it is possible to use different strategies for the computations.
Namely, for small toughness is is natural to use the representation (\ref{L_int_KGD}), for large toughness -- \eqref{L_tough_w0},
while a strategy based on two first terms in the asymptotic expansion (\ref{w_roz}) and formulae
\eqref{L_int_PKN}, \eqref{LC_KGD_toughness} can be adopted for any $K_I$.

\vspace{3mm}

\noindent
{\sc Remark 2.} It follows from the foregoing analysis that numerical evaluation of the leading coefficient (or two first multipliers for the toughness driven KGD model) in the asymptotic representation \eqref{w_roz}
 plays a crucial role in computing the crack length. This will be discussed in detail in Section 6.

\vspace{3mm}

\noindent
\noindent
{\sc Remark 3.} There is another way to determine the crack length by employing the balance condition (\ref{global_balance_n}). Indeed, this relation
represents the Volterra integral equation of the second kind with respect to the crack length $L(t)$. Such an equation has always a unique solution and there are effective numerical methods
 to solve it \citep{Linz}. The kernel of the equation, which is an indefinite integral itself, should be, generally speaking,
 modified at every time step and thus this approach is not cost effective.  However, in case of impermeable rock ($q_l=0$) the integral equation degenerates and the crack length can be found
explicitly regardless of the elasticity operator and fracture propagation regime.

\section{Problem reformulation in terms of reduced particle velocity}
\label{modified}

In this section we reformulate the problem by replacing the particle velocity with a new dependent variable, the \emph{reduced particle velocity}, and demonstrate its advantages.

\noindent
Let us introduce a new dependent variable called, from now on, the {\it reduced particle velocity}:
\begin{equation}
\label{Phi}
{\phi}(t,x)=v(t,x)-xv_0(t).
\end{equation}
Its asymptotic behaviour can be described qualitatively as (compare with Table~\ref{T1}):
\begin{equation}
\label{fi_rozw}
\phi(t,x)=\phi_0(t)(1-x^m)^{\zeta_0}+\phi_1(t)(1-x^m)^{\zeta_1}+O\Big((1-x^m)^{\zeta_2}\Big),\quad x \to 1.
\end{equation}
In numerical implementation $\phi$ possess all the advantages of the particle velocity, $v$,
allowing simultaneously to set a new boundary condition
\begin{equation}
\label{phi_tip}
\phi(t,1)=0,
\end{equation}
instead of $v(t,1)=v_0(t)$, with the unknown speed of the crack, $v_0(t)$. The latter is not easy to be used as it is directly defined by the product of two terms where one tends to zero while the other to infinity (Compare (\ref{v_0}), (\ref{lim_der}) and (\ref{q_0n})$_2$).
Qualitatively the new variable, $\phi$, exhibits similar asymptotic behaviour to the crack opening - it tends to zero at the fracture tip. Thus, we can define a respective boundary condition for this variable (see (\ref{phi_tip})).

Combining \eqref{Phi} with \eqref{v_norm}, and substituting the result into the continuity equation \eqref{cont_norm}, one obtains its modified form:
\begin{equation}
\label{Phi_cont}
\frac{\partial w}{\partial t}+\frac{1}{L}\frac{\partial }{\partial x}(w \phi)+\frac{v_0}{L}w+q_l=0.
\end{equation}

\noindent
The boundary condition \eqref{q_0n} can be now replaced by:
\begin{equation}
\label{q0_phi}
w(t,0)\phi(t,0)=q_0(t),
\end{equation}
while the crack tip conditions are \eqref{q_0n}$_2$  and \eqref{phi_tip}.

Note, that the following solvability condition for the equation (\ref{Phi_cont}) should be satisfied:
\begin{equation}
\label{Phi_solv}
\int_0^1\left(\frac{\partial w}{\partial t}(t,x)+q_l(t,x)\right)dx+\frac{v_0(t)}{L(t)}\int_0^1w(t,x)dx=\frac{1}{L(t)}q_0(t).
\end{equation}
It constitutes a local equivalent of the global balance condition \eqref{global_balance_n} and, since $w>0$ for any $x\in[0,1)$, it allows one to uniquely define the value of the crack velocity, $v_0$, at every time step.
As a result, equation (\ref{Phi_cont}) always has a unique solution with respect to the \emph{reduced particle velocity}, $\phi$.

From (\ref{v_norm}), the pressure derivative can be expressed as:
\begin{equation}
\label{p_prim}
\frac{\partial p}{\partial x}=-\frac{L}{w^2}\left(\phi+xv_0\right).
\end{equation}
In the case of PKN model ($p=w$), equation (\ref{p_prim}) can be transformed to a functional equation with respect to the crack opening, $w$,
which can be written in a symbolic manner:
\begin{equation}
\label{p_prim_1}
w=B_w(\phi,L,w,v_0).
\end{equation}
Here the right-hand side is be represented by one of two equivalent relationships:
\begin{equation}
\label{p_prim_1a}
w(t,x)=L(t)\int_x^1\frac{\phi(t,\eta)+\eta v_0(t)}{w^2(t,\eta)}d\eta.
\end{equation}
or
\begin{equation}
\label{w_B1_PKN}
w(t,x)=\left[3L(t)\int_x^1(\phi(t,\eta)+\eta v_0(t))d\eta\right]^{1/3}.
\end{equation}
The latter representation corresponds to the {\it proper variable} approach discussed in \cite{Linkov_3,M_W_L} and is usually  more effective in computations. In particular, it does not require iterations in numerical computing. However, in the case of the P3D model,  equation (\ref{p_prim_1}) takes the form
\begin{equation}
\label{w_B1_P3D}
w(t,x)=\frac{L(t)}{F(w(t,x),x)}\int_x^1\frac{\phi(t,\eta)+\eta v_0(t)}{w^2(t,\eta)}d\eta,
\end{equation}
where $F(w,x)$ is the proportionality coefficient in the elasticity relation, $p=F(w,x)w$, (see \cite{Warpinski,Brisbane}). For this single reason, we stay with both representations
(\ref{p_prim_1a}) and (\ref{w_B1_PKN}). Respective implications for the numerics shall be discussed later on.

In case of the KGD model, equation (\ref{p_prim}) should be combined with the transformed form of elasticity relation \eqref{inv_KGD_n_1} to eliminate the pressure gradient, $p'_x$, and to find the crack opening, $w$, again in form the (\ref{p_prim_1})
\begin{equation}
\label{w_to_finda}
w(t,x)= \frac{4}{\pi}L^2( t)\int_0^1 \frac{\phi(t,\eta)+\eta v_0(t)}{w^2(t,\eta)}
K(\eta,x) d\eta+4\sqrt{\frac{L( t)}{\pi}} K_I\sqrt{1- x^2},
\end{equation}
It is obvious that the boundary condition \eqref{q_0n}$_2$ holds automatically. It can be proved that for the KGD model, also condition  \eqref{BC_3a}$_2$ is fulfilled by definition.

As a  result, the basic system of equations now utilizes two dependent variables: the crack opening, $w(t,x)$, and the reduced particle velocity, $\phi(t,x)$.
We will be looking for a solution to the lubrication equation \eqref{Phi_cont}, under boundary conditions  \eqref{phi_tip}, \eqref{q0_phi},
initial conditions \eqref{IC_n} and the speed equation \eqref{v_0} written in one of the transformed forms (\ref{L_int_PKN}) -- (\ref{L_tough_w0}).
Finally, the pressure in the KGD model can be computed by integrating equation \eqref{p_prim}:
\begin{equation}
\label{p_prim_2}
p(t,x)=-L(t)\int_0^x\frac{\phi(t,\eta)+\eta v_0(t)}{w^2(t,\eta)}d\eta+p_*(t).
\end{equation}
where the constant $p_*$ is defined from (\ref{A_norm_PKN_0}) and (\ref{A_norm_KGD})
\begin{equation}
\label{p_0}
p_*(t)= -\frac{1}{2\pi L(t)} \int_0^1\frac{\partial  w( t,\eta)}{\partial \eta} \frac{d\eta}{\eta}.
\end{equation}
Note that the integral converges due to condition (\ref{BC_3a})$_2$. The physical interpretation of \eqref{p_0} is that the value of particle velocity at the fracture inlet is bounded.

\section{Self-similar solution}

Let us search for a solution of the problem described in the previous section in the following manner:
\begin{equation}
\label{self_norm}
w(t,x)=\psi(t)\hat w(x), \quad p(t,x)=\frac{\psi(t)}{L^{m-1}(t)}\hat p(x),
\end{equation}
\[
q(t,x)=\frac{\psi^4(t)}{L^m(t)}\hat q(x), \quad
v(t,x)=\frac{\psi^3(t)}{L^m(t)}\hat v(x), \quad
\phi(t,x)=\frac{\psi^3(t)}{L^m(t)}\hat \phi(x).
\]

As shown in Appendix A, such a separation of variables enables one to reduce the problem to the time-independent form
in the case when $\psi(t)$ is a power law or exponential function of time. Such a formulation will be called henceforth the self-similar formulation.
Respective spatial components of the solution (depending exclusively on  $x$) are marked by 'hat'-symbol.

Below we demonstrate the basic assumptions and features of the universal algorithm, considering first the self-similar formulation of the problem.

\subsection{Problem formulation}

As follows from the self-similar formulation given in Appendix A, a set of governing equations can be
written in general form as:
\begin{itemize}
\item{equation determining the reduced velocity following from \eqref{Phi_cont} (self-similar equivalents: \eqref{cont_exp} or \eqref{cont_pow})}
\begin{equation}\label{ss_1}
\hat \phi(x)=\frac{{\cal L}(\hat w)}{\hat w(x)}\int_x^1\Big(\chi \hat w(\xi)+ \varkappa \hat q_l(\xi)\Big)d\xi,
\end{equation}
\item{equation allowing the computation of the crack opening, which follows from \eqref{p_prim}} and the respective inverse elasticity operator (self-similar versions: \eqref{Am2_eq},\eqref{Am1_ss})
\begin{equation}\label{ss_2}
\hat  w ={\cal B}_w \left(\frac{1}{\hat w^2}\Big(\hat \phi +x{\cal L}(\hat w)C_{\cal A}\Big)\right), \quad
\end{equation}

\item
boundary conditions:
\begin{equation}\label{ss_bound}
\hat w(0)\hat \phi(0)=\hat q_0,\quad \hat w(1)=0,\quad \hat \phi(1)=0,
\end{equation}
\item{solvability condition allowing one to determine the value of the parameter $\hat v_0={\cal L}(\hat w)$, which follows immediately from (\ref{ss_1}) and (\ref{ss_bound}). It is
equivalent to the global fluid balance equation \eqref{global_balance_n} (self-similar
counterparts: \eqref{balance_exp} or \eqref{balance_pow}) }
\begin{equation}\label{fluid_gen}
\frac{1}{{\cal L}(\hat w)}\hat q_0-\chi \int_0^1 \hat w(x)dx- \varkappa \int_0^1\hat q_l(x)dx=0.
\end{equation}
\end{itemize}
For the PKN model operator ${\cal B}_w$ from \eqref{ss_2} has one of the following alternative forms:
\begin{equation}
\label{p_prim_1aaa}
{\cal B}_w\hat w(x)=\int_x^1\frac{\hat\phi(\eta)+\eta {\cal L}(\hat w)C_{\cal A}}{\hat w^2(\eta)}d\eta,
\end{equation}
or
\begin{equation}
\label{w_B1_PKNaa}
{\cal B}_w\hat w(x)=\left[3\int_x^1(\hat \phi(\eta)+\eta {\cal L}(\hat w)C_{\cal A})d\eta\right]^{1/3}.
\end{equation}
For the KGD variant of the problem it is directly defined by relation \eqref{Am2_eq} which is a self-similar equivalent of \eqref{inv_KGD_n_1}:
\begin{equation}
\label{w_to_findaaa}
{\cal B}_w\hat w(x)= \frac{4}{\pi}\int_0^1 \frac{\hat\phi(\eta)+\eta {\cal L}(\hat w)C_{\cal A}}{\hat w^2(\eta)}
K(\eta,x) d\eta+\frac{4}{\sqrt{\pi}} \hat K_I\sqrt{1- x^2},
\end{equation}
where the kernel $K(\eta,x)$ is defined in \eqref{kernel_KGD_inv}.

Note that the general form of the speed equation \eqref{v_0} and the additional boundary condition \eqref{BC_3a}$_2$ for the KGD model (\eqref{exp_ss_BC} in self-similar formulation) are satisfied by the system (\ref{ss_1}) -- (\ref{fluid_gen}) automatically.

 Positive constants $\chi$ and $\varkappa$ from equation (\ref{ss_1}) depend on
the type of the function $\psi$ defining the self-similar solution and given by \eqref{cont_pow} and \eqref{cont_exp},
respectively. All the values of different parameters used in the above equations are collected in Tables~\ref{T1}, \ref{T2}.

\begin{table}[h]
\begin{center}

\begin{tabular}{|c|c|c|}
  \hline
    Type of the self-similar law & $\chi$ & $\varkappa$ \\
  \hline\hline
  $\psi(t)=e^{\gamma t}$ &$ C_{\cal A}\frac{m+4}{3} {}^{}$ & $ C_{\cal A}\frac{m+1}{3\gamma}$ \\
  \hline
  $\psi(t)=(1+t)^{\gamma}$ & $ C_{\cal A}\left[\frac{(m+1)\gamma}{3\gamma+1}+1\right] $& $ C_{\cal A}\frac{m+1}{3\gamma+1}$ \\

  \hline
\end{tabular}
\end{center}
\caption{Values of the auxiliary parameters in the self-similar formulations reflecting different time dependent behaviours.}
\label{T2}
\end{table}

\subsection{Computational algorithm for the self-similar solution}

The solution of the self-similar problem formulated above is sought in the framework of the {\it universal algorithm}. The universality refers to the fact, that only some parameters in respective blocks should be changed to adjust the solver to work with different variants of the problem (PKN, KGD in both considered regimes).
The algorithm consists of the following iterative steps:
\begin{itemize}
\item
In the first stage we assume some initial approximation of the crack opening, $\hat w=\hat w^{(i-1)} $. Equations \eqref{fluid_gen} and \eqref{ss_1} are utilized to compute ${\cal L}^{(i)}(\hat w)$ and the reduced velocity $\hat \phi^{(i)}$. In general  \eqref{fluid_gen} yields ${\cal L}(\hat w)$ which, when substituted into \eqref{ss_1}, enables the integration of the latter to obtain $\hat \phi$. In the manner of the so-called $\varepsilon$-regularization technique \citep{Linkov_1,Kusmierczyk}
the integration is carried out over truncated spatial interval $x\in[0,1-\varepsilon]$, where $\varepsilon$ is a small parameter (a comprehensive description of this form of the $\varepsilon$ - regularization technique, together with methods for its implementation, can be found in \cite{Kusmierczyk}). The boundary condition \eqref{ss_bound}$_3$ is replaced by the condition resulting from the asymptotics \eqref{fi_rozw}, specified at $x=1-\varepsilon$. The regularized boundary condition is introduced in the form:
\begin{equation}
\label{fi_reg}
\hat \phi_N=s_1(\zeta_0,\zeta_1)\hat \phi_{N-1}+s_2(\zeta_0,\zeta_1)\hat \phi_{N-2},
\end{equation}
where the subscripts of $\hat \phi$ refer to the indices of nodal points of the spatial mesh (containing $N$ nodes). The values of multipliers $s_{1(2)}(\zeta_0,\zeta_1)$ depend on the particular asymptotic behaviour of the function $\hat \phi$ (see \eqref{fi_rozw} and Table~\ref{T1}) and the applied spatial meshing.\\  As a result, function $\hat \phi^{(i)}(x)$ and the constant
${\cal L}^{(i,1)}(\hat w)$ computed at this stage satisfy, together with predefined $\hat w^{(i-1)}$: i) the fluid balance equation \eqref{fluid_gen}, ii) the continuity equation \eqref{ss_1}, iii) the regularized boundary condition for $\phi$ \eqref{fi_reg}  (equivalent to \eqref{ss_bound}$_3$), iv) the influx boundary condition \eqref{ss_bound}$_1$ - indirectly, through the fluid balance equation.

\item
At the second stage of each iterative loop, the values of $\hat \phi^{(i)}$ obtained previously are utilized to compute the next iteration $\hat w^{(i)}$ from \eqref{ss_2}.
Note, that by the properties of the operator ${\cal B}_w$ corresponding to the KGD model, condition \eqref{sym_ss} is satisfied automatically.

While computing respective integral operators ${\cal B}_w$, it is crucial to preserve appropriate asymptotic behaviour of the integrands, resulting from
\eqref{w_roz}, \eqref{fi_rozw}. Moreover, at this stage ${\cal L}^{(i,2)}(\hat w)$ is considered a natural regularization parameter, chosen to satisfy
the influx boundary condition \eqref{ss_bound}$_1$. Hence, $\hat w^{(i)}$ computed at this stage satisfies respective elasticity relation
and boundary conditions: \eqref{ss_bound}$_2$ through the imposed asymptotics
\begin{equation}
\label{w_reg}
\hat w_N=s_1(\alpha_0,\alpha_1)\hat w_{N-1}+s_2(\alpha_0,\alpha_1)\hat w_{N-2},
\end{equation}
and \eqref{ss_bound}$_3$.
\item
The aforementioned two stages of the iterative loop are repeated until all components of the solutions $\hat \phi$, $\hat w$ and ${\cal L}(\hat w)$ have converged with the prescribed tolerances.
\end{itemize}

{\sc Remark 4}. As  shown in \cite{Kusmierczyk}, the
relationships  (\ref{fi_reg}) and (\ref{w_reg}) allow one to determine numerically the multipliers of the first two leading terms of the
asymptotics of $\hat\phi$ and $\hat w$.

{\sc Remark 5}. Similarly, as shown in \cite{solver_calkowy}, the performance of the algorithm improves significantly when instead of the dependent variables $\hat w$ and $\hat \phi$
one uses the difference between them and their leading asymptotic terms:
\[
\Delta \hat w=\hat w-w_0(1-x)^{\alpha_0},\quad \Delta \hat \phi=\hat \phi -\phi_0(1-x)^{\zeta_0}.
\]
Then the leading asymptotic terms in the left and right-hand sides of the equations (\ref{ss_1})
and (\ref{ss_2}) are canceled analytically and the functions $\Delta \hat w$ and $\Delta \hat \phi$ are computed in the iterative process.
Moreover, while searching for the regularization parameter  ${\cal L}(\hat w) $ we take into account  its relation to the respective coefficients in the
asymptotic expansion of the solution (compare
(\ref{LC_PKN}) -- (\ref{LC_KGD_toughness})). This, in turn, leads to nonlinear equations solved with the Newton �- Raphson method. Finally, the qualitative asymptotic behaviour of the new dependent variables
$\Delta \hat w$ and $\Delta \hat \phi$  is also known in advance and the respective exponents should be appropriately adopted in the $\varepsilon$-regularization technique.

{\sc Remark 6.} When computing the integral operators \eqref{ss_1} -- \eqref{ss_2}, we use the test (weight) functions coinciding with
the leading asymptotic terms of the integrands. This provides better accuracy and efficiency when integrating.

{\sc Remark 7.} Note, that, due to the modular algorithm architecture,
the subroutine for computing the crack opening can be easily replaced in accordance with the chosen elasticity operator.

\subsection{Algorithm performance in the self-similar formulation}

Since the self-similar variant of the algorithm constitutes
the integral part of the algorithm for the transient problem,
it is the first major step towards constructing the general numerical scheme.
In the following four subsections we investigate the properties of the
universal algorithm against several analytical benchmarks and compare the results to the data available in literature.

\subsubsection{Analysis of the algorithm - PKN model}

For the PKN model, we use two different benchmark solutions. The first of them uses formula \eqref{w_rep}
for three base functions of the type \eqref{h_PKN}. All the resulting quantities can be obtained by the way described in
Appendix B. Note that this benchmark solution, called hereafter {\it{benchmark I}}, assumes a predefined non-zero leak-off function.
The second benchmark example is taken from \cite{M_W_L} (p.7-8, eqs.(38) -- (39)). In this case there is no leak-off in the formulation.
This solution is called from now on {\it{benchmark II}}.

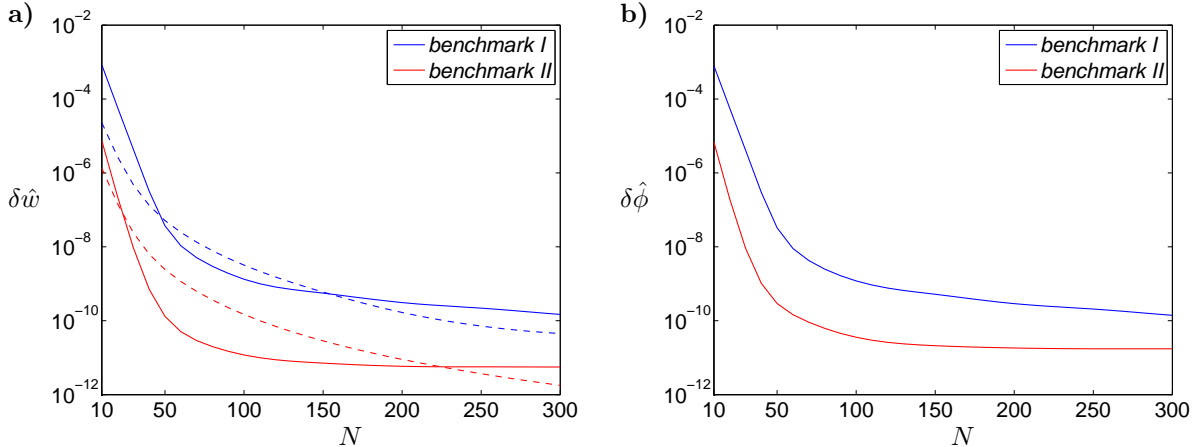
\begin{figure}[h!]

    \includegraphics [scale=0.40]{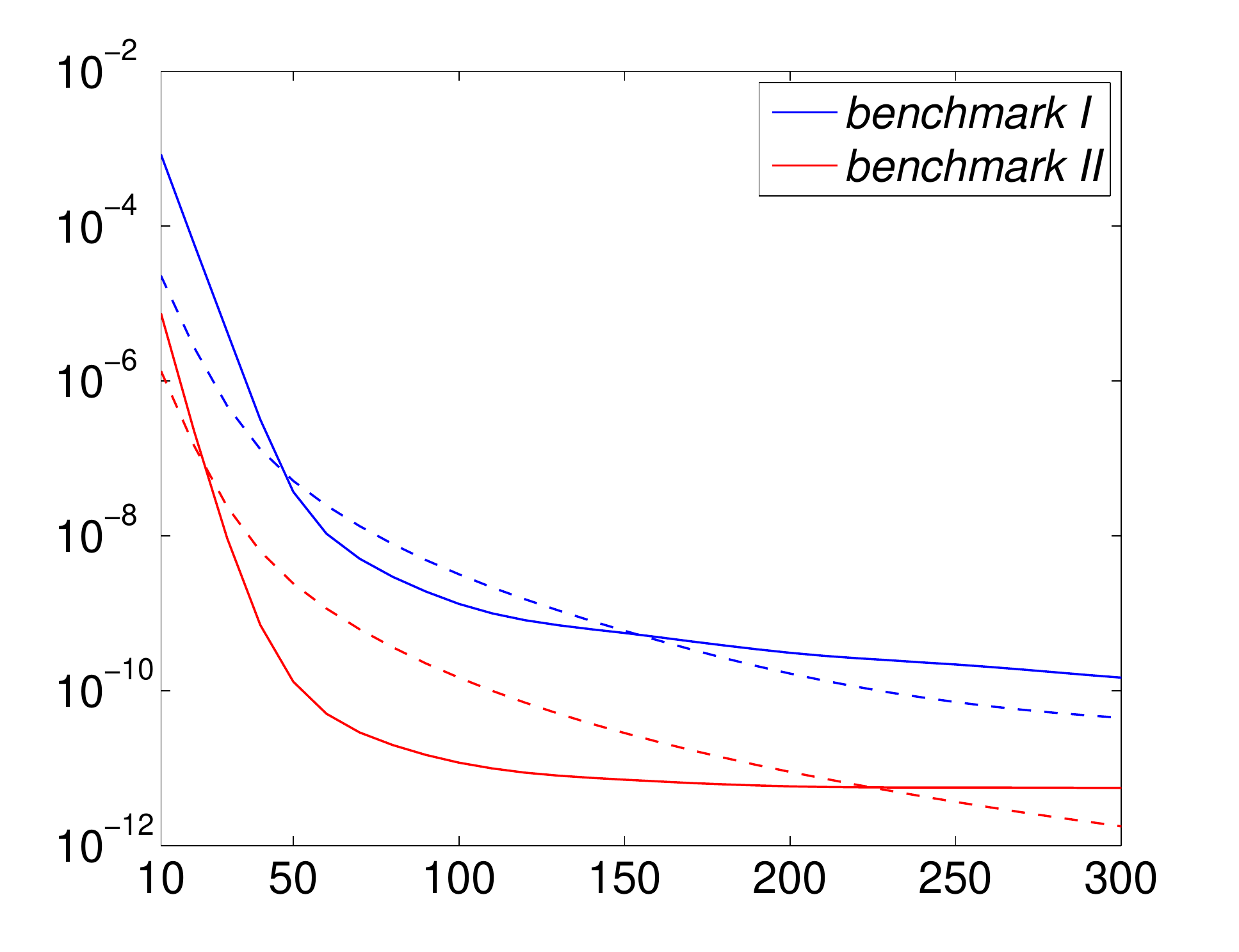}
    \put(-105,0){$N$}
    \put(-230,90){$\delta \hat w$}
    \put(-230,160){$\textbf{a)}$}
    \hspace{2mm}
    \includegraphics [scale=0.40]{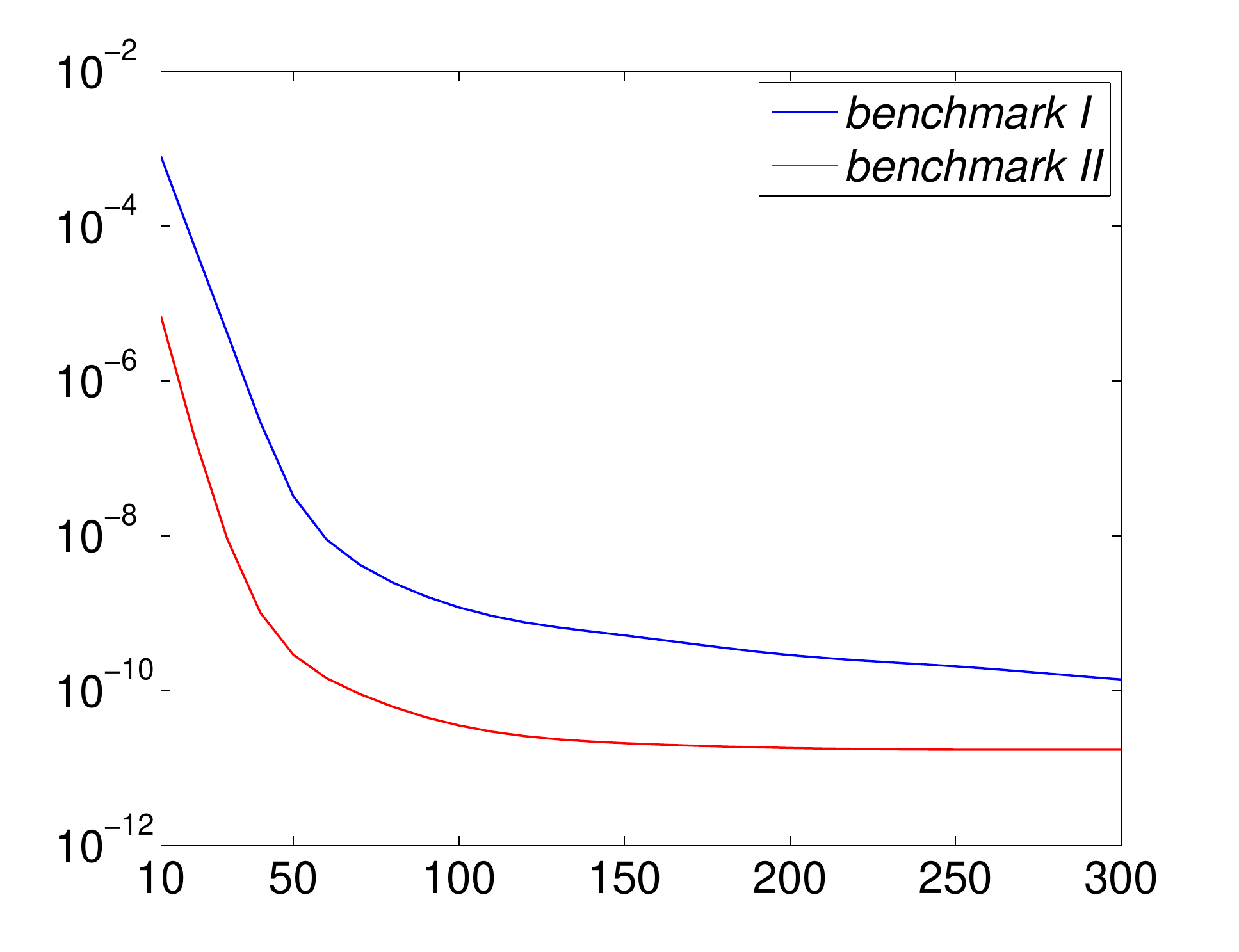}
    \put(-105,0){$N$}
    \put(-230,90){$\delta \hat \phi$}
    \put(-230,160){$\textbf{b)}$}

    \caption{\emph{PKN model.} Relative errors of the self-similar solution for: a) the crack opening $\hat w$, b) the reduced particle velocity $\hat \phi$.
    Dashed lines in a) correspond to the results obtained by integral solver described in \cite{solver_calkowy}. }

\label{bld_w_h_PKN}
\end{figure}

In the following we analyze the accuracy of computations described by two parameters: $\delta \hat w$ and $\delta \hat \phi$  - the maximal
relative errors of the crack opening and the reduced velocity, respectively. The corresponding results are presented in Fig.\ref{bld_w_h_PKN} a) and
Fig.\ref{bld_w_h_PKN} b).
These errors were estimated for a number of nodal points varying from 10 to 300, where
the spatial mesh density was increased at both sides of the interval (in a manner described in \cite{solver_calkowy}).
For comparison, we additionally show in Fig.\ref{bld_w_h_PKN} a) the accuracy of computations performed by the extremely efficient
integral solver proposed  in \cite{solver_calkowy} for the PKN model. The latter algorithm does not utilize the reduced velocity.
Its computation needs additional post-processing, and thus $\delta \hat \phi$ is not shown in this case.

The results presented in Fig.\ref{bld_w_h_PKN} demonstrate that the solution error depends on the type of benchmark. In this particular case there is a clear explanation for this fact, as {\it{benchmark II}} excludes an error introduced by the numerical integration of the leak-off function.
In the analyzed range of $N$, a clear trend of accuracy improvement with growing  mesh density can be observed. However, the error reduction becomes slower with  $N$ growth.

The comparison of the results with those for the integral solver shows that the latter can provide better accuracy for smaller $N$.
It also gives greater potential for the solution improvement for large values of $N$. On the other hand, for both benchmarks,
there exists an intermediate interval where the solver based on the universal algorithm gives better accuracy.
Summarizing, for both benchmarks the overall accuracy is extremely high, comparable with that  provided by the integral solver and much better that the level of accuracy reported in
\cite{Kovalyshen}.

{\sc Remark 8}. Computations shown in this subsection were done with the operator ${\cal B}_w$ defined in (\ref{w_B1_PKNaa}) as the respective algorithm provided better accuracy.
When using the alternative operator \eqref{p_prim_1aaa}, the accuracy was up to one order of magnitude worse for both the crack opening,  $w$, and the reduced partial velocity, $\phi$.
However, when moving forward to the transient state scheme, we observed no difference in accuracy between these approaches. The reason for this is rather clear:
the major error in the computations (and thus the accuracy limiting factor) in this case comes from the FD representation of the temporal derivative, even though we take its more accurate approximation
than that used in the standard hydrofracturing algorithms (compare \cite{Adachi-et-Al-2007}).

\subsubsection{Analysis of the algorithm - accuracy of computations for the KGD model}

In this subsection we estimate the solution accuracy for two analytical benchmarks (for the fluid and toughness driven regimes respectively)
given in Appendix B.

The benchmark solution for the fluid driven regime is based on the representation \eqref{h_KGD fluid} composed of three terms. For the toughness driven mode
we apply the basis \eqref{h_KGD tough} for four terms.   The computations for different numbers of nodal points $N$ ranging from
20 to 500 were performed. The lower limit of $N$ was set to 20 instead of 10, as the proper numerical computation of the inverse elasticity operator \eqref{Am2_eq} necessitates finer
meshing than its equivalent in the PKN model (identity operator). Again, the mesh density was increased at both ends of the spatial interval.

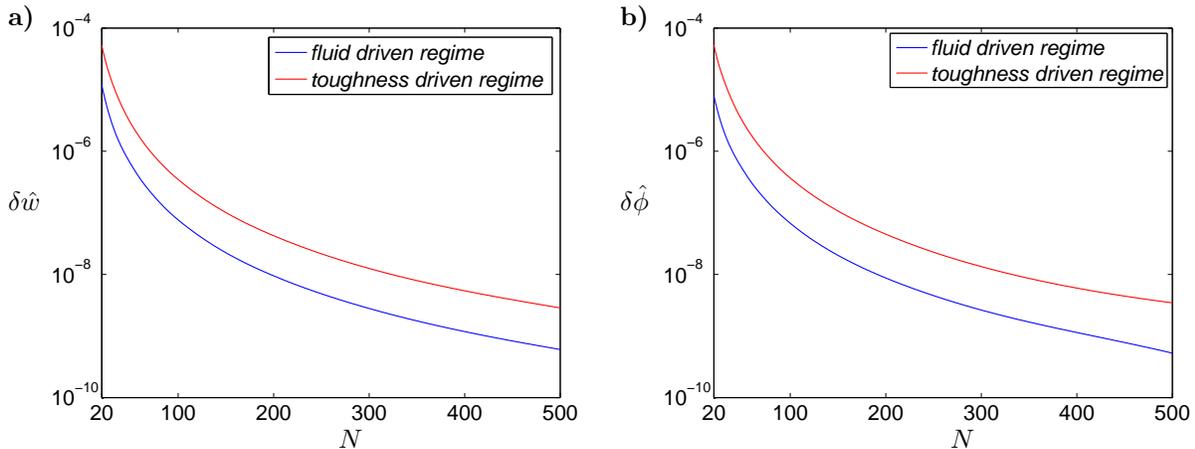
\begin{figure}[h!]

    \includegraphics [scale=0.40]{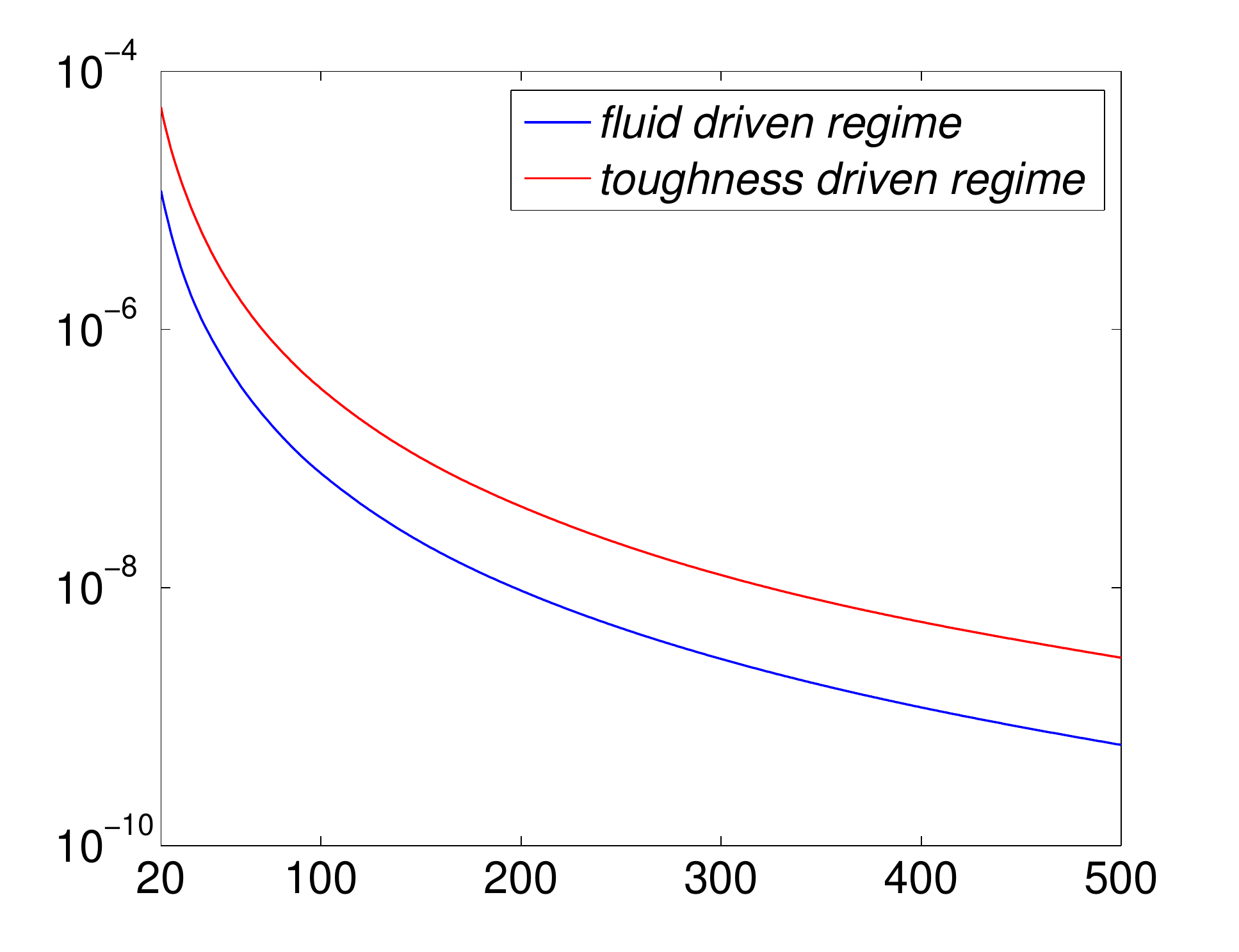}
    \put(-105,0){$N$}
    \put(-230,90){$\delta \hat w$}
    \put(-230,160){$\textbf{a)}$}
    \hspace{2mm}
    \includegraphics [scale=0.40]{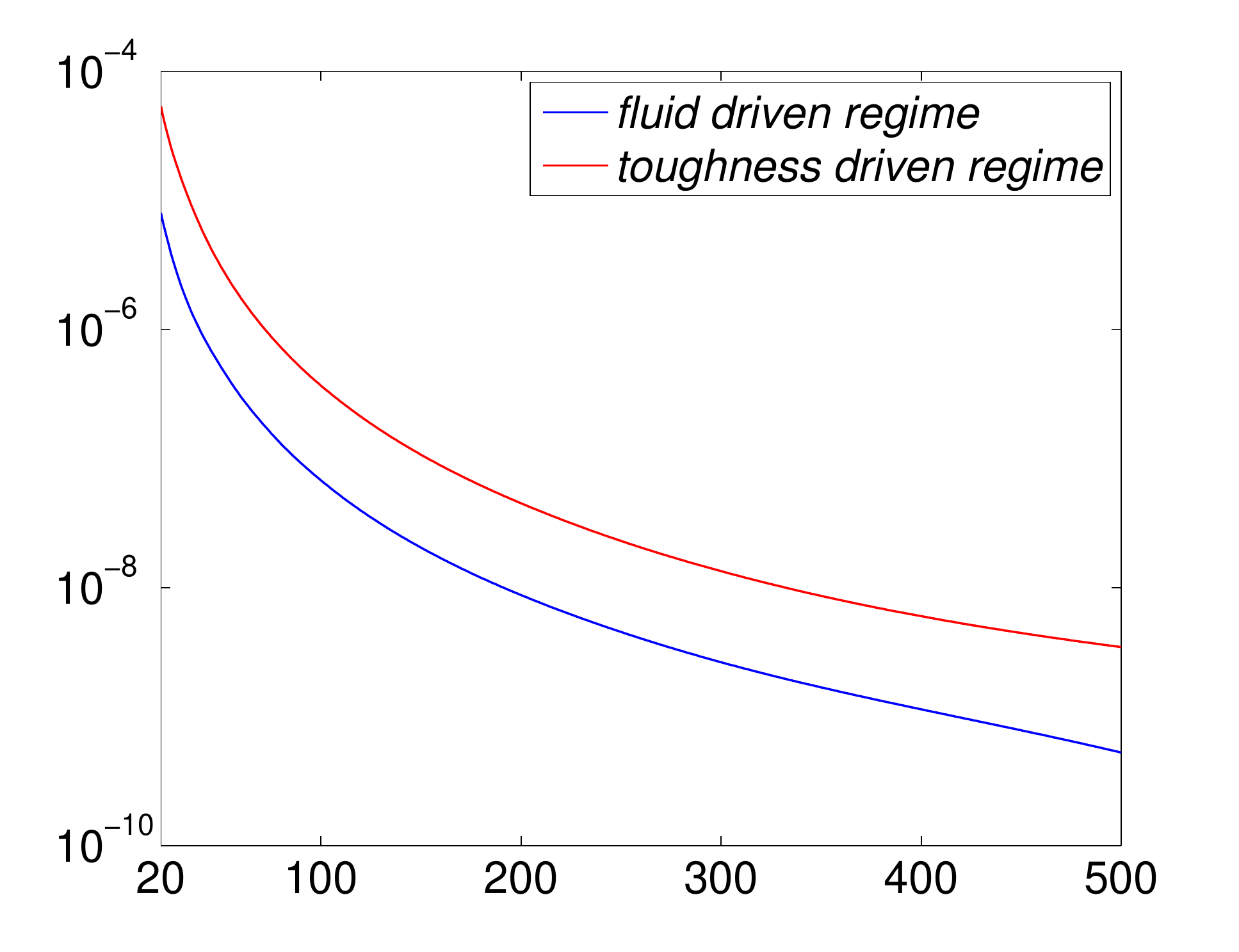}
    \put(-105,0){$N$}
    \put(-230,90){$\delta \hat \phi$}
    \put(-230,160){$\textbf{b)}$}

    \caption{\emph{KGD model.} Relative errors of the self-similar solution for: a) the crack opening $\hat w$, b) the reduced particle velocity $\hat \phi$.}

\label{bld_w_h_KGD}
\end{figure}

Similarly to the PKN model, the accuracy depends on the type of the benchmark.
 In the analyzed range of $N$ the increase of mesh density gives monotonic reduction of the error. This trend attenuates with growing $N$, however for $N<500$ it is still far away from the stabilization level.
In the benchmarks under consideration it is sufficient to take merely 20 nodal points to have the accuracy of the level $10^{-5}$ for both, the crack aperture
and the reduced velocity. For $N=100$ the error of computations does not exceed the level of $10^{-7}$ for both regimes.
Thus, the universal algorithm shows the same quality of performance in terms of the accuracy and efficiency as both the KGD and  PKN models (the latter being slightly less challenging for computations).
The toughness driven benchmark considered here does not refer to any of the boundary cases, namely neither to the small or to the large toughness regimes.  These will be discussed later on.

\subsubsection{Fluid driven KGD model for impermeable rock - comparison with other results.}

Having identified the accuracy of the algorithm, we now perform similar comparison against various classical results (both numerical and semi-analytical) available in the literature. We start with the numerical solution given in terms of series approximation in \cite{Adachi_Detournay}. The authors analyzed the fluid driven regime (KGD model)
for a number of shear-thinning fluids. The constant influx and impermeability of the rock formation ($q_l=0$) were also assumed. In our case only the
data for the newtonian fluid (n=1) will be used. For the same problem, a semi-analytical approximate solution has been recently proposed in \cite{Linkov_3}. Finally,
 one can find in \cite{Garagash_small_toughness} a simple approximations for the crack opening and the net fluid pressure, originally introduced in \cite{Adachi_PhD}.

In the following
we compare our numerical results with those given in the mentioned papers.
In order to make sure that the accuracy of our computations is at least of the order $10^{-7}$, we take a mesh composed of 300 nodal points
(compare Fig.~\ref{bld_w_h_KGD}) whose density was increased at both ends of the interval.
We compare the results in terms of: i) self-similar crack opening, ii) self-similar fluid pressure, iii) self-similar particle velocity and (iv) self-similar fluid flow rate.
The explicit formulae for the first three dependent variables are given in \cite{Linkov_3}. Although in \cite{Adachi_Detournay} there is no data for the particle velocity, it can be easily obtained through the flux and the fracture opening ($\hat v=\hat q/ \hat w$) or the fracture opening and pressure derivative ($\hat v=-\hat w^2  \hat p'$). Similarly, the lacking data for the fluid flow rate in \cite{Linkov_3} can be recreated as $\hat q=\hat w \hat v$. Unfortunately, there are no direct formulae in \cite{Garagash_small_toughness} for either the fluid flow rate or the fluid pressure derivative. Thus, aiming at a fair comparison, we do not show here results for $\hat v$ and $\hat q$ (which could theoretically be obtained by pressure differentiation).
As for the results by the universal algorithm, the particle velocity is retrieved directly from the reduced particle velocity, $\hat \phi$, while the computation
of the fluid pressure necessitates additional post-processing (integration).

\begin{figure}[h!]

    \includegraphics [scale=0.40]{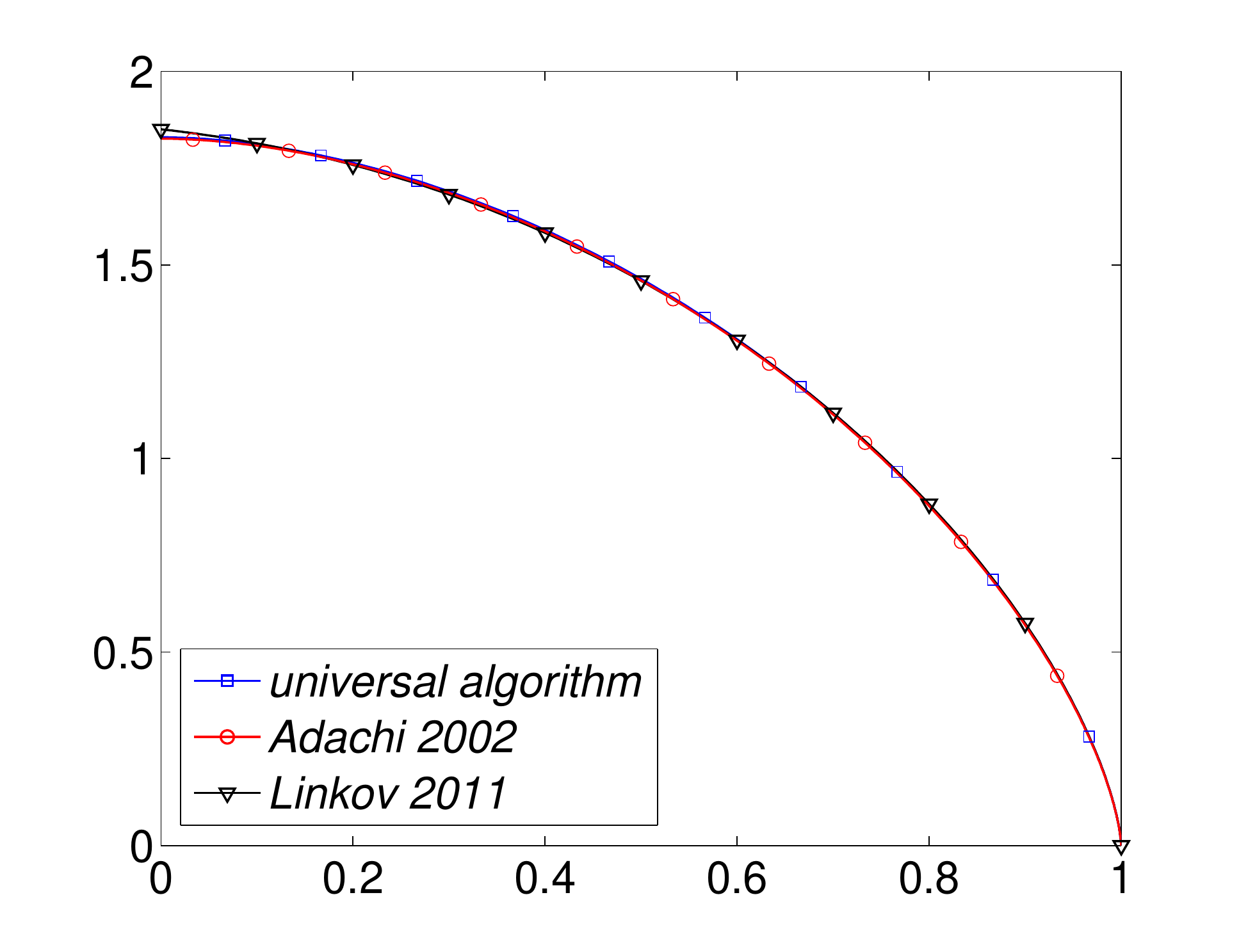}
    \put(-105,0){$x$}
    \put(-220,90){$ \hat w$}
    \put(-230,160){$\textbf{a)}$}
    \hspace{2mm}
    \includegraphics [scale=0.40]{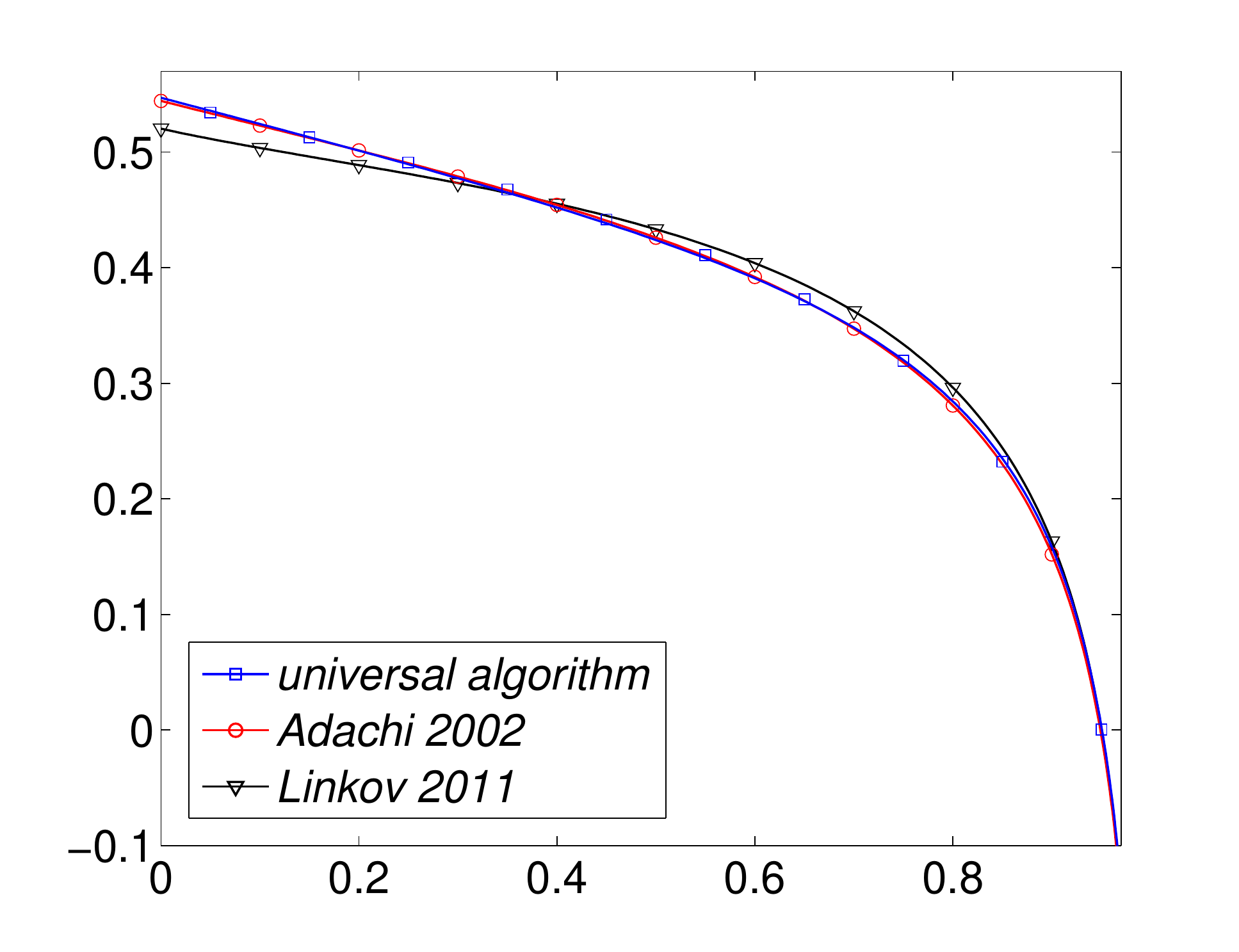}
    \put(-105,0){$x$}
    \put(-220,90){$\hat p$}
    \put(-230,160){$\textbf{b)}$}

    \caption{\emph{KGD model.} Results for the self-similar problem: a) the crack opening $\hat w$, b) the fluid pressure $\hat p$.}

\label{wyniki_Ad_1}
\end{figure}

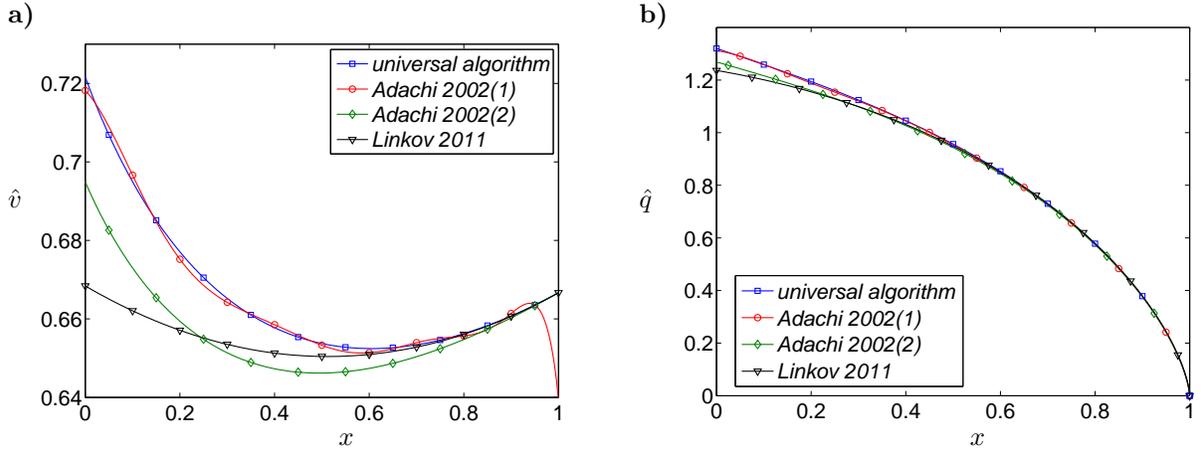
\begin{figure}[h!]
    \includegraphics [scale=0.40]{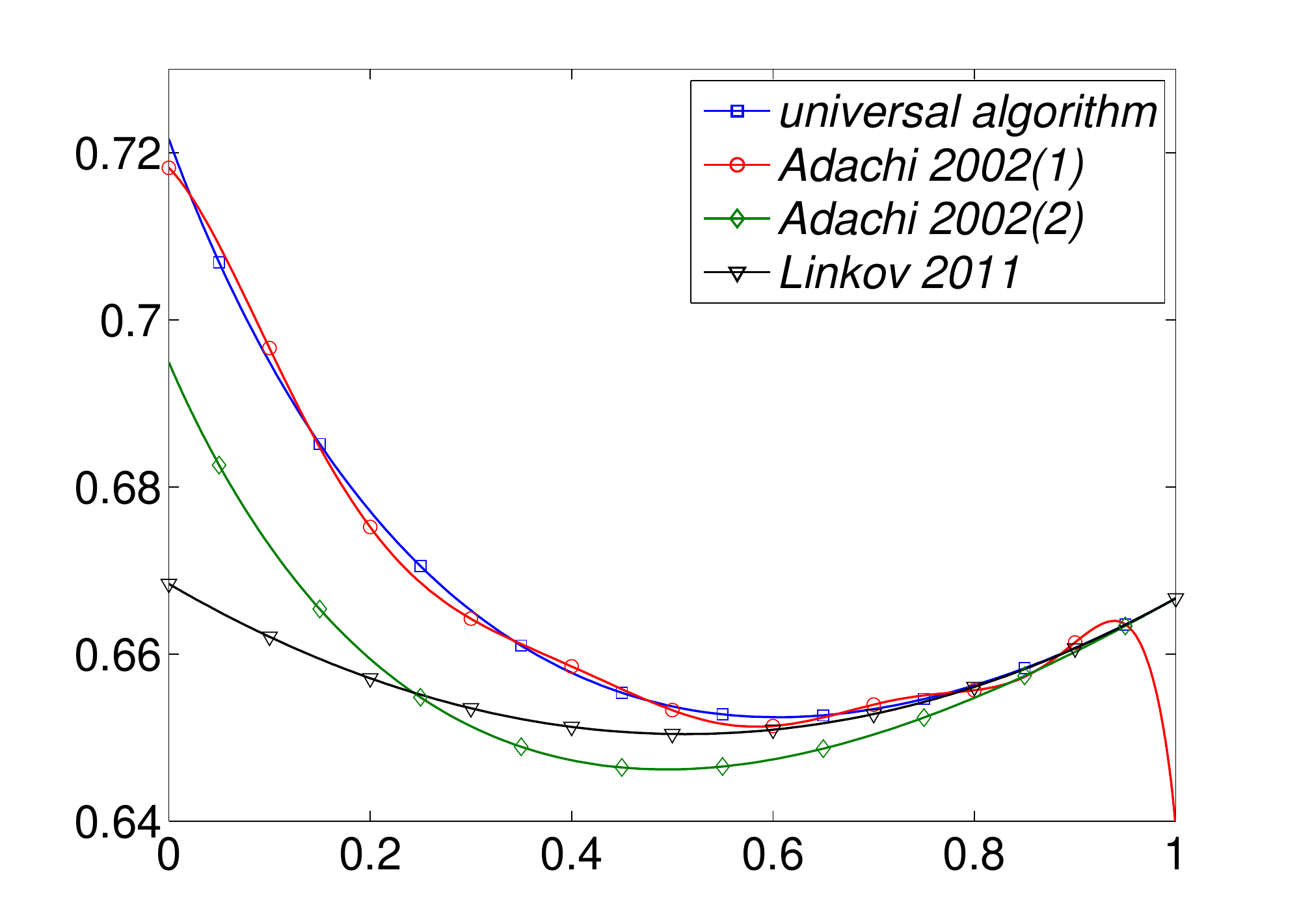}
    \put(-105,0){$x$}
    \put(-230,90){$ \hat v$}
    \put(-230,160){$\textbf{a)}$}
        \hspace{2mm}
    \includegraphics [scale=0.40]{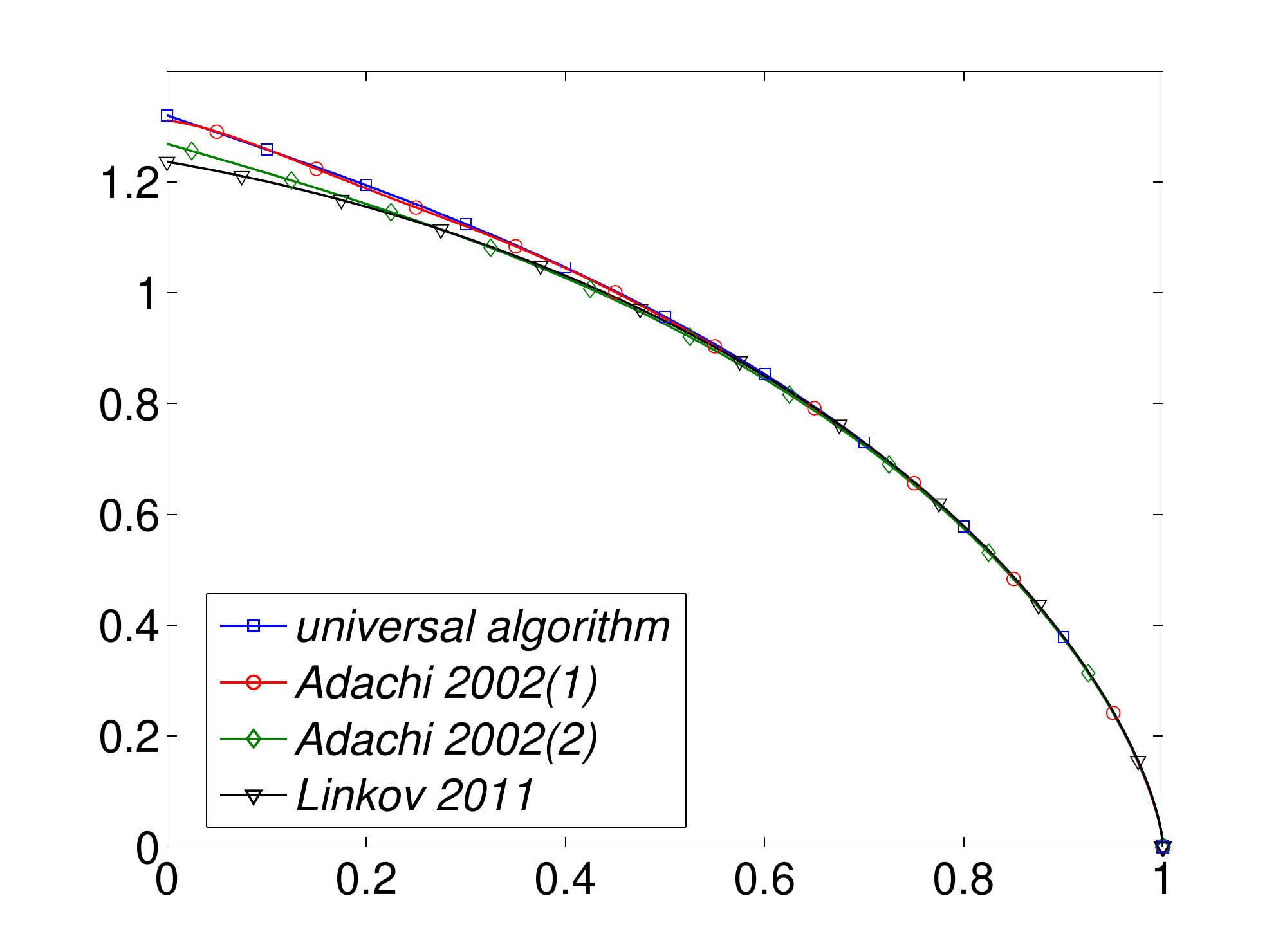}
    \put(-105,0){$x$}
    \put(-230,90){$\hat q$}
    \put(-230,160){$\textbf{b)}$}
    \caption{\emph{KGD model.} Results for the self-similar problem: a) particle velocity $\hat v$, b) fluid flow rate $\hat q$.}
    \label{wyniki_Ad_2}
\end{figure}

The respective results are shown in Fig.~\ref{wyniki_Ad_1} -- Fig.~\ref{wyniki_Ad_2}. To make the graphs more legible,
we do not depict here the approximation from \cite{Adachi_PhD}, which is discussed later on. Results by \cite{Adachi_Detournay} are denoted in the figures as \textit{Adachi 2002}.

Considering the crack opening and fluid pressure, one can see that
the curves corresponding to the universal algorithm and the solution presented in \cite{Adachi_Detournay} are indistinguishable from each other in the used scale (the $x$ interval in
Fig.~\ref{wyniki_Ad_2} is truncated,
since the pressure tends to infinity for $x \to 1$). The solution found in \cite{Linkov_3} provides a
very good approximation for the fracture aperture, however for the fluid pressure it deviates from other results.
When analyzing the particle velocity, we decided to use two methods of $\hat v$ computation for the data from \cite{Adachi_Detournay}, as each of them produces a slightly different result. The method refereed to as {\it{Adachi 2002(1)}} utilizes $\hat w$ and $\hat p'$ ($\hat v=-\hat w^2  \hat p'$), while for {\it{Adachi 2002(2)}} we employed $\hat q$ and $\hat w$ ($\hat v=\hat q / \hat w$). Both methods are equivalent in the case of an exact solution. It shows that the values given by the universal algorithm are in a good agreement
with those by {\it{Adachi 2002(1)}} except for the fracture tip, where the apparent deterioration of the latter solution takes place. However, in this region {\it{Adachi 2002(2)}} turns out to be perfectly consistent with our results.
The solution from \cite{Linkov_3} is hardly distinguishable from that by the universal algorithm for $x>0.8$, but deviates when $x$ decreases.

Finally, for the fluid flow rate, $\hat q$, we observe a good agreement between our data and that by {\it{Adachi 2002(1)}} (series approximation of $\hat q$ given in the paper) over the whole interval. Results by {\it{Adachi 2002(2)}} (flux recreated as and $\hat q=-\hat w^3 \hat p'$) and \cite{Linkov_3} diverge from ours for decreasing $x$.

The above analysis confirms the credibility of our solution, which together with the previous accuracy estimation allows us to treat it now as a numerical benchmark in and of itself. Following the idea from \cite{Adachi_PhD} and \cite{Linkov_3}, we propose a new {\it improved approximation} of the dependent variables analyzed above,
which provides higher accuracy than other known semi-analytical formulae and can be treated as the reference data when
testing other numerical algorithms.

Namely, we express the fracture opening, $\hat w$, the fluid pressure, $\hat p$, and the
particle velocity, $\hat v$, in the following manner:
\begin{equation}
\label{w_AD_ap}
\hat w(x)=\sqrt{3}(1-x^2)^{2/3}+0.3\left[\sqrt{1-x^2}-\frac{2}{3}(1-x^2)^{3/2}-x^2\ln \frac{1+\sqrt{1-x^2}}{x}\right],
\end{equation}
\begin{equation}
\label{p_Ad_ap}
\hat p(x)=\frac{B(1/2,2/3)}{\sqrt{3}\pi} {_2}F_1(1/6,1;1/2;x^2)+\sum_{i=1}^6p_i(1-x)^{(6-i)/2},
\end{equation}
\begin{equation}
\label{v_Ad_ap}
\hat v(x)=\frac{2}{3}\left[1+\sum_{i=1}^5v_i(1-x)^i\right],
\end{equation}
where $B$ is the beta function, $_2F_1$ denotes the Gauss hypergeometric function, and
the respective multipliers from \eqref{p_Ad_ap} and \eqref{v_Ad_ap} assume values: $p_1=0.274395$, $p_2=-0.56408$, $p_3=0.547395$,
 $p_4=-0.15621$, $p_5=-0.02495$, $p_6=-0.007285$, $v_1=-0.1$, $v_2=0.10542$, $v_3=0.02875$, $v_4=-0.02739$, $v_5=0.0752$.
Respective approximation for the fluid
flow rate, $\hat q$, can be easily obtained from the product of the fracture opening \eqref{w_AD_ap} and particle velocity \eqref{v_Ad_ap}.
Note that the term multiplied by 0.3 in
\eqref{w_AD_ap} is exactly the special term, $\hat w_N$, used in  representation \eqref{h_KGD fluid} for the pressure derivative correction.

\begin{figure}[h!]
    \includegraphics [scale=0.40]{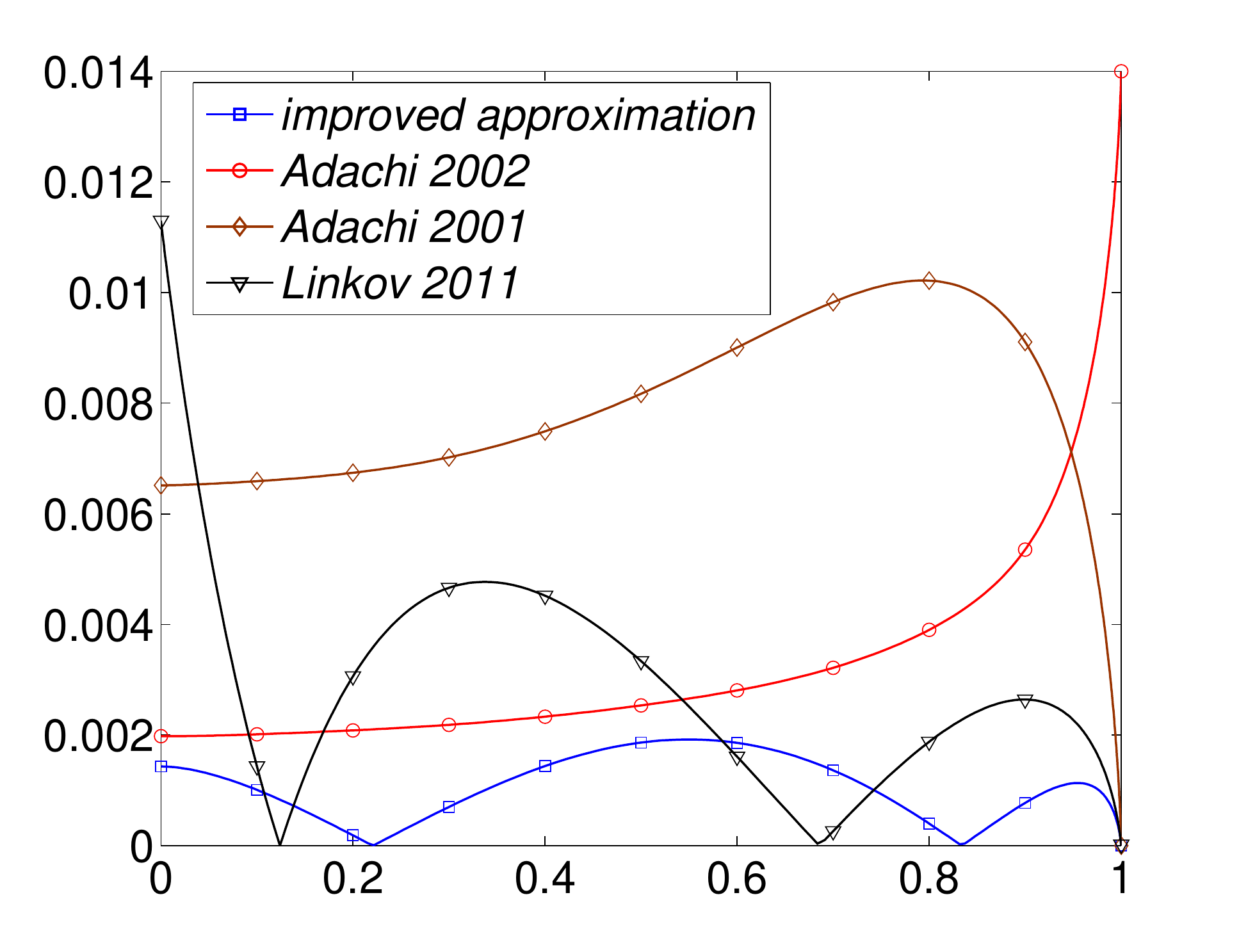}
    \put(-105,0){$x$}
    \put(-230,90){$ \delta \hat w$}
    \put(-230,160){$\textbf{a)}$}
    \hspace{2mm}
    \includegraphics [scale=0.40]{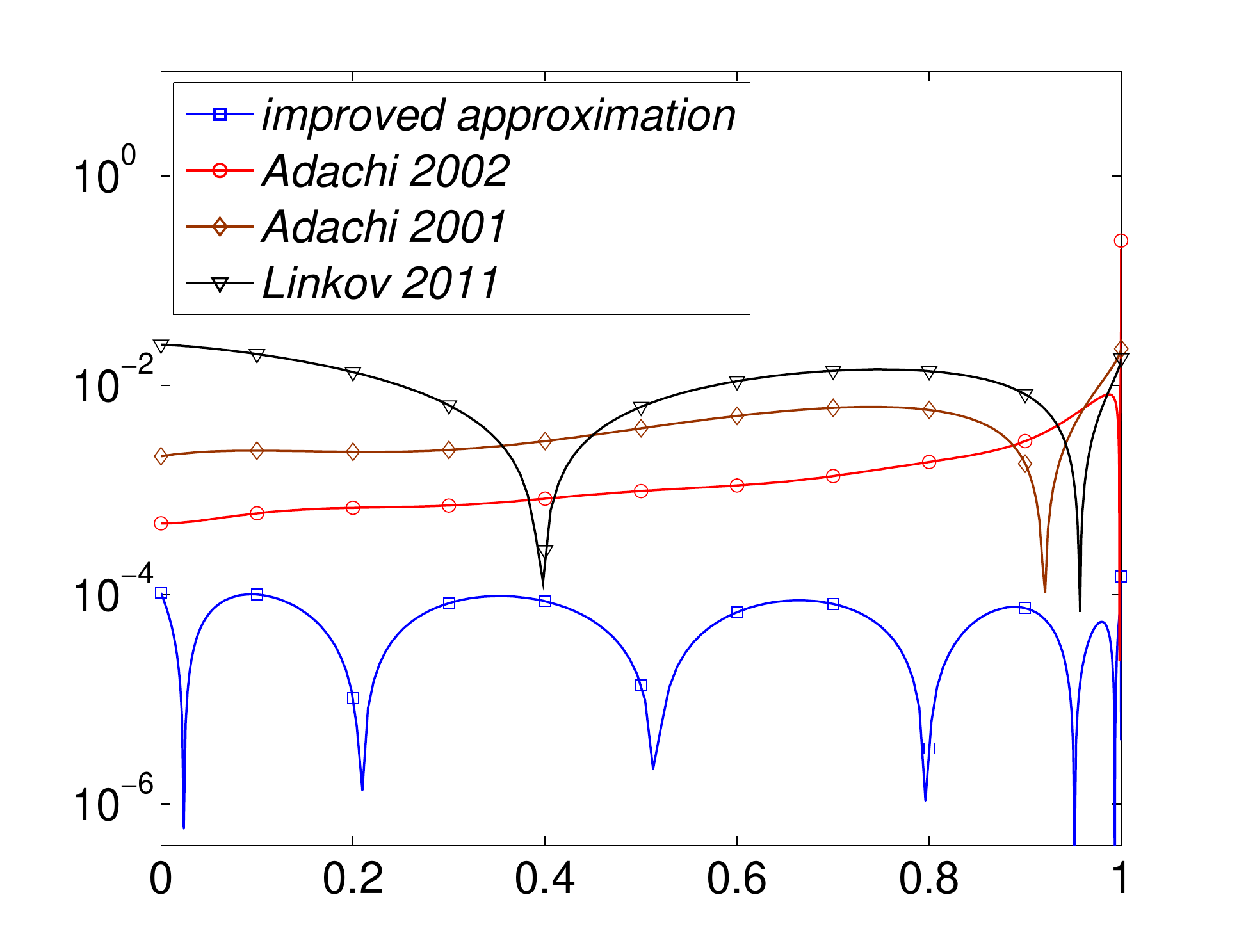}
    \put(-105,0){$x$}
    \put(-230,90){$\Delta \hat p$}
    \put(-230,160){$\textbf{b)}$}

    \caption{\emph{KGD model.} Comparison of various approximate formulae for the accurate numerical solution: a) the relative deviation of the crack opening $\hat w$,
    b) the absolute deviation of the fluid pressure $\hat p$.}

\label{wyniki_aproks_1}
\end{figure}

\begin{figure}[h!]

    \includegraphics [scale=0.40]{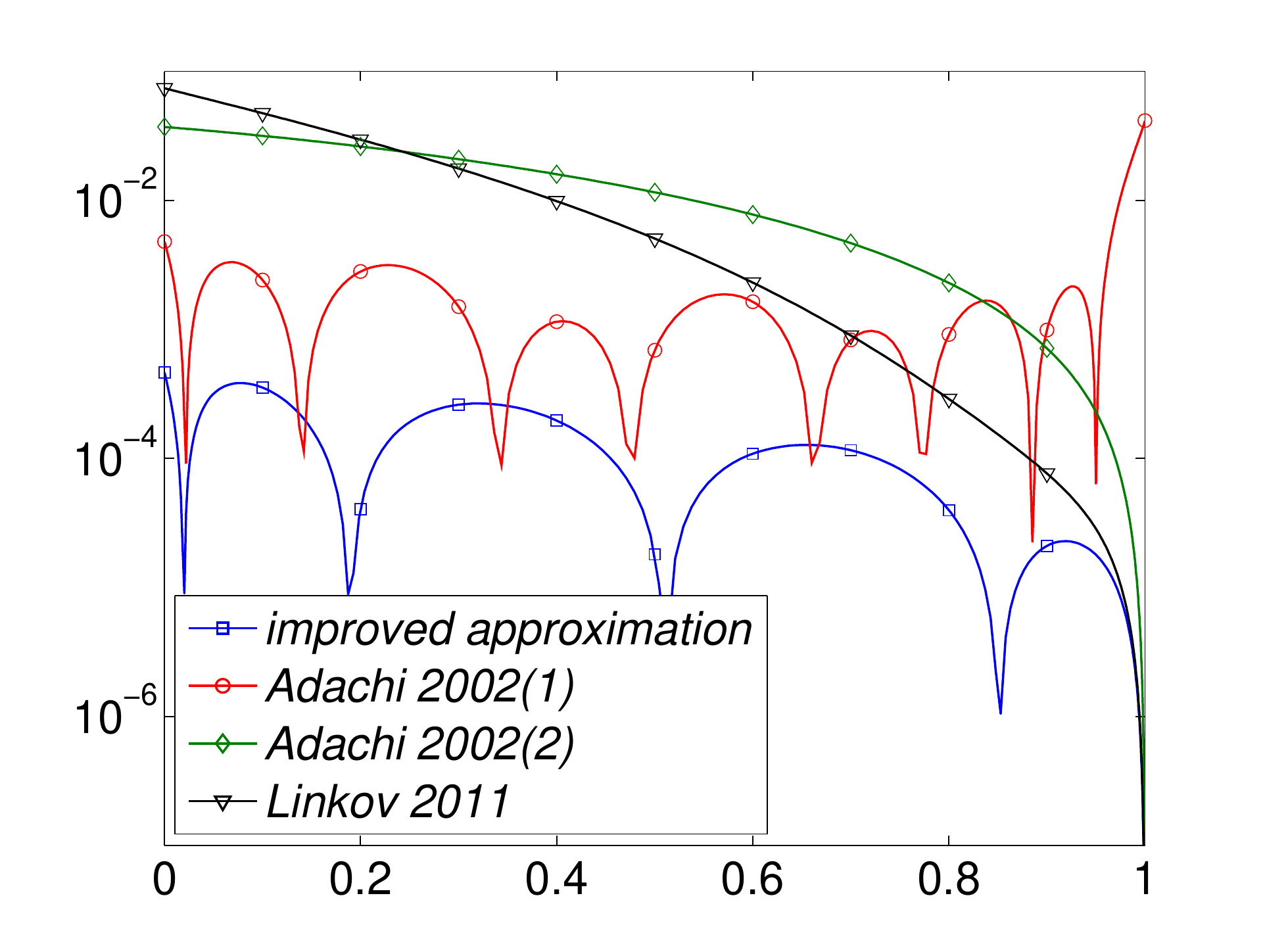}
    \put(-105,0){$x$}
    \put(-230,90){$ \delta \hat v$}
    \put(-230,160){$\textbf{a)}$}
    \hspace{2mm}
    \includegraphics [scale=0.40]{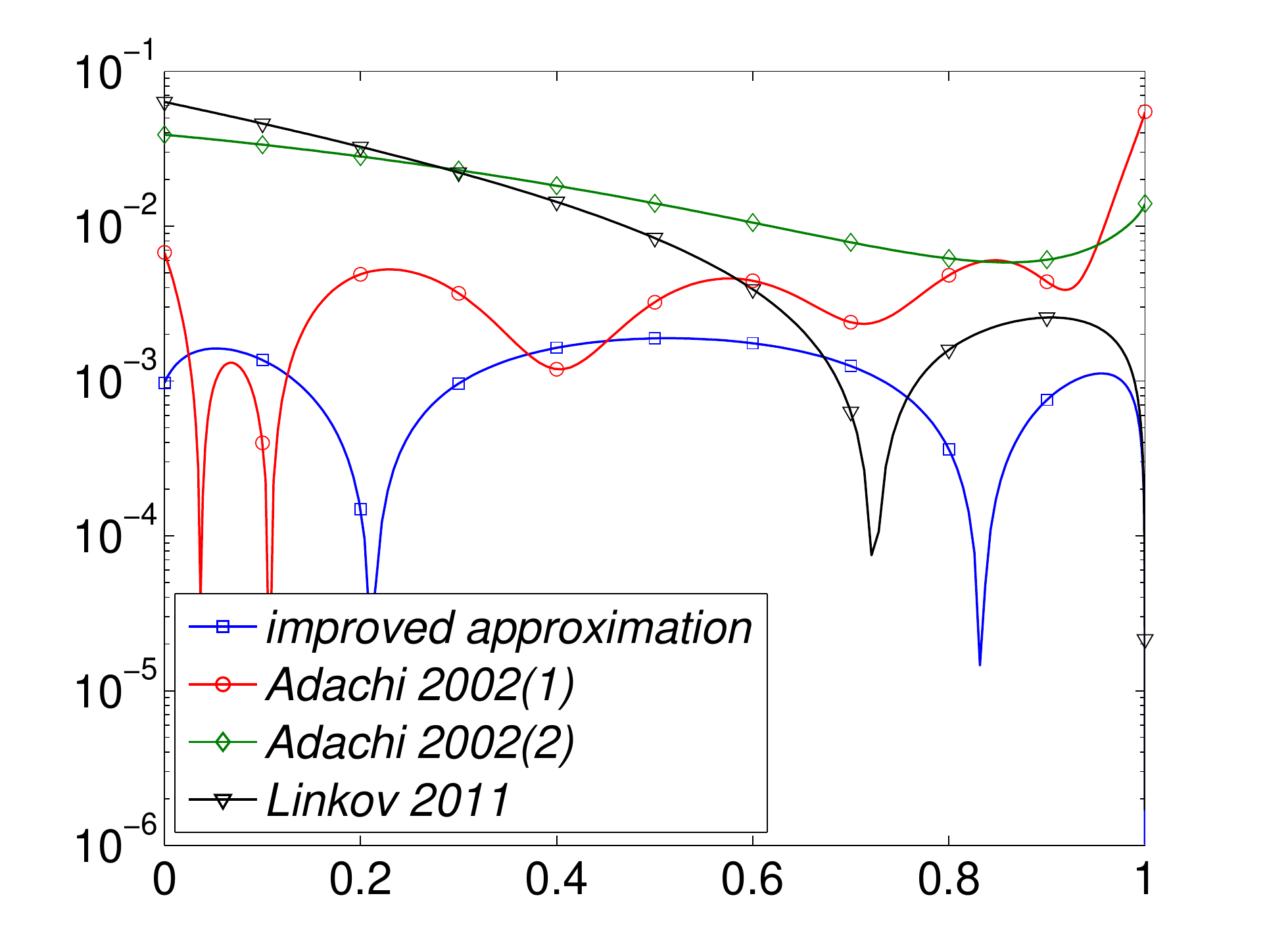}
    \put(-105,0){$x$}
    \put(-230,90){$\delta \hat q$}
    \put(-230,160){$\textbf{b)}$}

    \caption{\emph{KGD model.} Comparison of various approximate results against accurate numerical computations: a) the relative deviation of the particle velocity $\hat v$, b) the relative deviation of the fluid flow rate $\hat q$}
\label{wyniki_Ad_2a}
\end{figure}
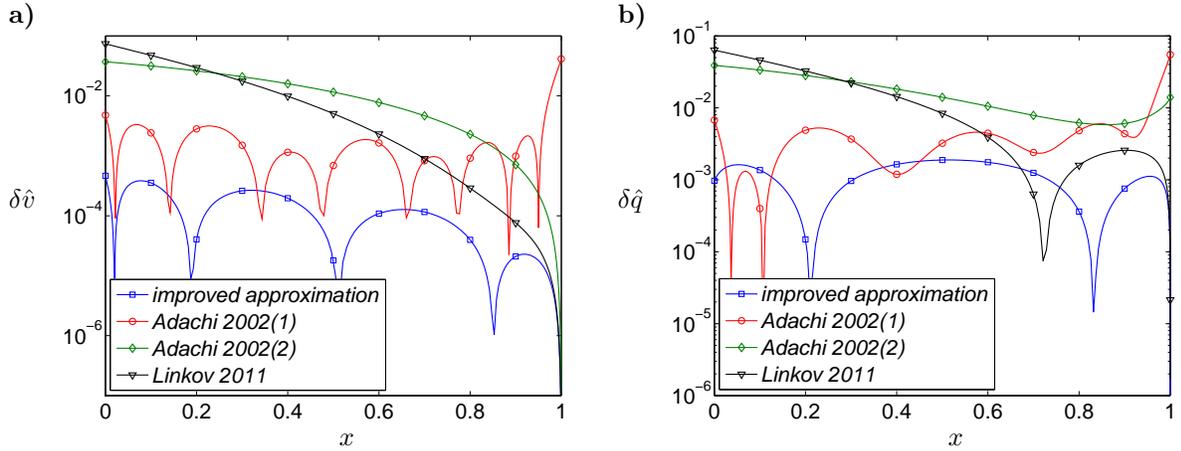

In Fig.~\ref{wyniki_aproks_1} -- Fig.~\ref{wyniki_Ad_2a}
we show comparisons between the {\it improved approximation} and the results known from \cite{Adachi_Detournay}, \cite{Linkov_3}, \cite{Adachi_PhD},
referring them all to our numerical reference solution. For the
crack opening, $\hat w$, the particle velocity, $\hat v$, and the fluid flow rate, $\hat q$, their relative deviations $\delta \hat w$, $\delta \hat v$ and $\delta \hat q$ from the  numerical solution are given. Again, $\hat v$ and $\hat q$ from \cite{Adachi_Detournay} are computed in two alternative ways described above.
For the fluid pressure we show the absolute difference, $\Delta \hat p$, as the pressure curve intersects the $x$-axis.

As can be seen, the new representation \eqref{w_AD_ap}
imitates the crack opening with an accuracy of the order $10^{-3}$ (the maximal value of deviation is less than 0.2\%). The same level of error of approximation is obtained for the fluid flow rate. For the fluid pressure, {\it the improved approximation}
gives a maximal deviation from the numerical benchmark of the order $10^{-5}$. Finally, for the particle velocity, the error of the approximation is of order $10^{-4}$.

For all the considered parameters, {\it the improved approximation} gives a better agreement with the accurate numerical solution than any of the known results.
Note, that this analysis also reveals the level of accuracy of all previously reported approximations. It shows that the solution from \cite{Adachi_Detournay} is better than the one proposed in
\cite{Linkov_3} in an average sense. However, its quality deteriorates near the fracture tip, where the latter exhibits the correct tip asymptotics. For the particle velocity and fluid flow rate, the solution from \cite{Adachi_Detournay} could be improved by merging two representations obtained by two different ways of defining $\hat q$: series approximation near the fracture tip and $\hat q=-\hat w^3 \hat p'$ on the rest of the interval. On the other hand,
the approximation from \cite{Adachi_PhD} preserves the tip asymptotic behaviour much better than the series approximation \cite{Adachi_Detournay}. However, it gets worse when moving away from the fracture tip.

\vspace{1mm}

{\sc Remark 9}.
Respective formulae defining {\it the improved approximation} should be treated independently, i.e.
when trying to recreate the particle velocity by using \eqref{w_AD_ap} and differentiating
\eqref{p_Ad_ap}, one cannot expect the same accuracy as for \eqref{v_Ad_ap}.

\subsubsection{Toughness driven KGD model for impermeable rock -- comparison with other results.}

For the toughness driven regime of the KGD model, it is difficult to find formulae in the literature which could be directly compared
with our solution in the same way as done above.
Such solutions are either not complete, in a sense they do not describe all the components analyzed above (rather only the crack opening and the fluid pressure), or one can find only the
values of a few first multipliers of the respective base functions approximating the solution. As a result, it is practically impossible to provide
a fair and credible comparison. In some cases, in order to rebuilt the solution given by the author, one needs also to repeat the respective numerical algorithm.
Then, the quality of the comparison would essentially depend on the algorithm implementation.

\begin{figure}[h!]

    \includegraphics [scale=0.40]{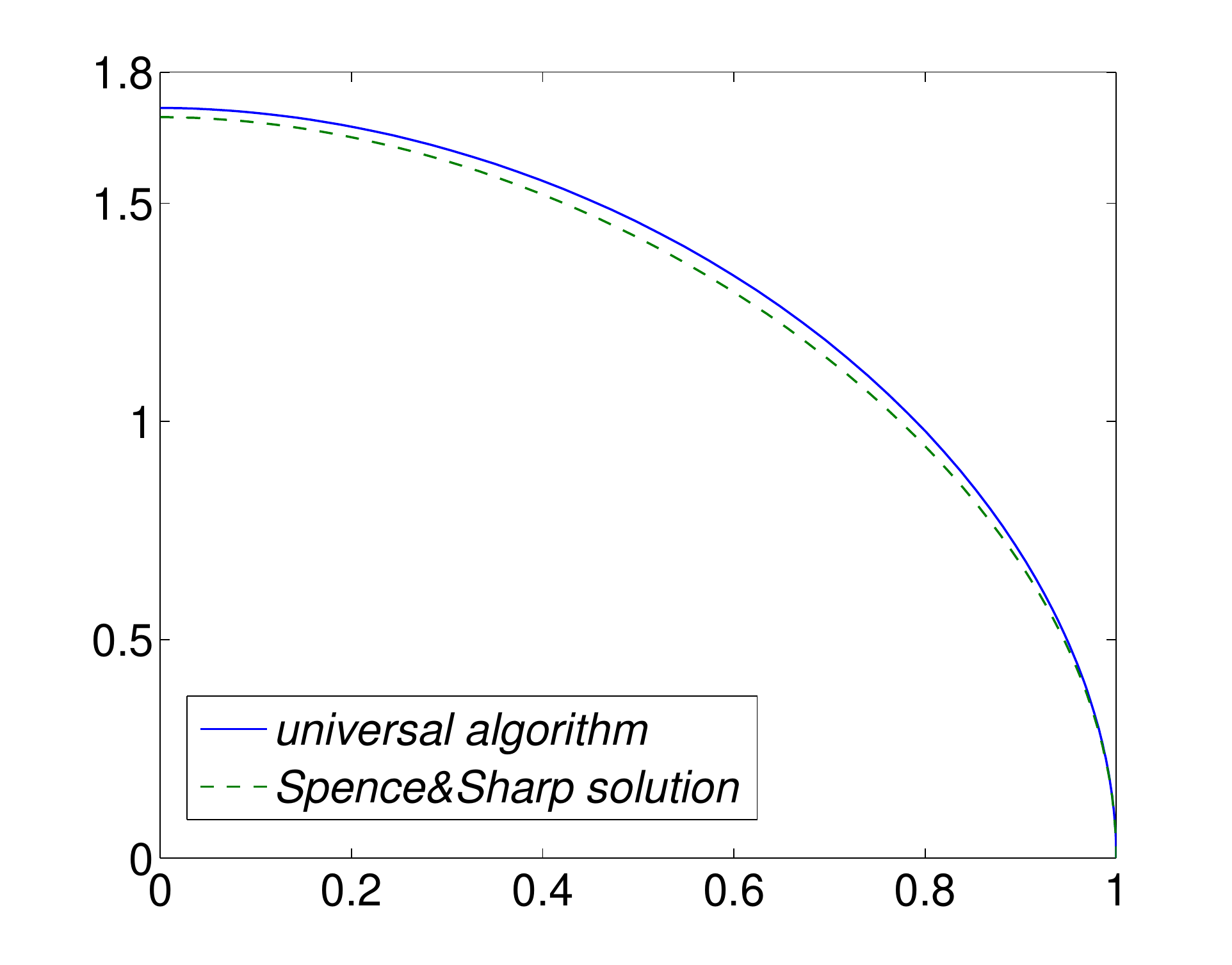}
    \put(-105,0){$x$}
    \put(-220,90){$\hat w$}
    \put(-230,160){$\textbf{a)}$}
    \hspace{2mm}
    \includegraphics [scale=0.40]{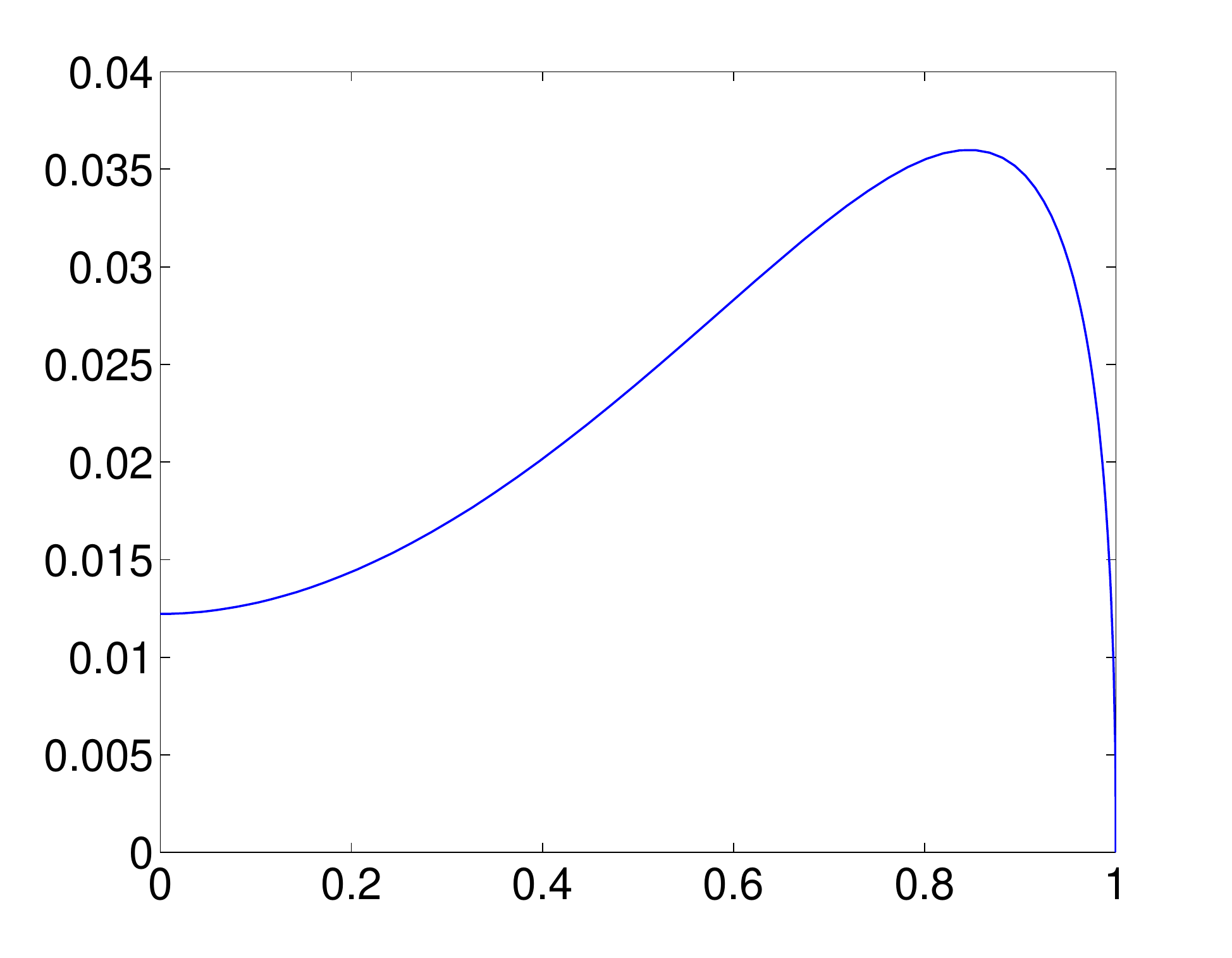}
    \put(-105,0){$x$}
    \put(-230,90){$\delta \hat w$}
    \put(-230,160){$\textbf{b)}$}

    \caption{\emph{KGD model.} $Spence$\&$Sharp$ solution - comparison of the numerical solution with data provided in \cite{Spence&Sharp}: a) the crack opening, $\hat w$, b) relative deviation between crack openings. }

\label{SS_solution_1}
\end{figure}
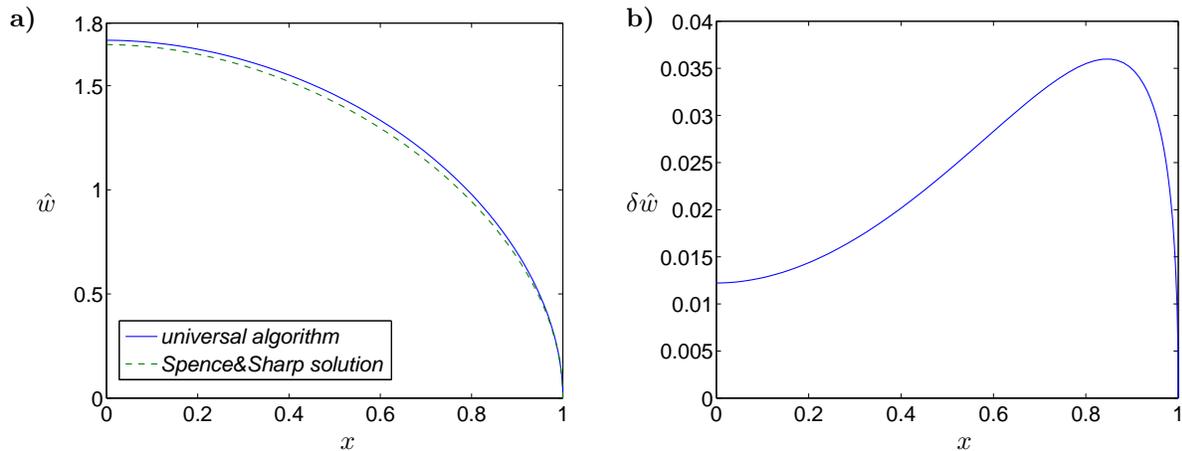

We compare our numerical results with those few sources available. First, we start with the classical example from \cite{Spence&Sharp}, where the authors provide a number of numerical results obtained by the method of series approximation for different variants of the problem (p.300, Table~1). Unfortunately, the format of the data records (two or three decimal digits),
and the fact that multipliers for merely three leading terms are given, allow us to consider the reconstructed solution rather as a rough approximation.
We recreated that solution for the data corresponding to the first case ($\alpha=1$, $\lambda=2/3$) from  the aforementioned Table~1.
The self-similar stress intensity factor $\hat K$ was computed in accordance with the given coefficients. The utilized spatial mesh was composed of 300 nodal points with the density refined near the end points (compare Fig. \ref{bld_w_h_KGD}). The comparison of our results with those from \cite{Spence&Sharp} is given in In Fig.~\ref{SS_solution_1}.

Here only the crack opening, $w$, and its relative deviation are shown. The representation of the net fluid pressure based on only three terms is far from completeness (and does not reflect the proper asymptotic behaviour). Thus the comparison would be unfair in this case.
One can see that the maximal relative error of the solution from \cite{Spence&Sharp} amounts to 3.5\% (compare Fig.~\ref{SS_solution_1}b))
and is located at some distance away from the crack tip. The asymptotic behaviours of both solutions in the near-tip region coincide with each other, being directly embedded into the respective numerical schemes.

Another reference solution  used in this subsection is the one given in \cite{garagash_large_toughenss} for the large toughness (small viscosity) regime.
The author provides the data sufficient to recreate the crack opening and the net fluid pressure (Table~II - pp. 1458, with respective formulae for the opening and pressure approximations), for three different values of the fluid behaviour index. Since in this paper we consider Newtonian fluids,  only the data corresponding to the case $n=1$ was taken (the correct value of the first coefficient should be two times smaller than that reported: $6.05\cdot10^{-3}$ \citep{Garagash_2014}). As previously, we carried out the computations on a mesh composed of 300 nodal points. This, according to the characteristic given in Fig.~\ref{bld_w_h_KGD}, should provide an error of the level $10^{-7}$.

\begin{figure}[h!]

    \includegraphics [scale=0.40]{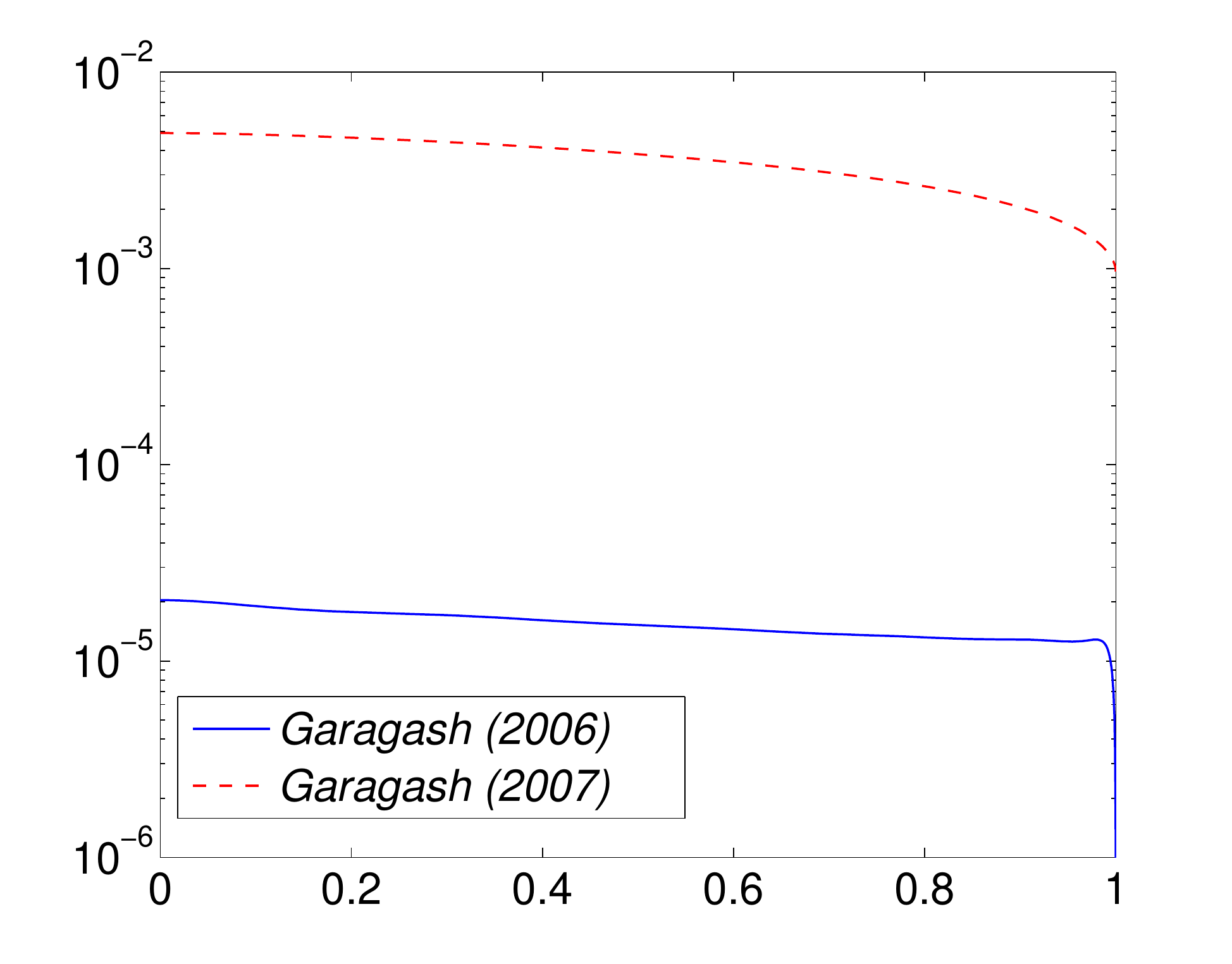}
    \put(-105,0){$x$}
    \put(-230,90){$\delta \hat w$}
    \put(-230,160){$\textbf{a)}$}
    \hspace{2mm}
    \includegraphics [scale=0.40]{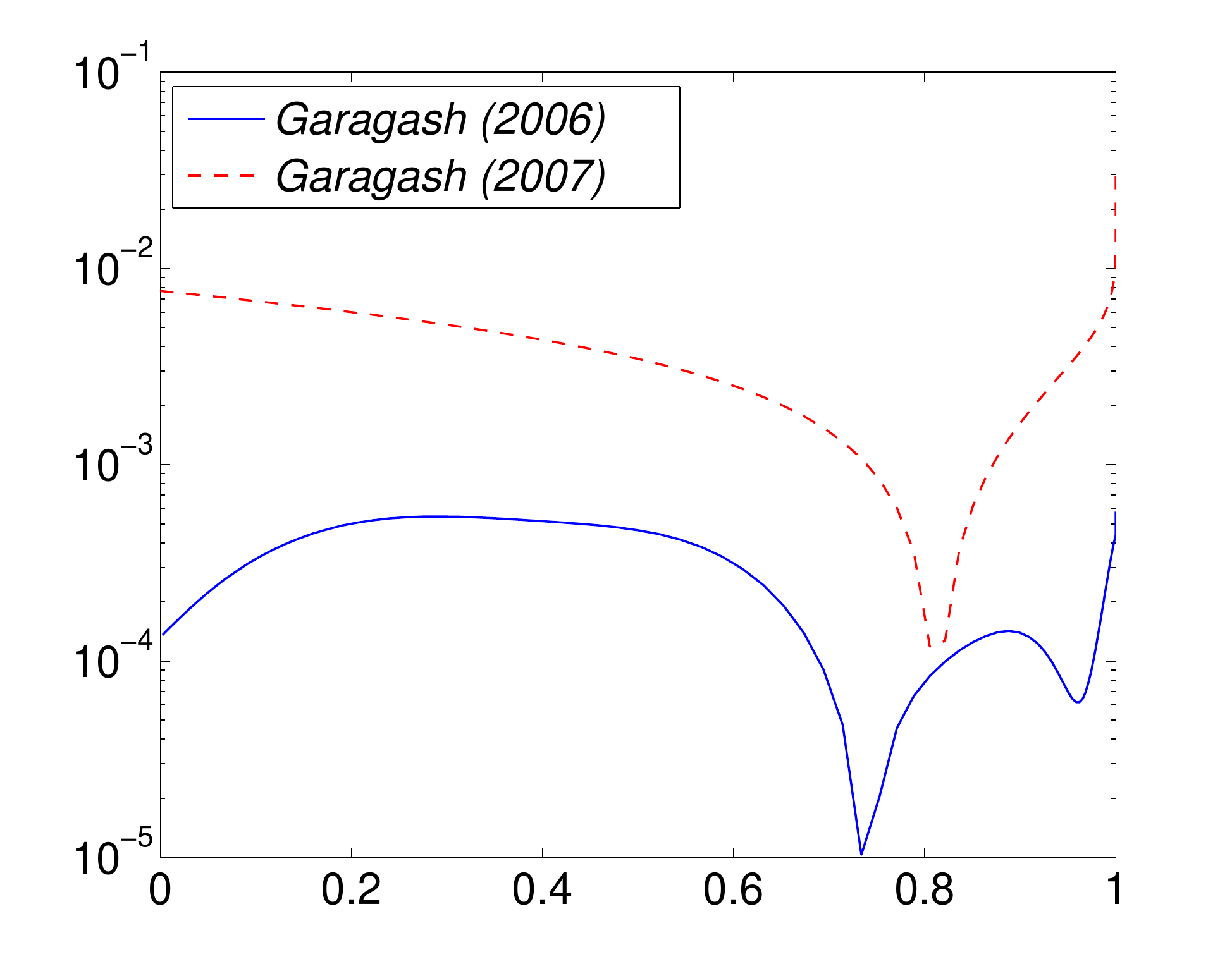}
    \put(-105,0){$x$}
    \put(-230,90){$\delta \hat p$}
    \put(-230,160){$\textbf{b)}$}

    \caption{\emph{KGD model.} Comparison of the numerical results with data provided in \cite{garagash_large_toughenss} and \cite{Garagash_shut_in}: a) relative deviation of the crack opening, $\hat w$, b) relative deviation of the net fluid pressure, $\hat p$.}

\label{Gar_porownanie}
\end{figure}
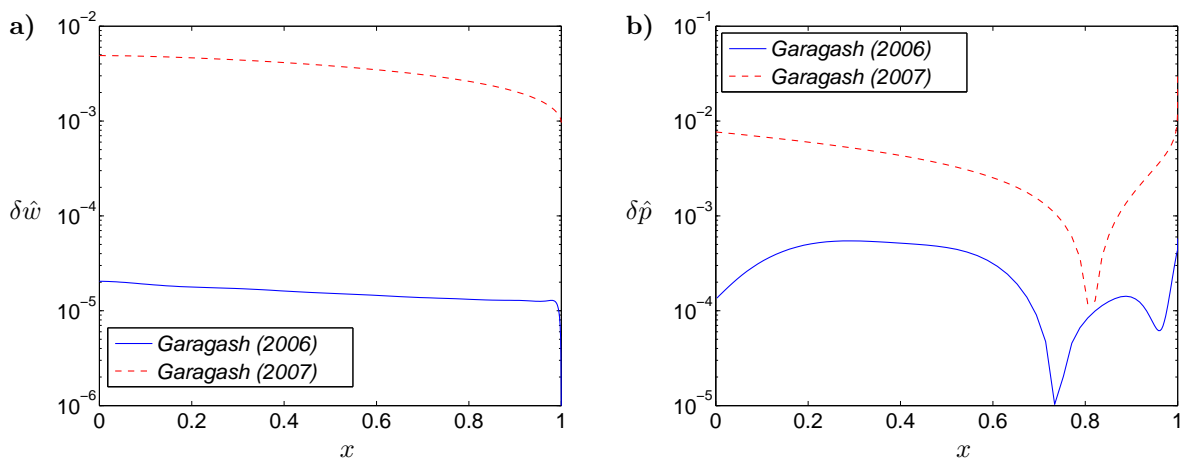

The relative difference between our numerical solution and the one reconstructed from the data given in \cite{garagash_large_toughenss} (based on 13-terms approximation) is shown in Fig.~\ref{Gar_porownanie}. For comparison we present also the less accurate two-terms approximate solution available in \cite{Garagash_shut_in}.

It shows that the solution based on 13 terms yields a very good accuracy. For the fracture aperture, $\hat w$, its relative deviation from our solution is of the order $10^{-5}$. The lowest discrepancy between respective solutions can be observed in the near-tip region, which demonstrates high quality of our numerical computations. Indeed the solution from \cite{garagash_large_toughenss} represents the accurate asymptotics there.
For the net fluid pressure, $\hat p$, the deviation is of one order greater. Similar analysis for the two-terms approximation from \cite{Garagash_shut_in} gives the level of $10^{-3}$ for the crack opening, and $10^{-2}$ of the net fluid pressure. We believe that Fig.~\ref{Gar_porownanie} reveals the accuracy of semi-analytical solutions
solutions proposed in \cite{garagash_large_toughenss} and \cite{Garagash_shut_in}. It is worth mentioning that the error measure introduced in \cite{garagash_large_toughenss}, the quadratic global error $e^{(N)}$ (equation (E1), pp.1471 therein), is  of the level $10^{-6}$ for $N=10$ (see figure E2, pp.1472 therein). This establishes, in this particular case, a relationship between the recalled measure and the relative error of the solution.

In the foregoing analysis, we have compared our numerical solution with known results confirming its high accuracy. Now, we analyse the performance of our algorithm for different values of the normalised self-similar stress intensity factor. Special attention is paid to the problem of small toughness as it constitutes the most challenging case for the computations, which was underlined in \cite{Lecampion_Brisbane}. In such a limiting variant of the problem, it is convenient to have a reference to the fluid-driven solution discussed in Fig.~\ref{wyniki_Ad_1} -- Fig.~\ref{wyniki_Ad_2}.

\begin{figure}[h!]
    \center
    \includegraphics [scale=0.40]{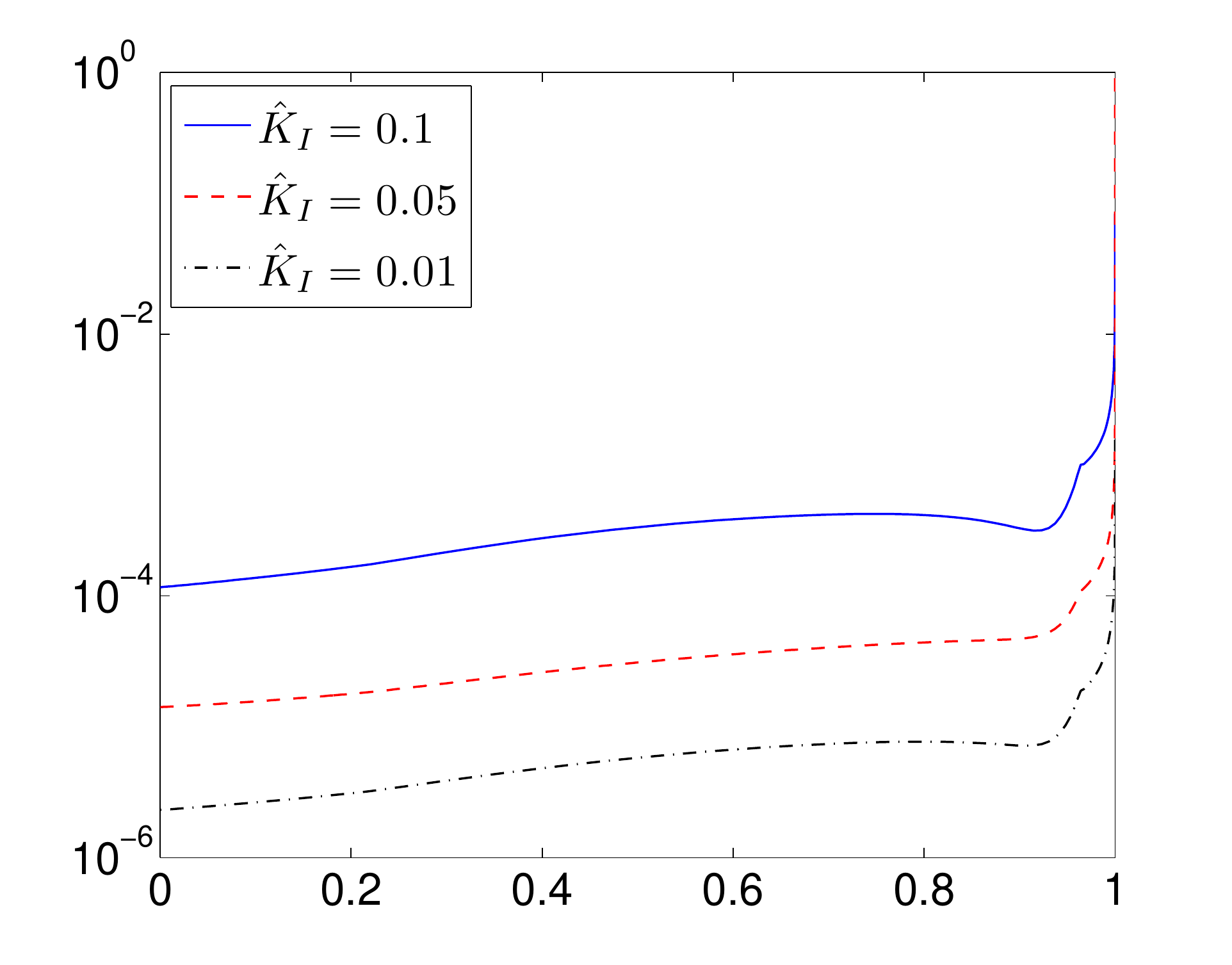}
    \put(-110,0){$x$}
    \put(-230,80){$\delta \hat w$}
    \caption{\emph{KGD model.} Relative deviation of the toughness driven numerical solution from the fluid driven reference solution for different values of the self-similar stress intensity factor, $\hat K_I$.}
\label{K_I_dev_fluid}
\end{figure}

For
this reason let us impose the same case as recalled for the influx magnitude. Then, by changing (decreasing) the values of $\hat K_I$, we investigate different variants of the problem.
It is not a surprise that for sufficiently small $\hat K_I$ the numerical results converge to the fluid driven reference solution. The relative deviations between the latter and the solutions, $\delta \hat w$, for three variants of $\hat K_I$: 0.1, 0.05, 0.01, are shown in Fig.~\ref{K_I_dev_fluid}, where:
$\delta \hat w=|\hat w_T-\hat w_F|/|\hat w_T|$,
for $\hat w_T$ being the solution of the toughness driven variant and $\hat w_F$ refereing to the fluid driven case.

 As can be seen, the greatest discrepancies can be observed in the vicinity of the crack tip, where the respective asymptotics do not correspond to each other. It is enough to take $\hat K_I=0.01$ to have the relative deviation from the fluid driven solution of the order $10^{-6}$ along almost the entire spatial interval. This fact calls into question the sensibility of conducting computations for the toughness driven regime with such small values of $\hat K_I$, especially as the efficiency of computations deteriorates as $\hat K_I$ decreases.

{\sc Remark 10.} Note that in case of very small toughness, the parameter $\varepsilon$ in the $\varepsilon$-regularization technique should be extremely small to capture the tip asymptotics. Indeed, in our computations it was $\varepsilon=10^{-9}$. Moreover, when decreasing $\hat K_I$ to very low values, we arrive at a situation where the relative difference between the small toughness solution and the fluid driven one is of the same order as the error of computation.

\begin{figure}[h!]

    \includegraphics [scale=0.40]{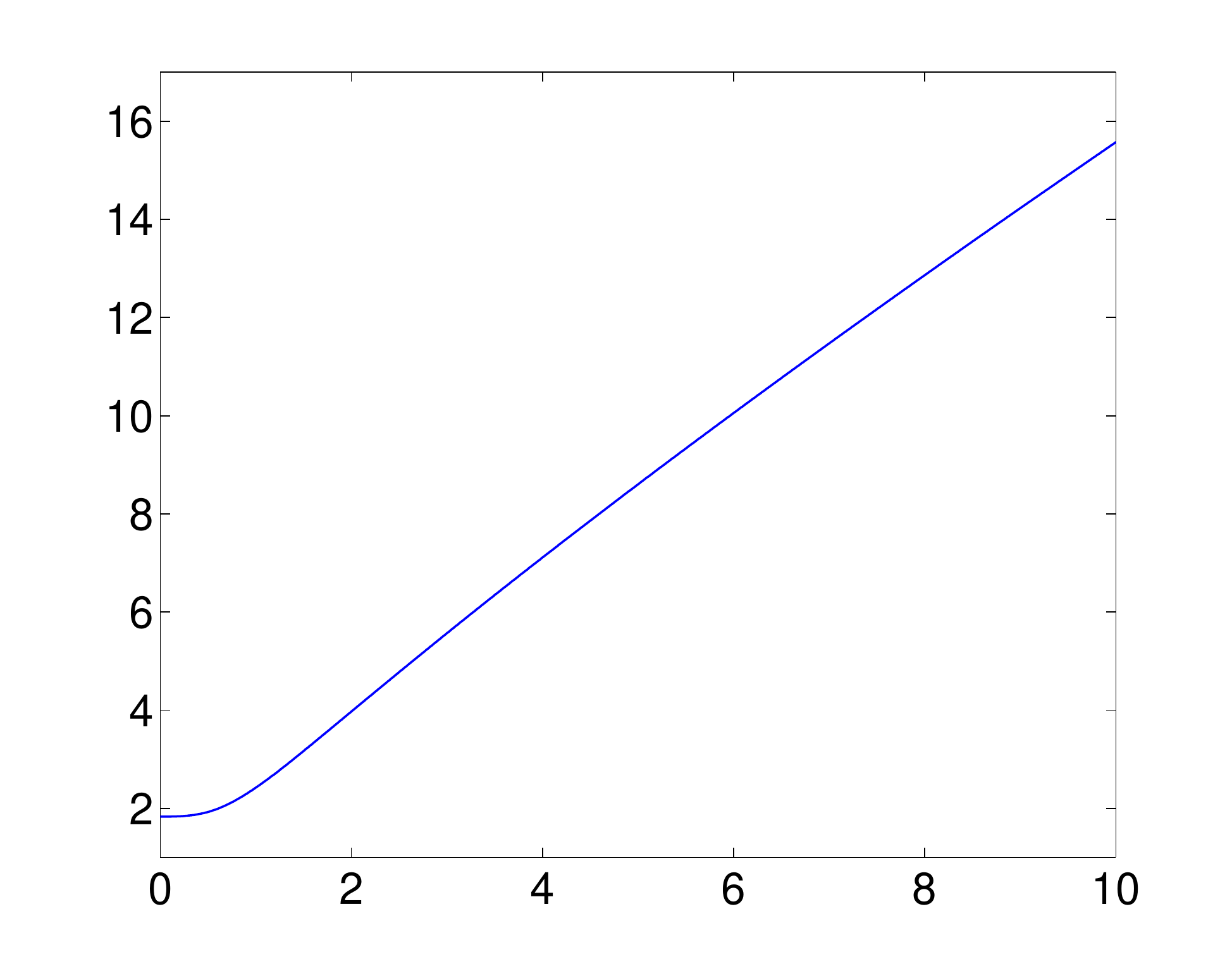}
    \put(-105,0){$\hat K_I$}
    \put(-230,90){$\hat w(0)$}
    \put(-230,160){$\textbf{a)}$}
    \hspace{2mm}
    \includegraphics [scale=0.40]{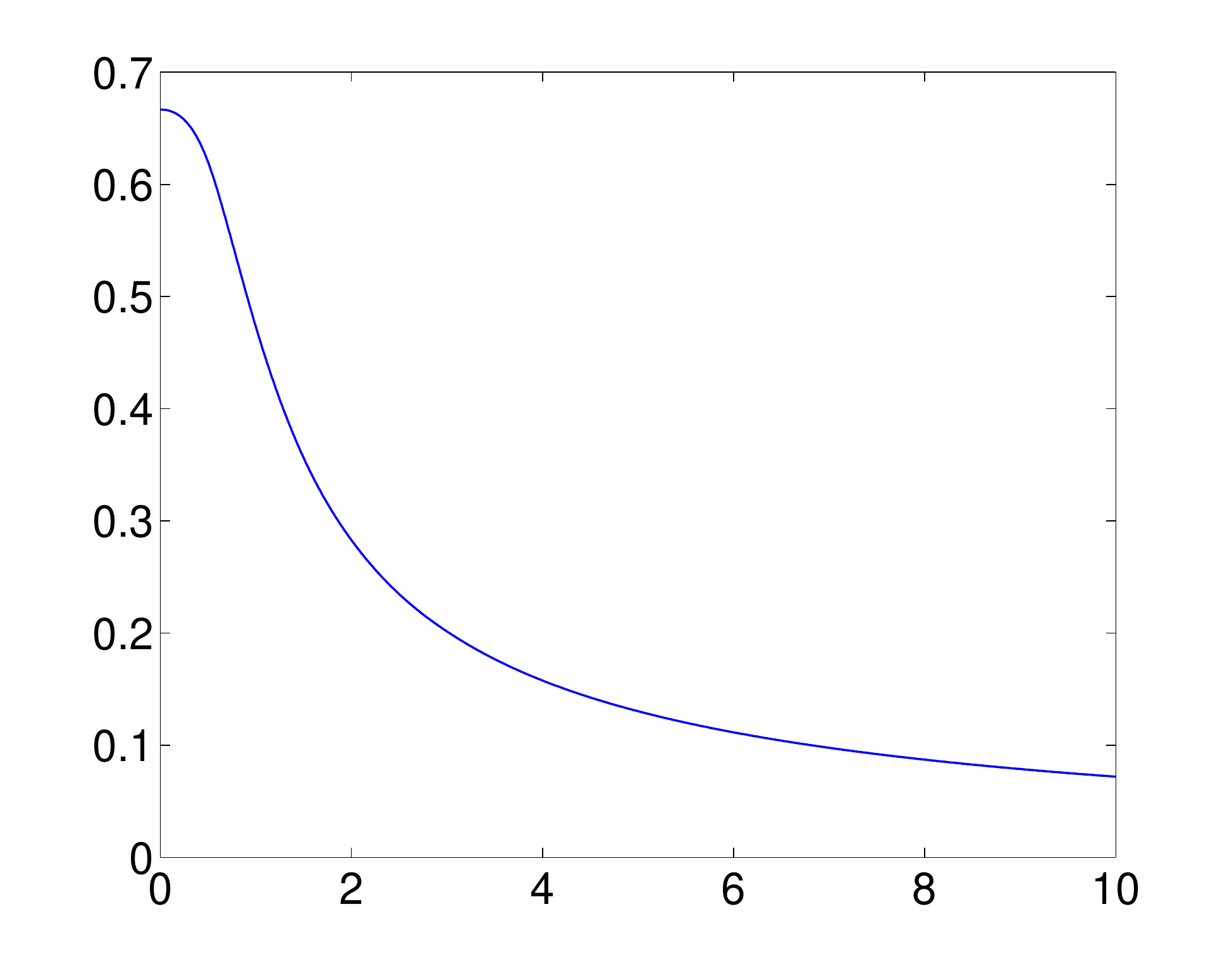}
    \put(-105,0){$\hat K_I$}
    \put(-230,90){$\hat v(1)$}
    \put(-230,160){$\textbf{b)}$}

    \caption{\emph{KGD model.} Results for different values of $\hat K_I$ ($\hat K_I=0$ corresponds to the fluid driven regime): a) crack opening at the inlet, $\hat w(0)$, b) crack propagation speed, $\hat v(1)$. }

\label{rozne_K_3}
\end{figure}

In Fig.~\ref{rozne_K_3} we present the evolution of the crack opening at inlet, $\hat w(0)$, and the crack propagation speed, $\hat v(1)$, for continuously growing $\hat K_I$, ranging from 0 (fluid driven solution) to 10 (large toughness solution). It shows that, for greater values of the self-similar stress intensity factor, $\hat w(0)$ increases as a linear function of $\hat K_I$, while the crack propagation speed is inversely proportional to this parameter:
\[
\hat w(0) =O(\hat K_I), \quad \hat v(1) =O(\hat K_I^{-1}), \quad \hat K_I\to\infty.
\]

\begin{figure}[h!]

    \includegraphics [scale=0.40]{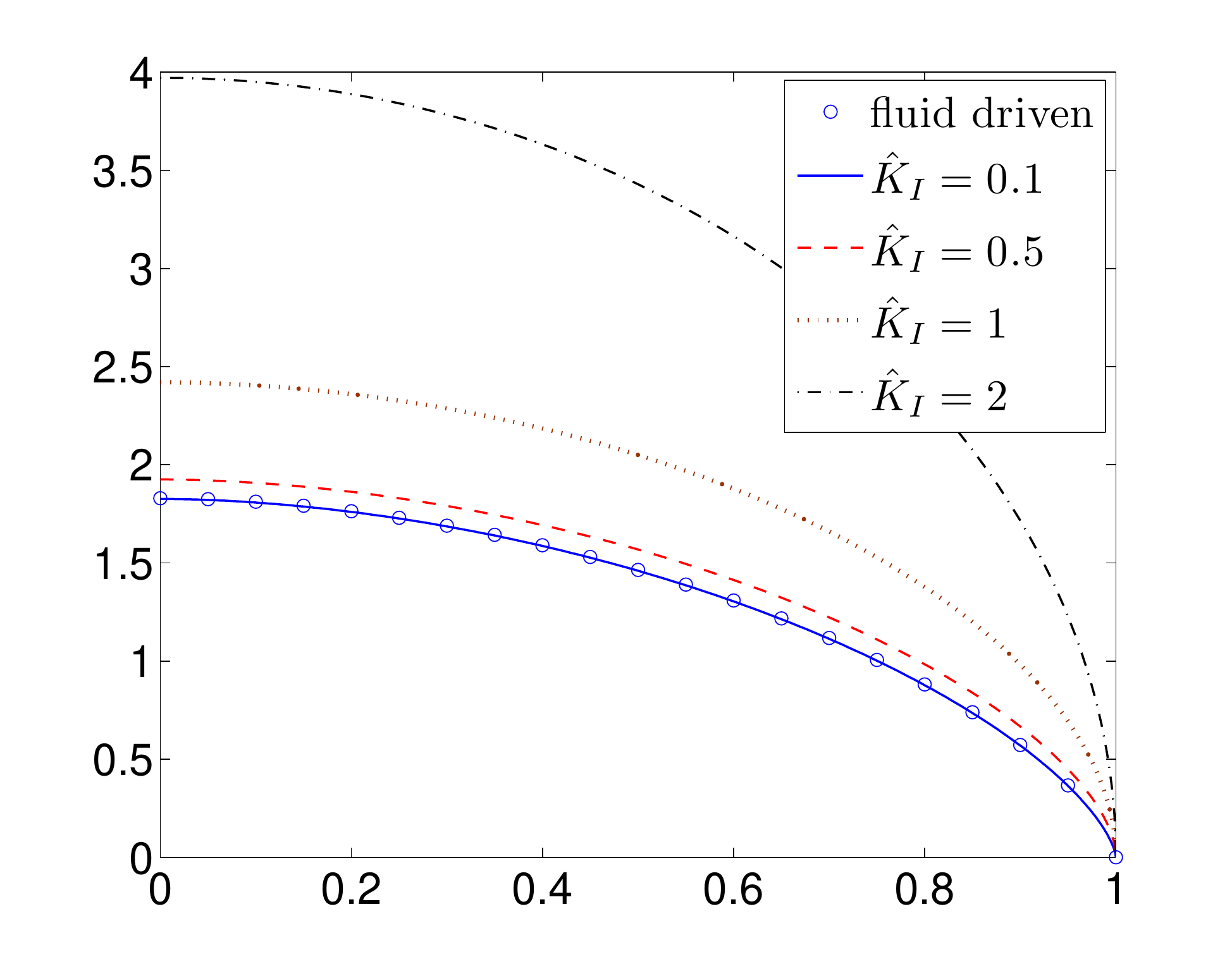}
    \put(-105,0){$x$}
    \put(-230,90){$\hat w$}
    \put(-230,160){$\textbf{a)}$}
    \hspace{2mm}
    \includegraphics [scale=0.40]{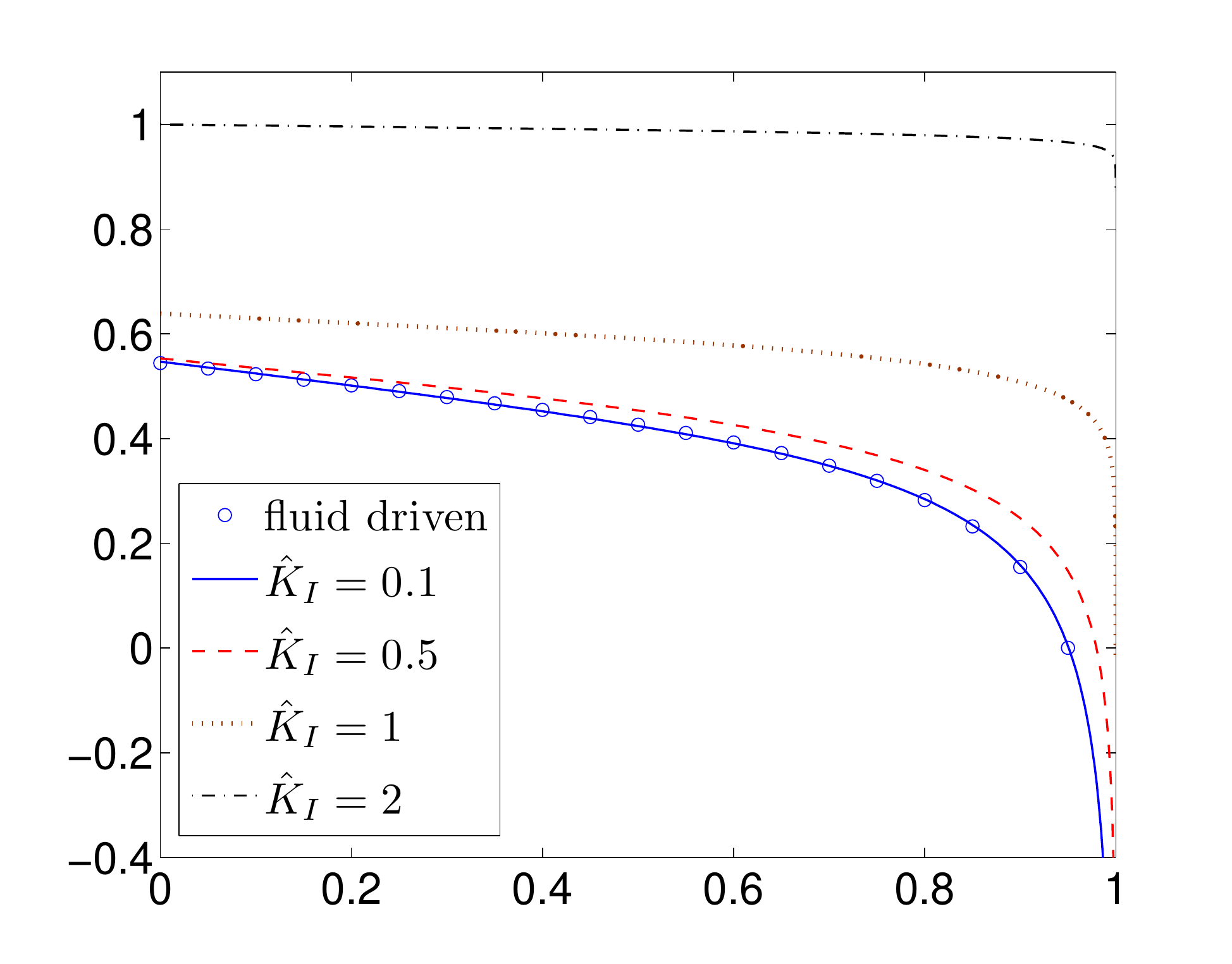}
    \put(-105,0){$x$}
    \put(-230,90){$\hat p$}
    \put(-230,160){$\textbf{b)}$}

    \caption{\emph{KGD model.} Solutions for different values of the self-similar stress intensity factor, $\hat K_I$: a) the crack opening, $\hat w$, b) the net fluid pressure, $\hat p$. }

\label{rozne_K_1}
\end{figure}

\begin{figure}[h!]

    \includegraphics [scale=0.40]{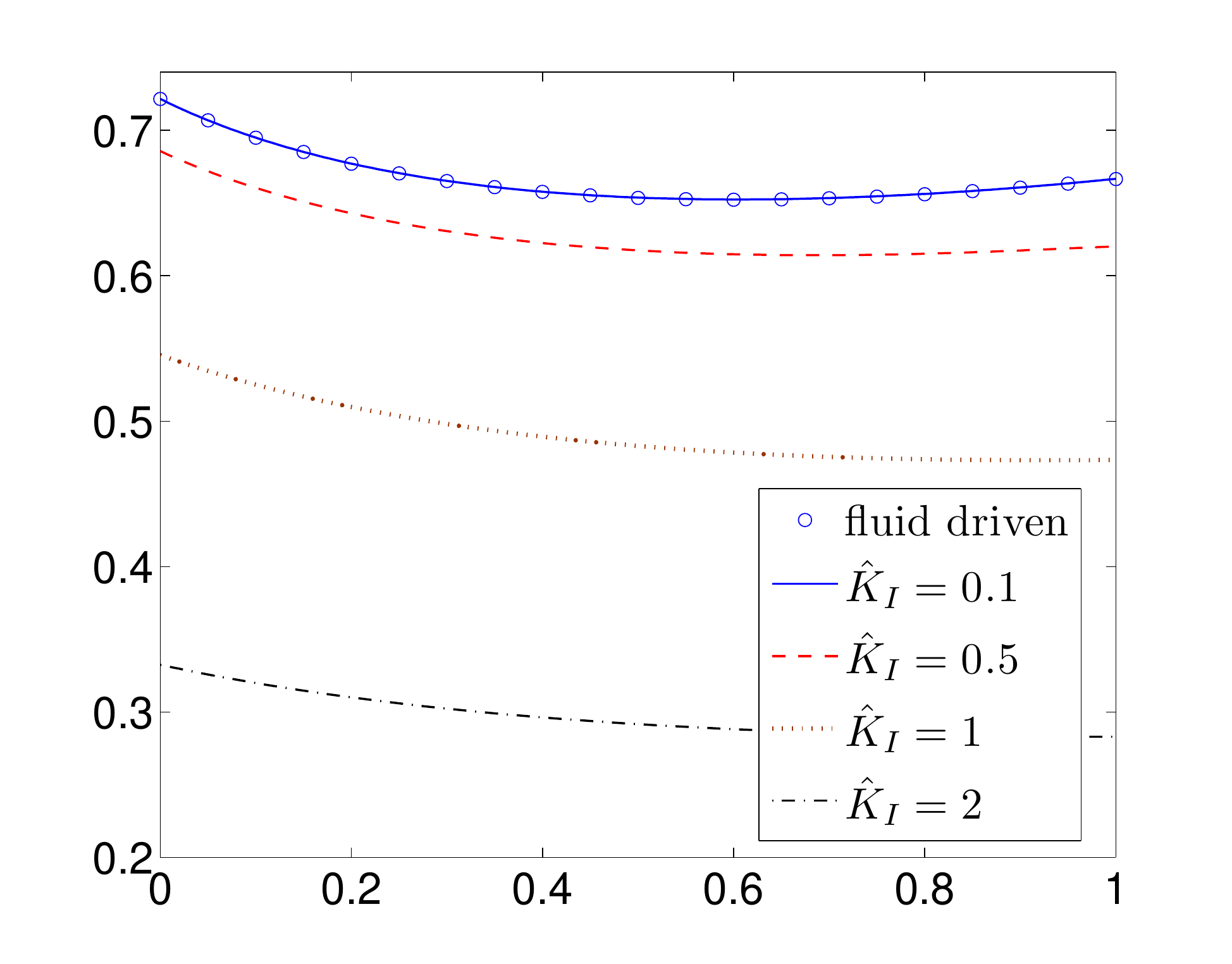}
    \put(-105,0){$x$}
    \put(-230,90){$\hat v$}
    \put(-230,160){$\textbf{a)}$}
    \hspace{2mm}
    \includegraphics [scale=0.40]{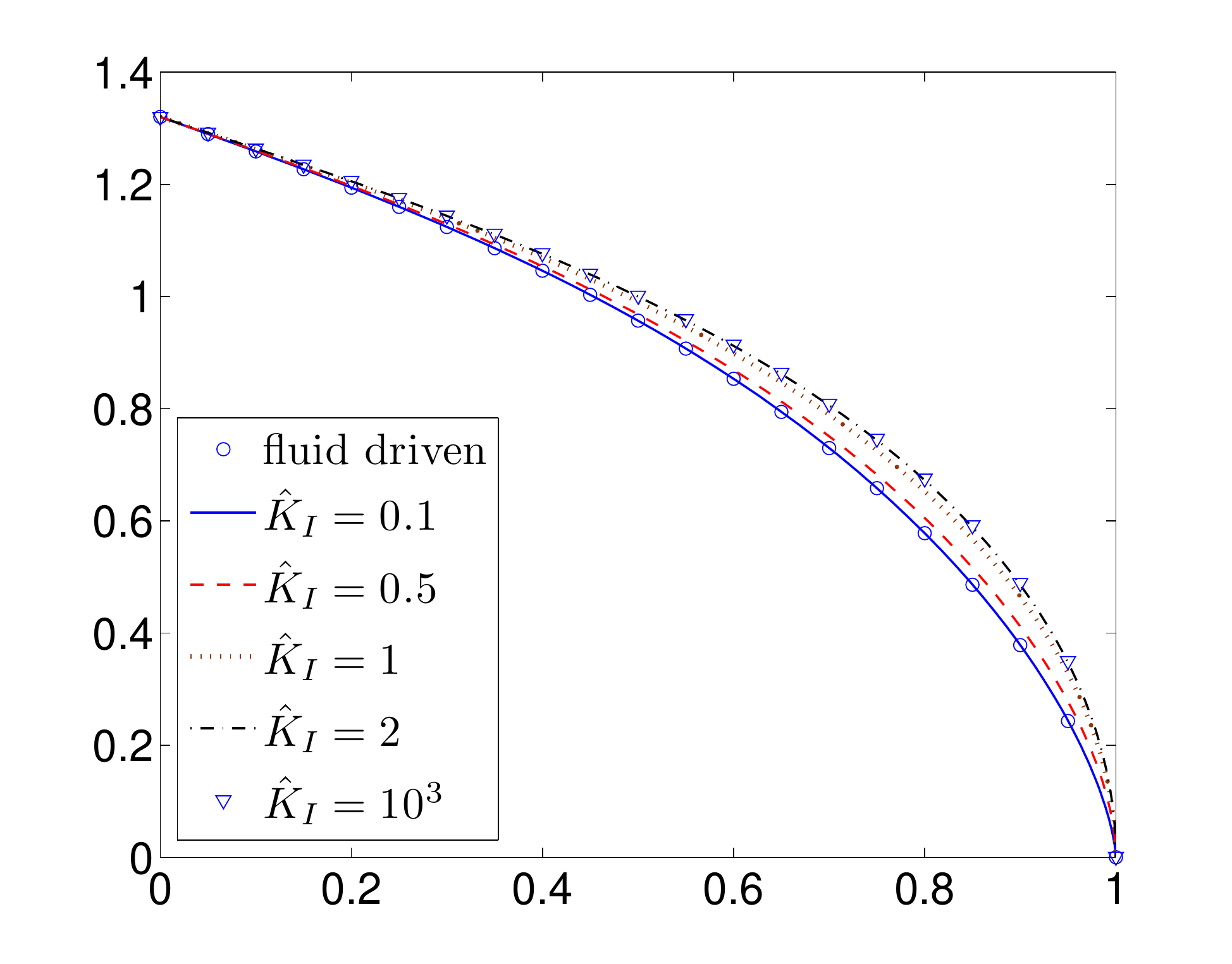}
    \put(-105,0){$x$}
    \put(-230,90){$\hat q$}
    \put(-230,160){$\textbf{b)}$}

    \caption{\emph{KGD model.} Solutions for different values of the self-similar stress intensity factor, $\hat K_I$: a) the particle velocity, $\hat v$, b) the fluid flow rate, $\hat q$. }

\label{rozne_K_2}
\end{figure}

To complete the analysis, in Fig.~\ref{rozne_K_1} -- Fig.~\ref{rozne_K_2} we depict the data for $\hat K_I=0.1,0.5,1,2$. As one can expect, an increase in $\hat K_I$ entails growth in the fracture width with simultaneous deceleration of the crack. When analyzing the pressure graphs, one can see that the average value of $\hat p$ is growing, which is a result of the increasing contribution of the square root term in the crack  opening (this single term gives a constant over $x$ component of pressure). The values of pressure derivatives decrease with $\hat K_I$ growth. This, together with the counteracting trend for $\hat w$, enables us to reach one of two extreme regimes: i) fluid driven regime (zero toughness), ii) storage regime (infinite toughness). The latter is illustrated for the fluid flux in Fig.~\ref{rozne_K_2} b) by the data for $\hat K_I=10^3$.  One can see that the transient state between two boundary cases is relatively small.

\section{Solution of the problem in transient regime}

In this section we discuss an extension of the algorithm presented above to the time-dependent variant of the problem. The main assumptions and blocks of the algorithm remain the same.
The new features introduced here are subroutines for approximating the temporal derivative and the crack length computation.
Presented numerical examples demonstrate the performance of the general variant of the algorithm.

\subsection{Problem formulation}

Analogously as it was for the self-similar variant of the problem, let us write the basic system of equations, collecting them in the order of
employment in the algorithm. The fundamental difference between the scheme presented in the previous section and the one for the transient problem is the introduction of a mechanism for approximation of the temporal derivative, which shall be accounted for while computing the reduced velocity. Here the initial condition is also implemented.
Finally, the crack length should be estimated for both main stages of the algorithm (for $\phi$ and $w$).

We consistently use in the computations the following representation of the temporal derivative of the crack opening:
\begin{equation}
\label{dwdt}
\frac{\partial w}{\partial t}\big|_{t_{i+1}}=2\frac{w(x,t_{i+1})-w(x,t_i)}{t_{i+1}-t_i}-\frac{\partial w}{\partial t}\big|_{t_i}.
\end{equation}
This representation has already been utilized in \cite{solver_calkowy}, where its advantages were discussed.
The representation (\ref{dwdt}) yields an accuracy of $O((\Delta t)^2)$, while the standard
finite difference provides the order of $O(\Delta t)$ only. Note that at every time step the value of the derivative for the previous time instant is known. At the initial time it is taken from the initial conditions and the continuity equation. Then, by formula \eqref{dwdt} one can update $w'_t$ for $t_{i+1}$.

Below we itemize the basic set of equations employed in the universal algorithm for the transient regime:
\begin{itemize}
\item{equation \eqref{dwdt} to compute the temporal derivative,}
\item{equation defining the reduced velocity in the following form (compare \eqref{Phi_cont})}
\begin{equation}
\label{fi_trans}
\phi=\frac{L}{w}\int_x^1\left(\frac{\partial w}{\partial t}+\frac{{\cal{C}}_A {\cal{L}}(w)}{L^{m+1}}w+q_l\right)d\eta,
\end{equation}
\item
solvability condition resulting from \eqref{fi_trans} and \eqref{q0_phi}, equivalent to the global fluid balance condition \eqref{global_balance_n} utilized to compute the crucial  parameter ${\cal L}(w)$
\begin{equation}
\label{solv_trans}
\int_0^1 \frac{\partial w(t,x)}{\partial t}dx-\frac{q_0(t)}{L(t)}+{\cal{C}}_A\frac{{\cal{L}}(w)}{L^{m+1}(t)}\int_0^1w(t,x)dx+\int_0^1q_l(t,x)dx=0,
\end{equation}
\item{equation to compute the crack opening obtained by merging \eqref{p_prim} with the respective form of the inverse elasticity operator
${\cal{B}}_w$ (compare \eqref{p_prim_1a}, \eqref{w_B1_PKN} or \eqref{inv_KGD_n_1}) }
\begin{equation}
\label{B_2_dyn}
w ={\cal B}_w \left(\frac{L(t)}{ w^2}\Big( \phi +x{\cal L}( w)\frac{C_{\cal A}}{L^m(t)}\Big)\right),
\end{equation}
\item{boundary conditions: \eqref{q_0n}$_2$, \eqref{phi_tip} and \eqref{q0_phi},}
\item{initial conditions \eqref{IC_n},}
\item{relation to compute the crack length: \eqref{L_int_PKN}, \eqref{L_int_KGD} or \eqref{L_tough_w0}, respectively.}
\end{itemize}

\subsection{Computational algorithm for the transient regime}

The solution to the transient variant of the problem, described by the system of equations collected above, is sought in the framework of an iterative algorithm. The main idea and assumptions of the numerical scheme are the same as for the self-similar formulation. By analogy to the description given in subsection 5.2, we can define the following stages of computations when looking for an unknown solution at the time instant $t_{i+1}$:
\begin{itemize}
\item
\emph{Preliminary step.} The process is initiated by specifying the first approximation of the crack opening $w^{(j-1)}(t_{i+1},x)$. One can use here the preconditioning based on the temporal derivative $w'_t(t_i,x)$ and the initial condition $w(t_i,x)$. The first approximation of the crack length $L^{(j-1)}(t_{i+1})$ can be also easily computed by preconditioning based on the value of the parameter ${\cal{L}}(w_i)$ from the previous time step (or initial conditions).
     \item
    \emph{First step.} According to \eqref{dwdt} the temporal derivative of the crack opening is computed. Note, that when obtaining the final solution $w(t_{i+1},x)$, one automatically has its temporal derivative too.
    Next, equation \eqref{solv_trans} yields ${\cal{L}}^{(j)}(w_{i+1})$, which substituted into \eqref{fi_trans} gives the reduced particle velocity $\phi ^{(j)}_{i+1}$. The integration in \eqref{fi_trans} is carried out with application of the $\varepsilon$-regularization technique, where the regularized tip condition has the form of \eqref{fi_reg}. As a result, functions ${\cal{L}}^{(j)}(w_{i+1})$ and $\phi ^{(j)}_{i+1}$ computed at this stage satisfy: i) fluid balance equation \eqref{solv_trans}, ii) continuity equation \eqref{fi_trans}, iii) regularized boundary condition for $\phi$ (see \eqref{fi_reg}) which is an equivalent of \eqref{phi_tip}, iv) the influx boundary condition \eqref{q0_phi} indirectly through the fluid balance equation.
\item{\emph{Second step}. The crack length is updated by substituting ${\cal{L}}^{(j)}(w_{i+1})$ into one of \eqref{L_int_PKN} -- \eqref{L_tough_w0}. Then, the next approximation of the crack opening $w^{(j)}_{i+1}$ is obtained from \eqref{B_2_dyn}. The technique of numerical computation of the operator ${\cal B}_w$ is exactly the same as it was for the self-similar variant of the problem. Also here, ${\cal{L}}^{(j)}(w_{i+1})$ is considered a natural regularization parameter,
    used to satisfy the influx boundary condition \eqref{q0_phi}.}

\end{itemize}
The aforementioned \emph{two steps} of the algorithm are repeated until respective components of the solution have converged to within a prescribed tolerance.

{\sc Remark 11.} The modular algorithm architecture enables us to easily introduce the subroutine
for the crack length computation as an additional block. Naturally, this block was not present in the self-similar variant of the algorithm.

\subsection{Algorithm performance in the transient regime}

In this part of the paper we present a brief investigation into the performance of the universal algorithm for various hydraulic fracture models. The aim of this analysis is just to highlight its main peculiarities.

\subsubsection{Algorithm performance for the PKN model}

We utilize for the computations the benchmark solution described in subsection 2.3.1 as {\it{benchmark I}} for the time dependent term of the power law type (see \eqref{psi_pow}), where $\gamma=1/3$. For the transient regime, the accuracy and performance of the algorithm depends on discretization of both
independent variables. For this reason, four different
variants of spatial and temporal meshing are considered.
The number of spatial mesh points is denoted by $N$, while $M$ stands for the number of predefined time steps.
We analyze respective combinations for $N=20$, $N=100$ and $M=50$, $M=100$.
The time stepping strategy was taken from \cite{solver_calkowy} (p.162 formula (60)). The target time, $t=100$, is the same for all computations.
As previously for the self-similar problem, the spatial mesh density was increased at both ends of the interval.

In the analysis we use the following  measures of the  solution accuracy: i) the relative error of the crack opening, $\delta w$, ii) the relative error of the reduced particle velocity, $\delta \phi$, iii) the relative error of the crack length, $\delta L$,  iv) and finally the relative error of the temporal derivative of the crack opening, $\delta w'_t$.

\begin{figure}[h!]
\begin{center}
    \includegraphics [scale=0.35]{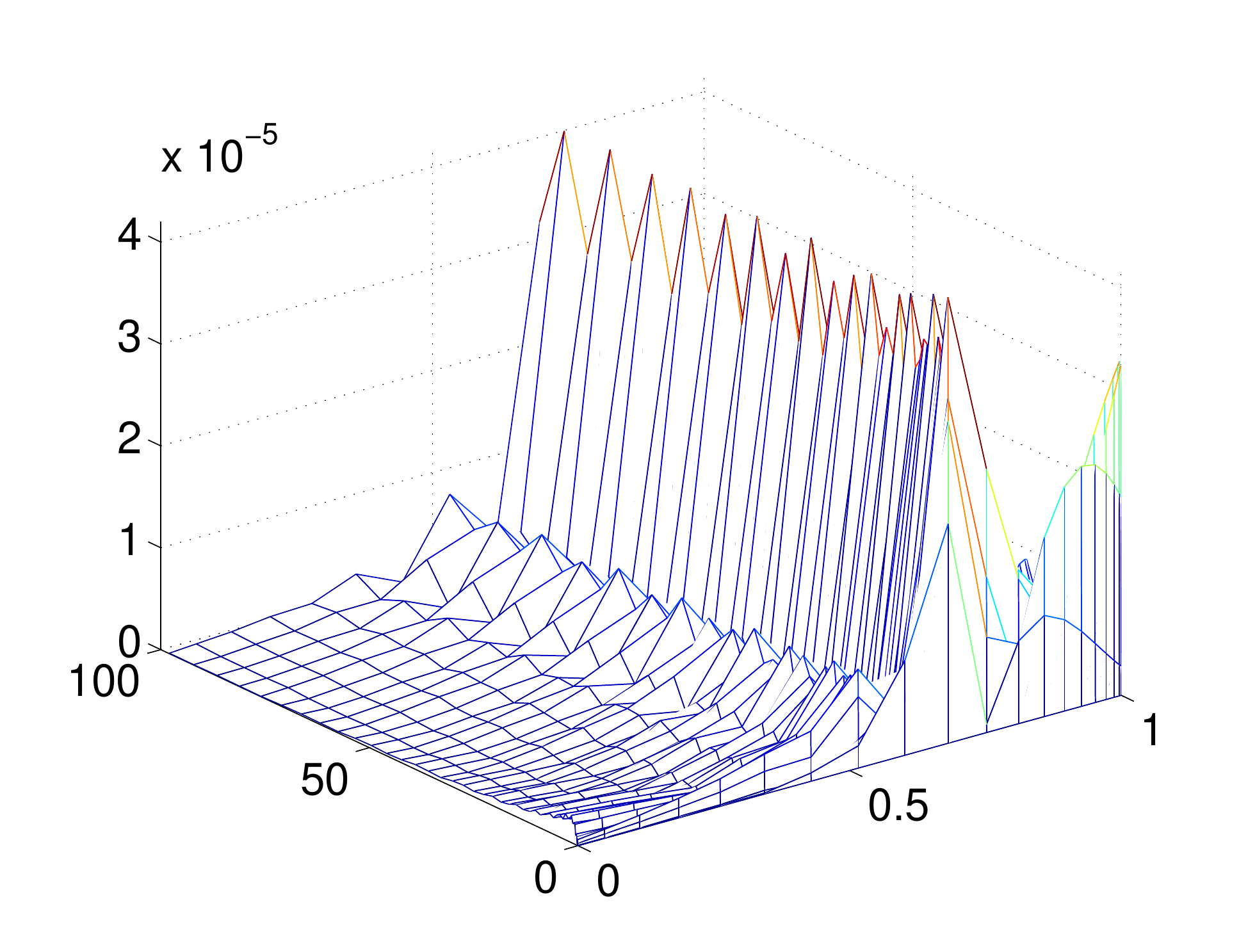}
    \put(-50,10){$x$}
    \put(-155,15){$t$}
    \put(-200,80){$\delta w$}
    \put(-220,130){$\textbf{a)}$}
    \hspace{2mm}
    \includegraphics [scale=0.35]{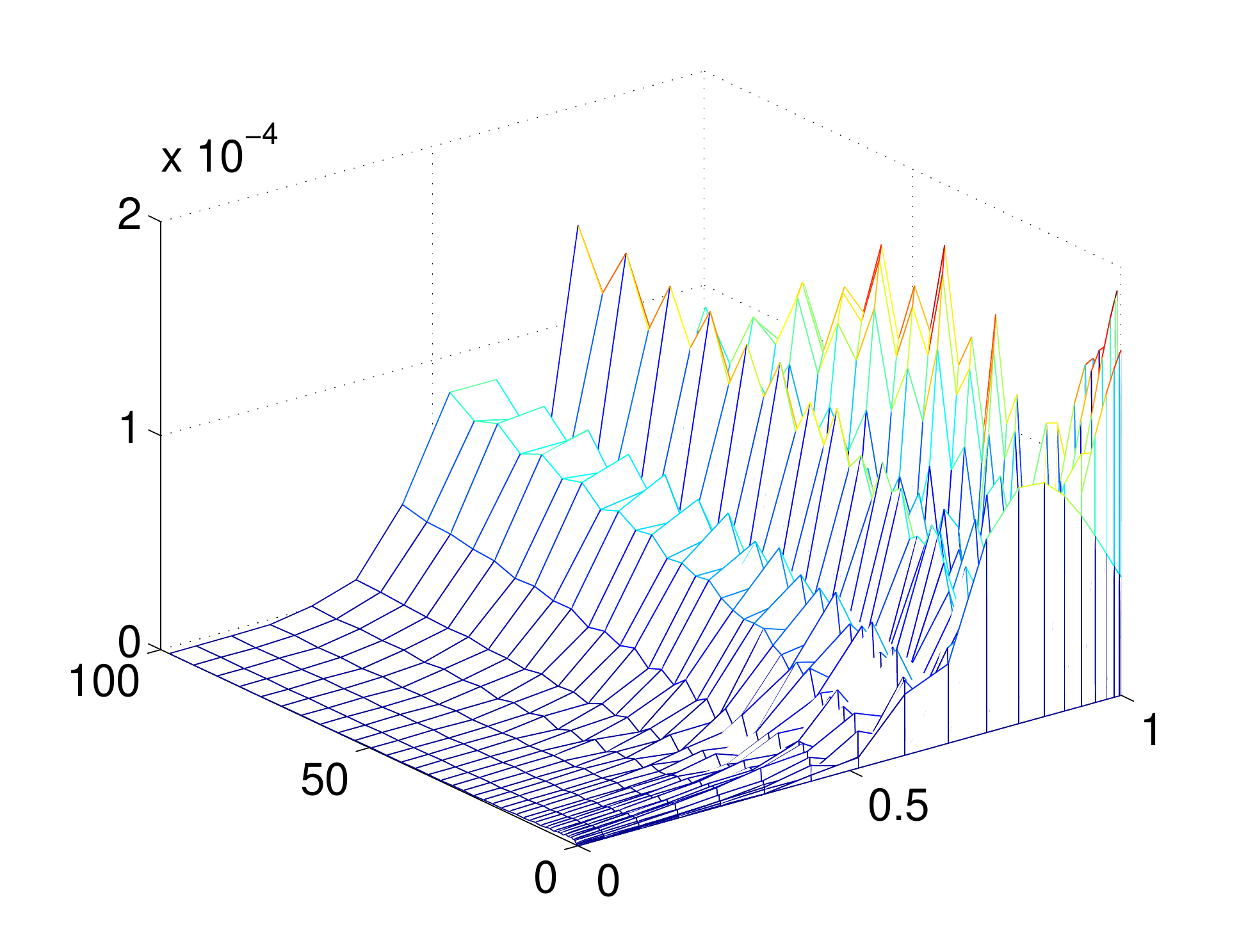}
    \put(-50,10){$x$}
    \put(-155,15){$t$}
    \put(-200,80){$\delta \phi$}
    \put(-220,130){$\textbf{b)}$}
\end{center}
    \caption{\emph{PKN model.} The relative errors of solution for $N=20$ (spatial mesh), $M=50$ (temporal mesh): a) the error of crack opening, $\delta w$, b) the error of reduced particle velocity, $\delta \phi$. }

\label{bledy_PKN_1}
\end{figure}

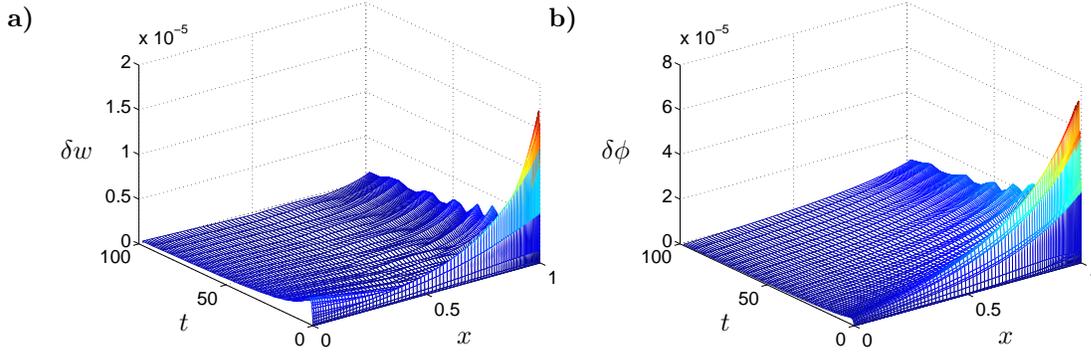
\begin{figure}[h!]
\begin{center}
    \includegraphics [scale=0.35]{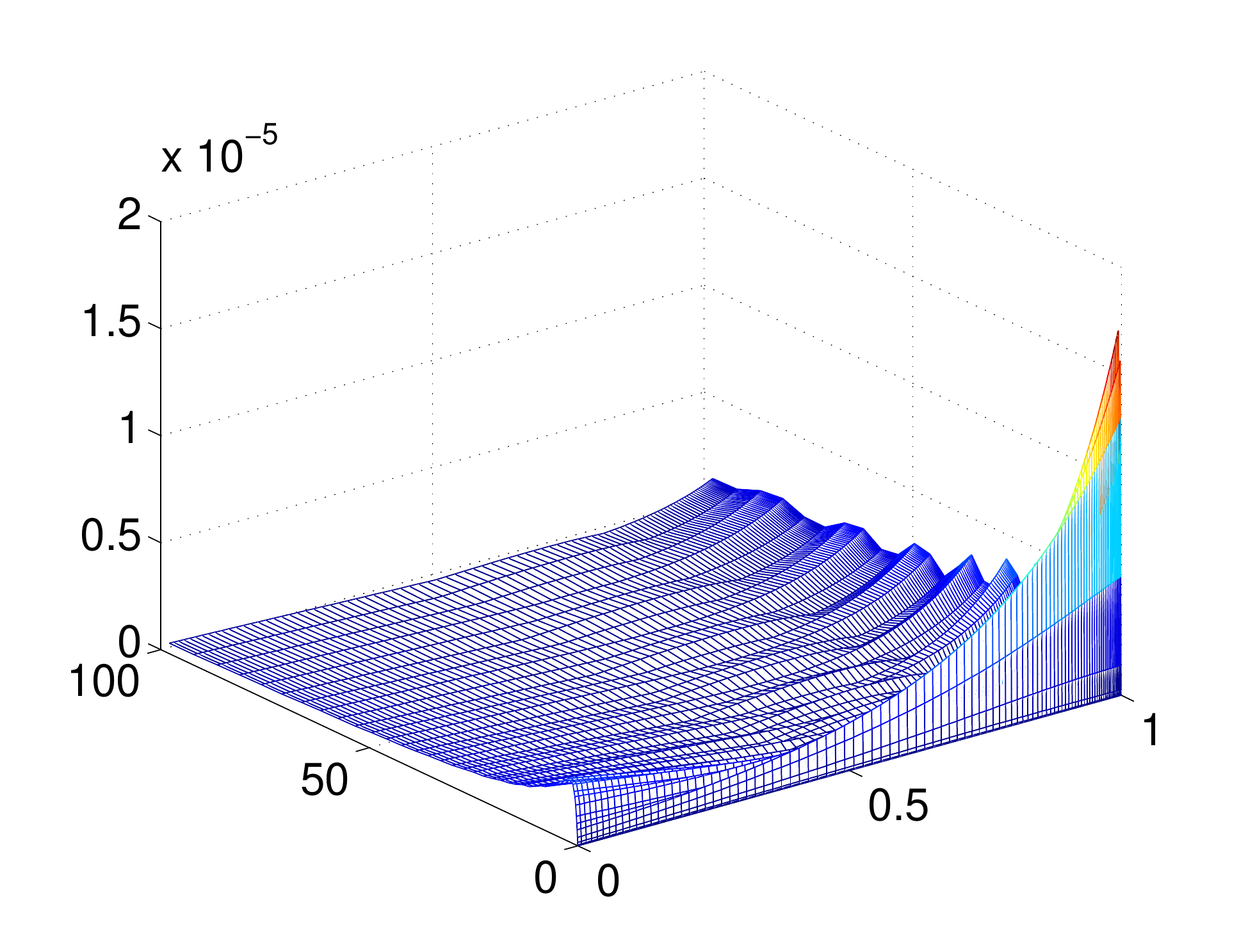}
    \put(-50,10){$x$}
    \put(-155,15){$t$}
    \put(-200,80){$\delta w$}
    \put(-220,130){$\textbf{a)}$}
    \hspace{2mm}
    \includegraphics [scale=0.35]{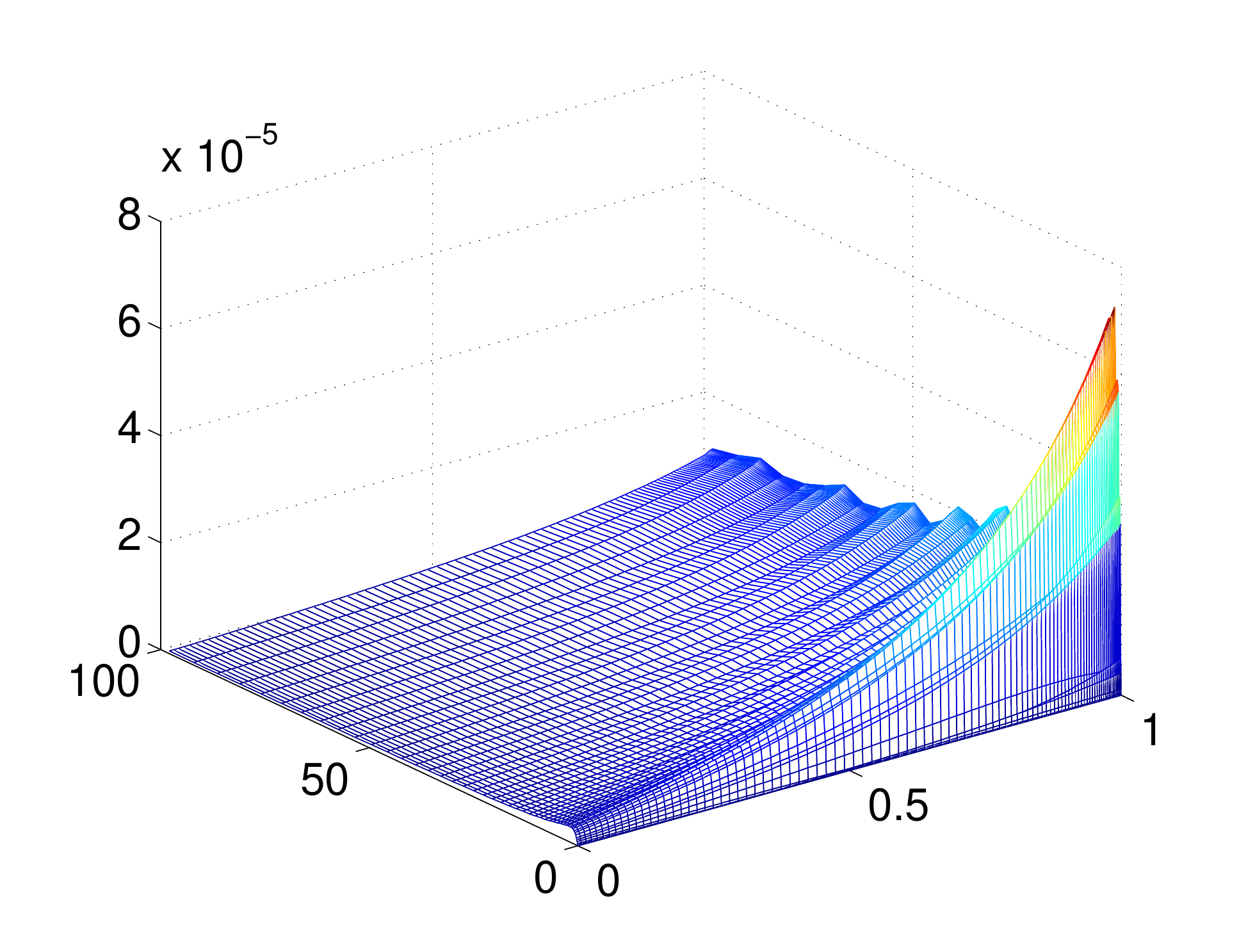}
    \put(-50,10){$x$}
    \put(-155,15){$t$}
    \put(-200,80){$\delta \phi$}
    \put(-220,130){$\textbf{b)}$}
\end{center}
    \caption{\emph{PKN model.}  The relative errors of solution for $N=100$ (spatial mesh), $M=50$ (temporal mesh): a) the error of crack opening, $\delta w$, b) the error of reduced particle velocity, $\delta \phi$. }

\label{bledy_PKN_2}
\end{figure}

\vspace{10mm}

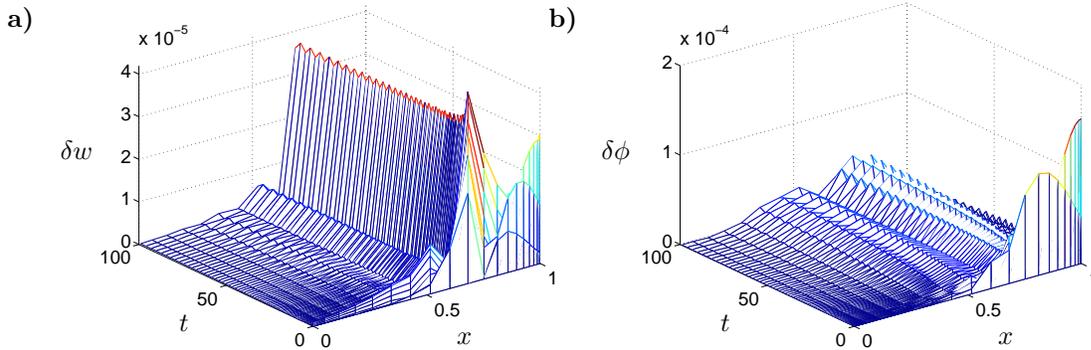
\begin{figure}[h!]
\begin{center}
    \includegraphics [scale=0.35]{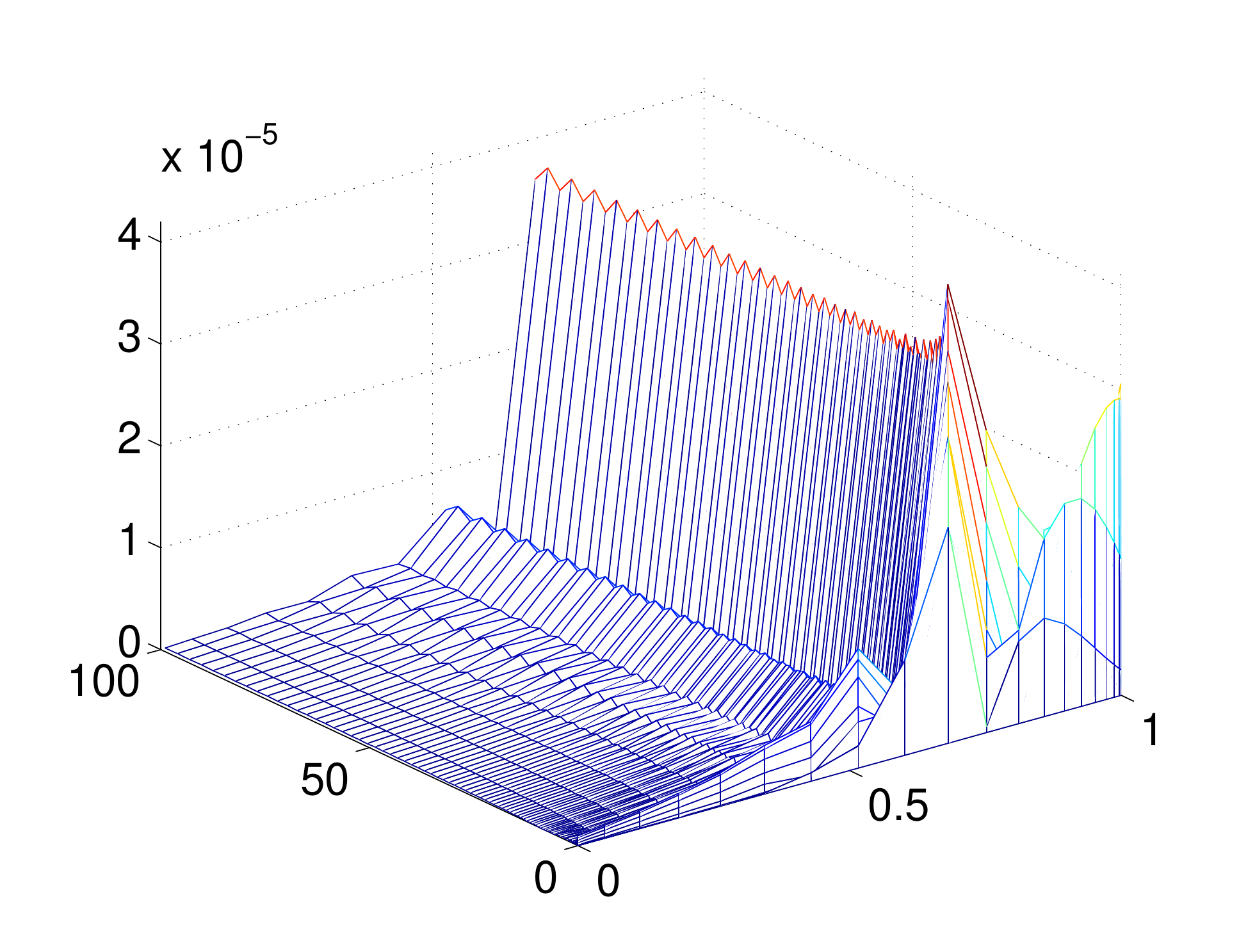}
    \put(-50,10){$x$}
    \put(-155,15){$t$}
    \put(-200,80){$\delta w$}
    \put(-220,130){$\textbf{a)}$}
    \hspace{2mm}
    \includegraphics [scale=0.35]{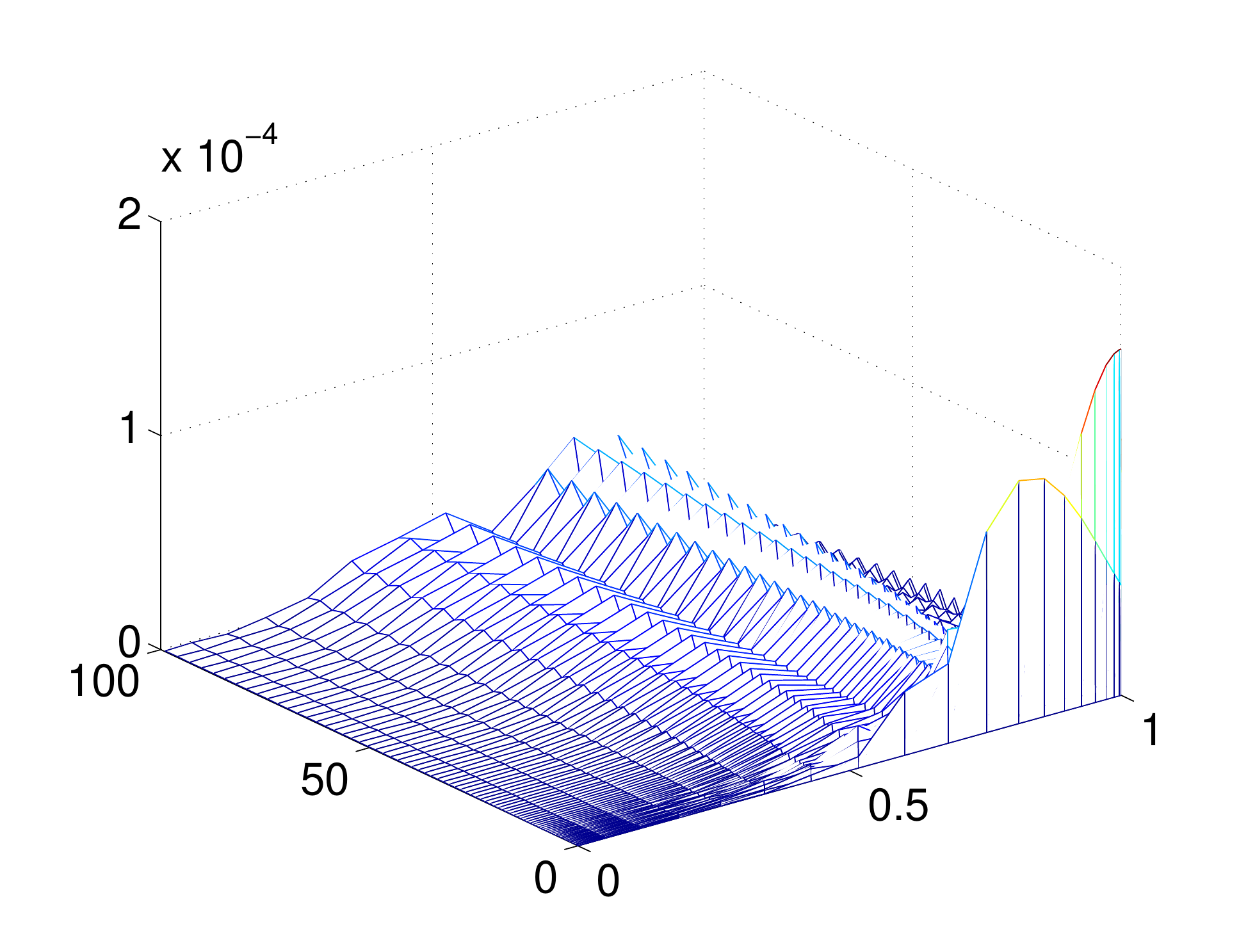}
    \put(-50,10){$x$}
    \put(-155,15){$t$}
    \put(-200,80){$\delta \phi$}
    \put(-220,130){$\textbf{b)}$}
\end{center}
    \caption{\emph{PKN model.} The relative errors of solution for $N=20$ (spatial mesh), $M=100$ (temporal mesh): a) the error of crack opening, $\delta w$, b) the error of reduced particle velocity, $\delta \phi$. }

\label{bledy_PKN_3}
\end{figure}

\begin{figure}[h!]
\begin{center}
    \includegraphics [scale=0.35]{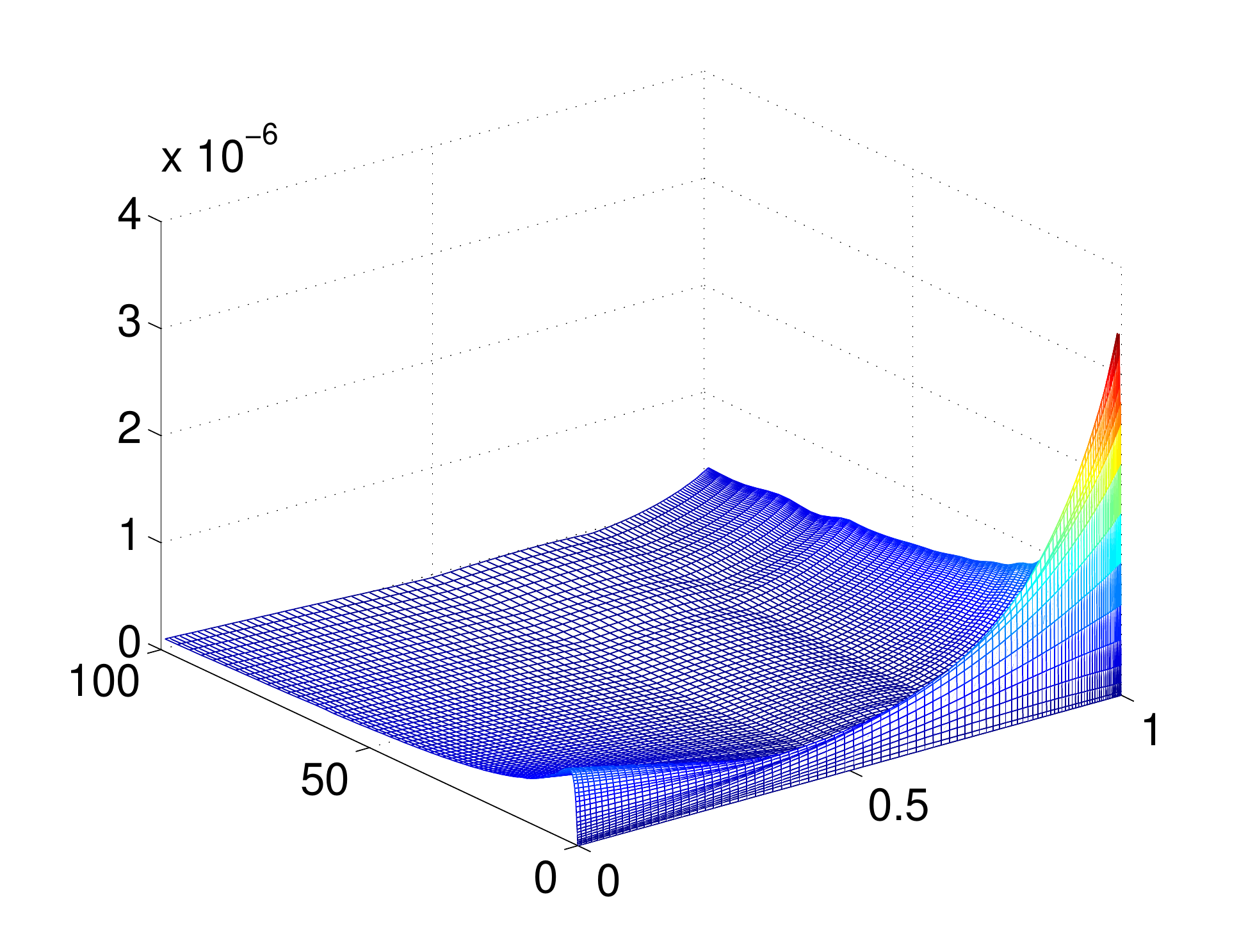}
    \put(-50,10){$x$}
    \put(-155,15){$t$}
    \put(-200,80){$\delta w$}
    \put(-220,130){$\textbf{a)}$}
    \hspace{2mm}
    \includegraphics [scale=0.35]{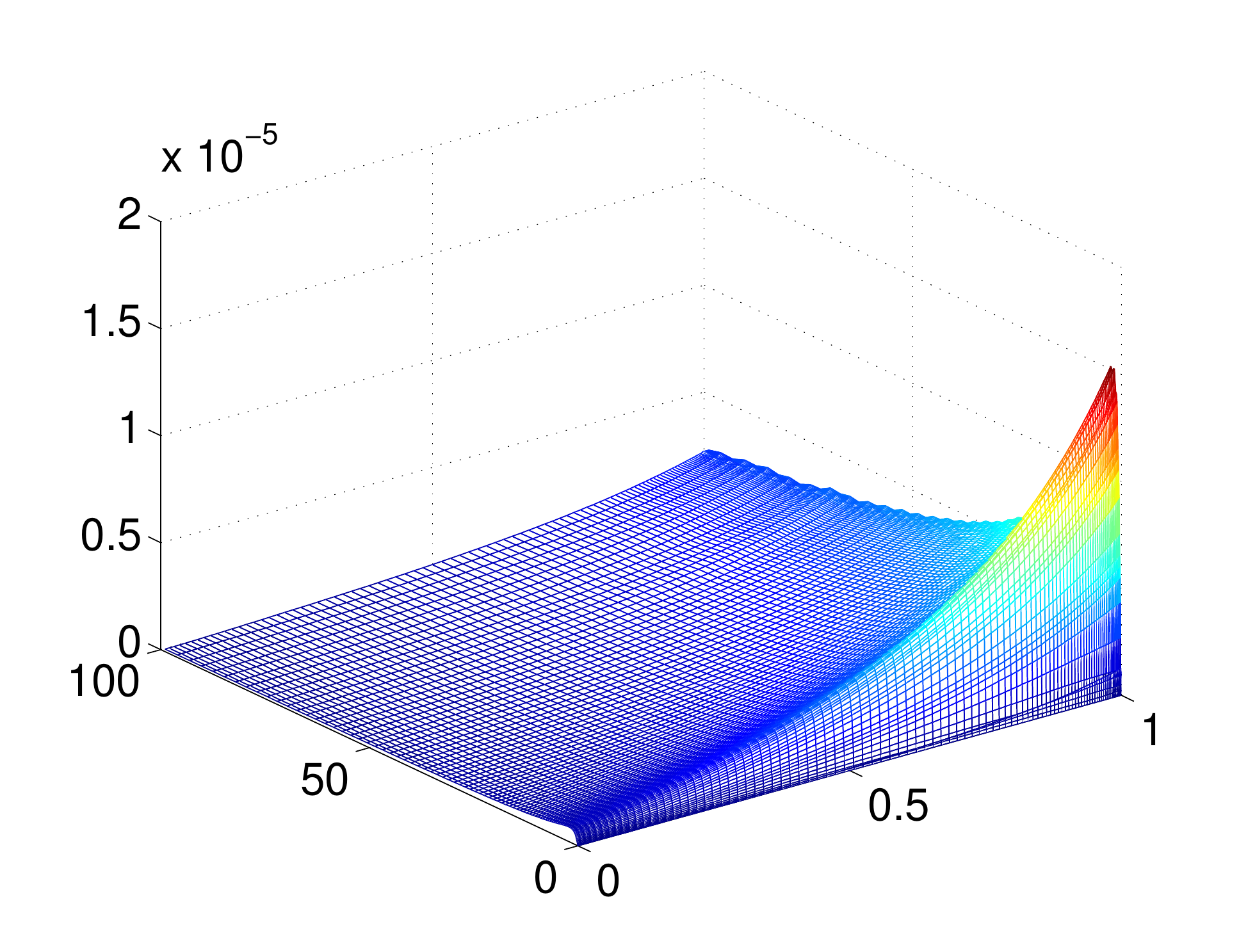}
    \put(-50,10){$x$}
    \put(-155,15){$t$}
    \put(-200,80){$\delta \phi$}
    \put(-220,130){$\textbf{b)}$}
\end{center}
    \caption{\emph{PKN model.} The relative errors of solution for $N=100$ (spatial mesh), $M=100$ (temporal mesh): a) the error of crack opening, $\delta w$, b) the error of reduced particle velocity, $\delta \phi$. }

\label{bledy_PKN_4}
\end{figure}

\begin{figure}[h!]
\begin{center}
    \includegraphics [scale=0.35]{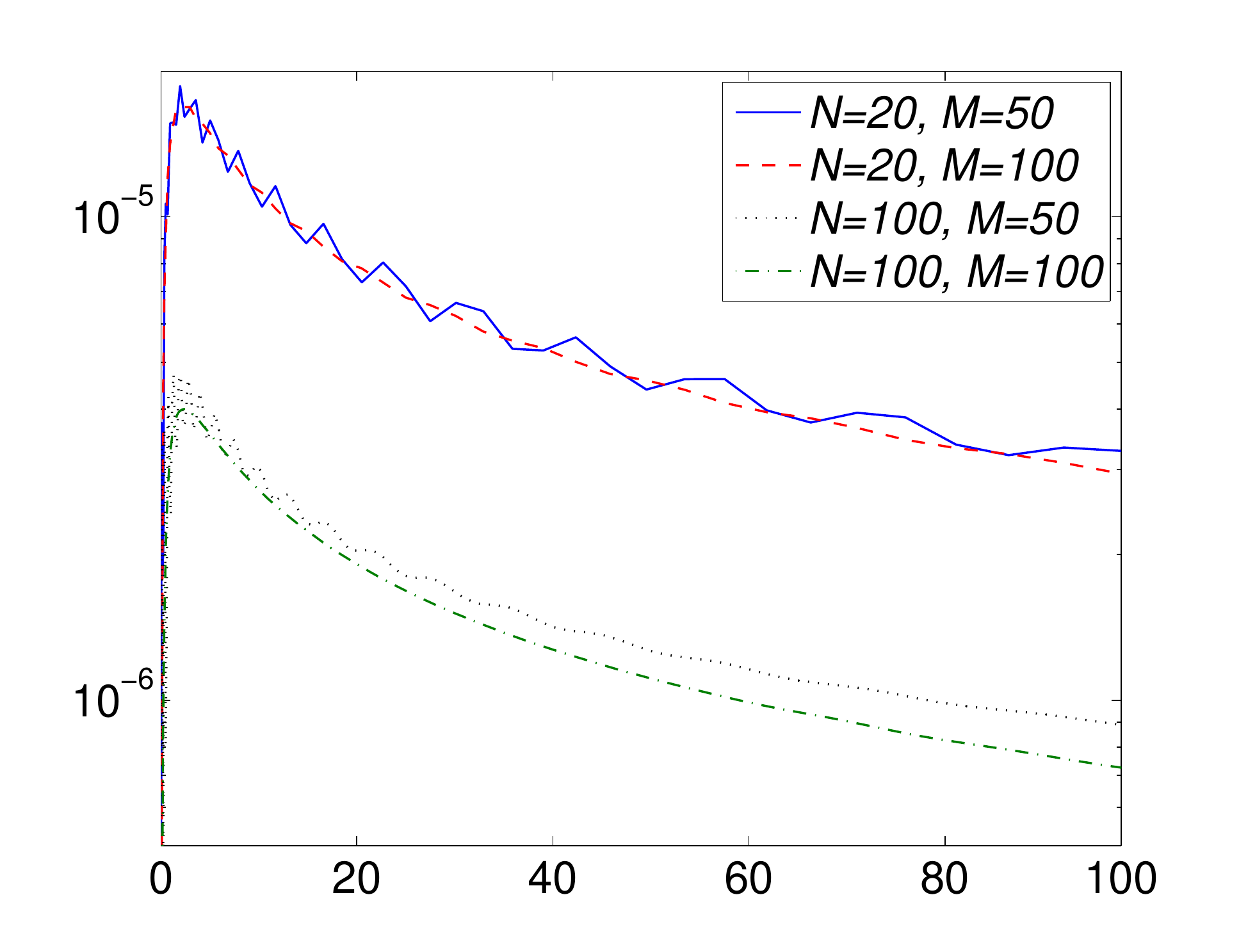}
    \put(-95,-5){$t$}
    \put(-200,70){$ \delta L$}
 \end{center}
    \caption{\emph{PKN model.} The crack length error, $\delta L$ for various spatial and temporal meshing. }
\label{bledy_PKN_5}
\end{figure}

\begin{figure}[h!]
\begin{center}
    \includegraphics [scale=0.35]{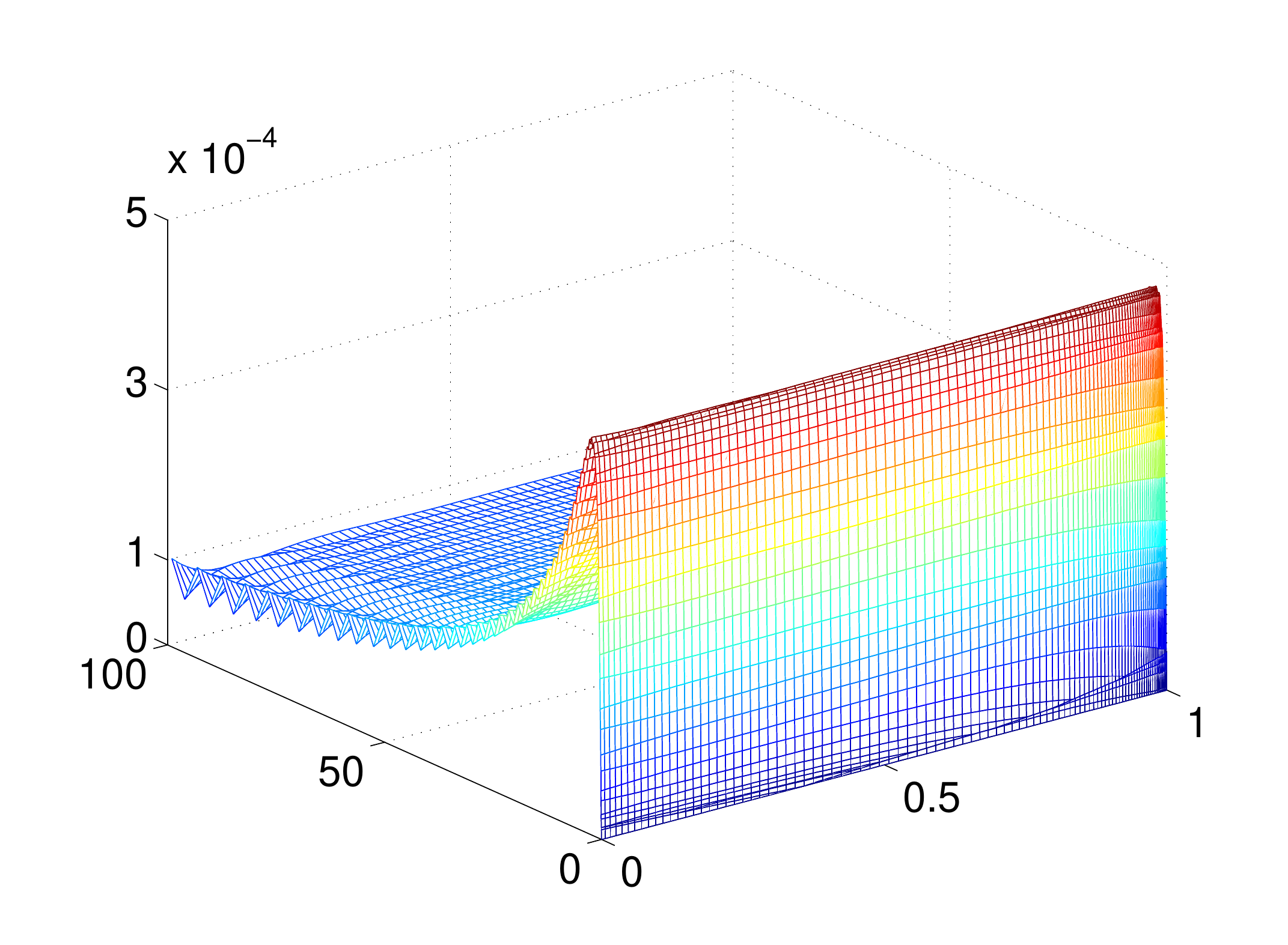}
    \put(-60,10){$x$}
    \put(-175,15){$t$}
    \put(-220,90){$\delta w'_t$}
    \put(-220,130){$\textbf{a)}$}
    \hspace{2mm}
    \includegraphics [scale=0.35]{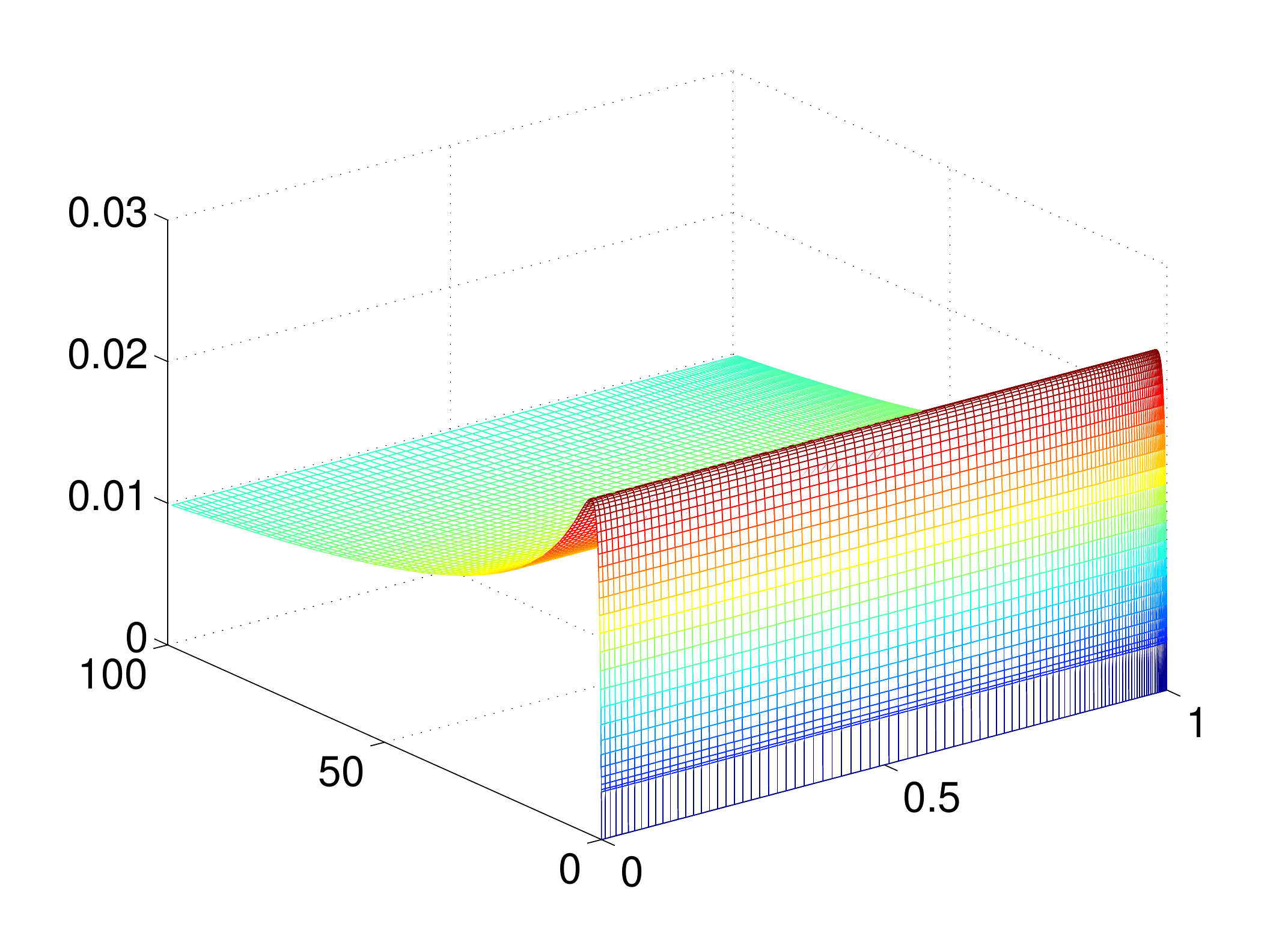}
    \put(-60,10){$x$}
    \put(-175,15){$t$}
    \put(-225,90){$\delta w'_t$}
    \put(-220,130){$\textbf{b)}$}
\end{center}
    \caption{\emph{PKN model.} The relative error of the solutions temporal derivative for $N=100$ (spatial mesh), $M=100$ (temporal mesh): a) improved temporal approximation b) ordinary finite difference. }

\label{dwt_PKN}
\end{figure}
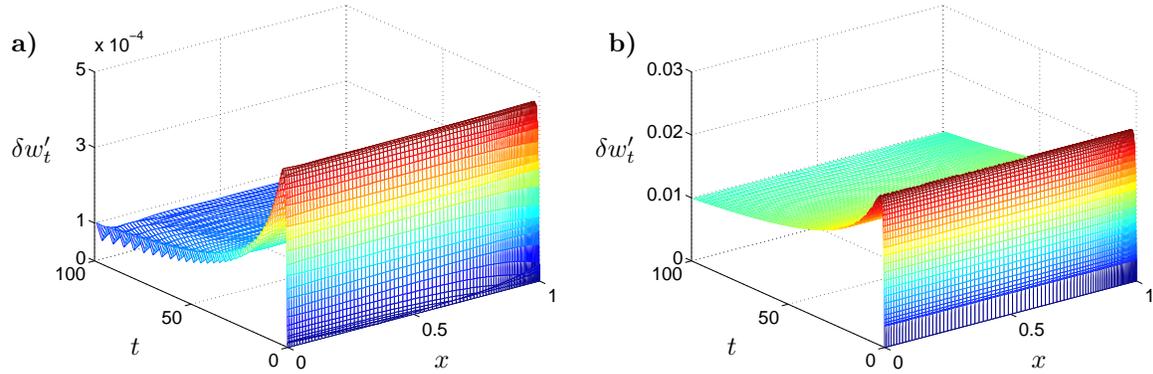

The computational errors for the crack opening, reduced velocity and the temporal derivative are shown in Fig.~\ref{bledy_PKN_1} -- Fig.~\ref{dwt_PKN}.

When analyzing the solution errors for $w$ and $\phi$, one can see that the relation between $N$ and $M$ is of crucial importance, while in general the finer meshing produces better results. For example, for $N=20$ there is almost no difference in accuracy between $M=50$ and $M=100$. In this case, the overall solution error is limited by the accuracy resulting from a coarse spatial meshing. However, better stabilization of the error in time can be seen for $M=100$ (compare Fig. \ref{bledy_PKN_1} and Fig. \ref{bledy_PKN_3}). On the other hand, when the spatial meshing is appreciably improved ($N=100$), the same change in the time step (from $M=50$ to $M=100$) yields results up to one order of magnitude better (see Fig. \ref{bledy_PKN_2} and Fig. \ref{bledy_PKN_4}).
In all the investigated cases one can see a very low error of $w$ and $\phi$ when $x=0$. It shows very good fulfillment of the influx boundary condition \eqref{q0_phi}.

The graphs for evolution of the crack length error, $\delta L$, over time are collected in Fig.~\ref{bledy_PKN_5} for all the considered  discretization variants. It shows that  the accuracy of the crack length is primarily determined by the quality of spatial meshing. It is directly related to the quality of computation of the parameter ${\cal{L}}(w)$ defining the crack propagation speed (compare \eqref{LC_PKN} -- \eqref{LC_KGD_toughness}), which in fact  is based on the leading asymptotic term(s) of the crack opening. Indeed, better accuracy for the crack opening $w$ near the crack tip corresponds to better accuracy for the crack length $L$. Note that a low sensitivity of the results to the time step density for a fixed $N$ is a direct consequence of using the general relation \eqref{v_0_univ} following from the speed equation.

In the end of this subsection let us discuss the issue of approximating the temporal derivative of the crack opening by formula \eqref{dwdt}. In Fig.~\ref{dwt_PKN} we show the relative errors of $w'_t$ for two ways of computing it: a) by formula \eqref{dwdt}, b) by the two-point finite difference (FD). The presented example involves $N=M=100$. As can be seen, although the character of the $\delta w'_t$ distribution is similar for both variants,  the first one gives an error two orders of magnitude lower than the second. For the coarser temporal meshing ($M=50$) we obtained approximately two times larger errors (we do not show this example in separate figures), however the mutual relation between both cases remained the same.

The advantages of using formula  \eqref{dwdt} in computations instead of the simplest FD scheme becomes less pronounced when one uses a rough spatial mesh. For example, when $N=20$ ($M=50, 100$) the errors for both approximations are of a similar order (2-4$\%$) to that shown in Fig.~\ref{dwt_PKN}b).

 In general, approximation \eqref{dwdt} is not worse than that the two-points FD, giving increasingly greater superiority when refining the computational mesh. Note that in our analysis the FD approximation $w'_t$ is in fact a post-processing. It is difficult to speculate to what degree it would deteriorate the overall solution accuracy when implemented in the algorithm instead of \eqref{dwdt}. Such a replacement, however, would neither introduce any simplification to the numerical scheme nor decrease the computational cost. The only benefit of this would be a minor saving in the memory (as there is no need to store the values of $w'_t$ from the previous time step).

\subsubsection{The algorithm performance for the fluid driven KGD model}

\begin{figure}[h!]

    \includegraphics [scale=0.40]{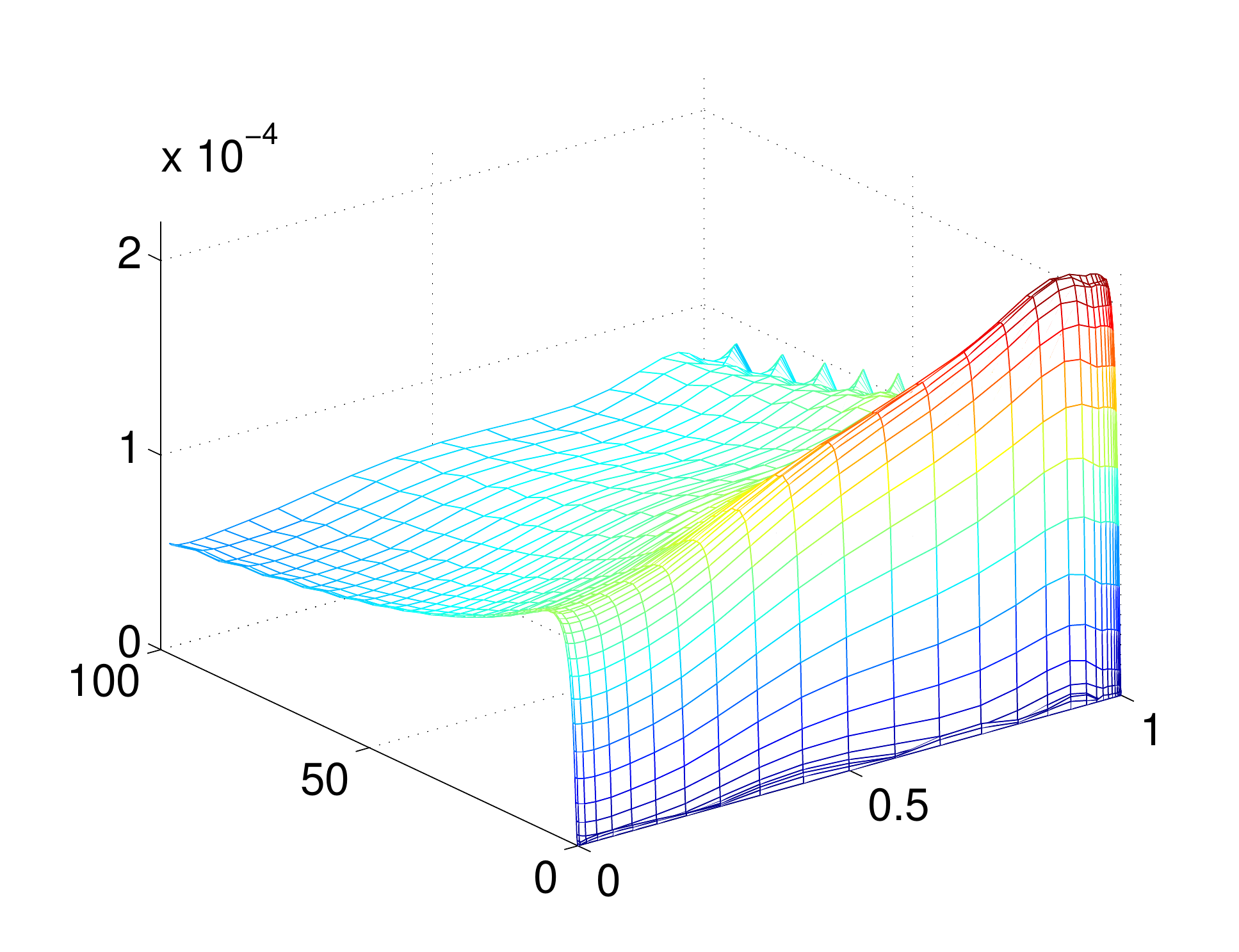}
    \put(-60,10){$x$}
    \put(-175,15){$t$}
    \put(-220,90){$\delta w$}
    \put(-220,160){$\textbf{a)}$}
    \hspace{2mm}
    \includegraphics [scale=0.40]{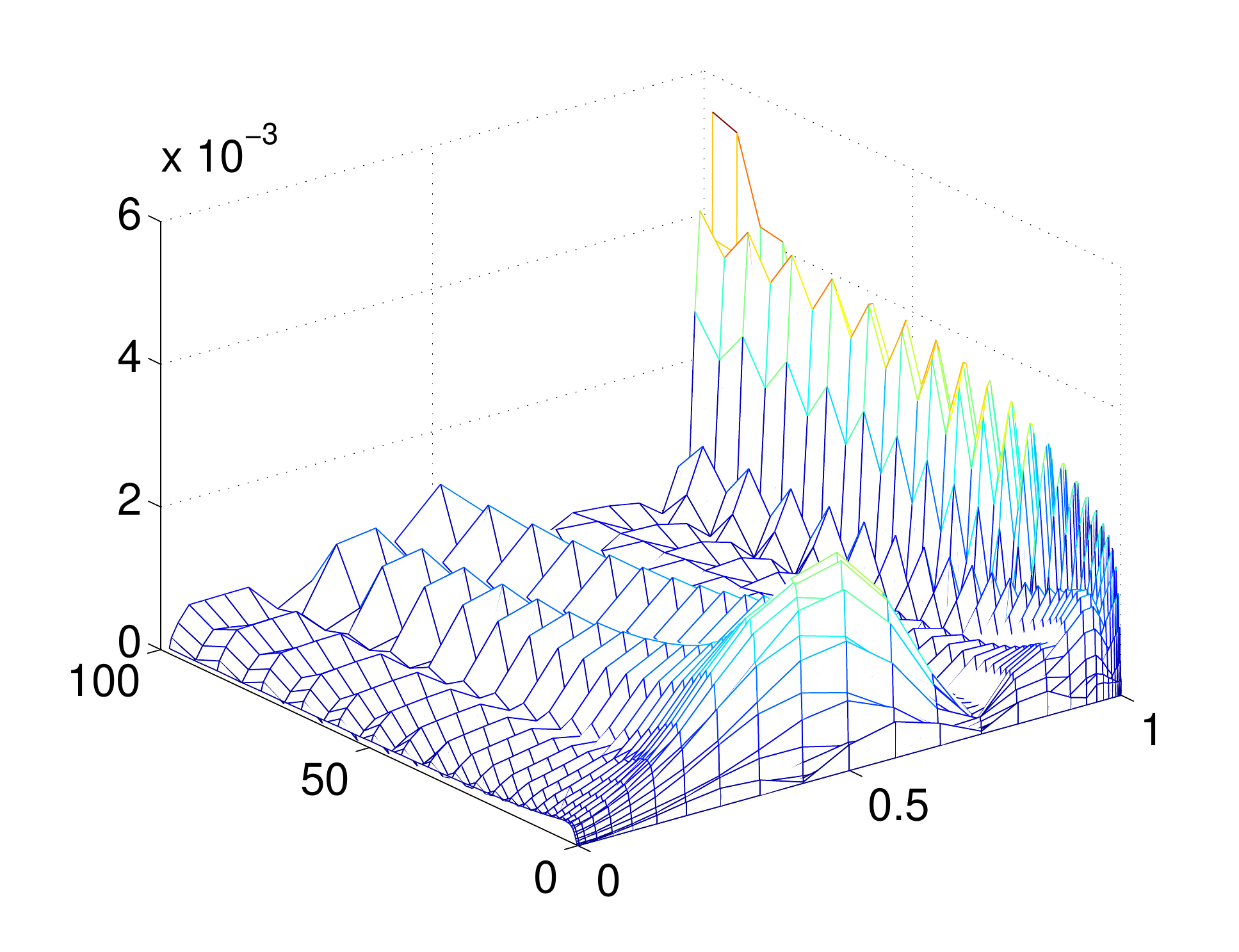}
    \put(-60,10){$x$}
    \put(-175,15){$t$}
    \put(-220,90){$\delta \phi$}
    \put(-220,160){$\textbf{b)}$}

    \caption{\emph{KGD model fluid driven regime.} The relative error of the solution for $N=30$ (spatial mesh), $M=50$ (temporal mesh): a) the error of crack opening, $\delta w$, b) the error of reduced particle velocity, $\delta \phi$. }

\label{bledy_KGD_f_1}
\end{figure}

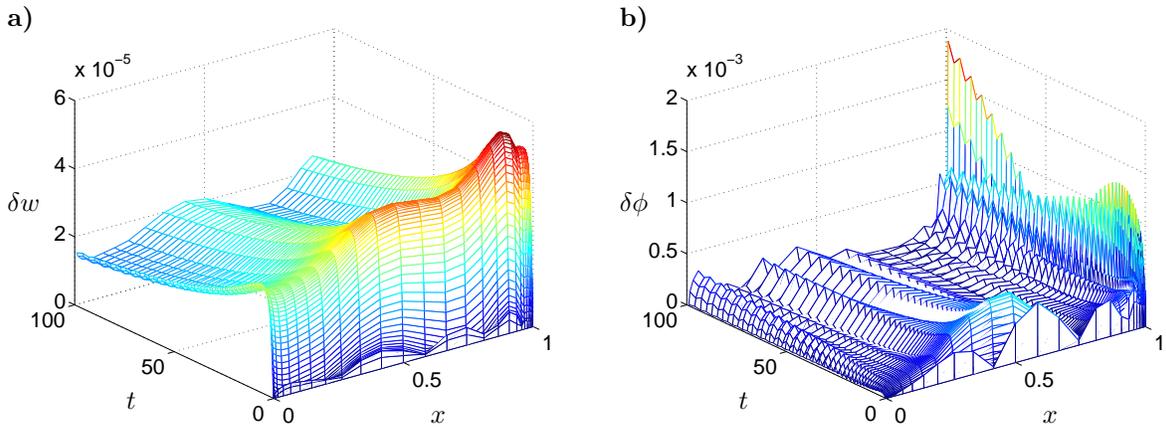
\begin{figure}[h!]

    \includegraphics [scale=0.40]{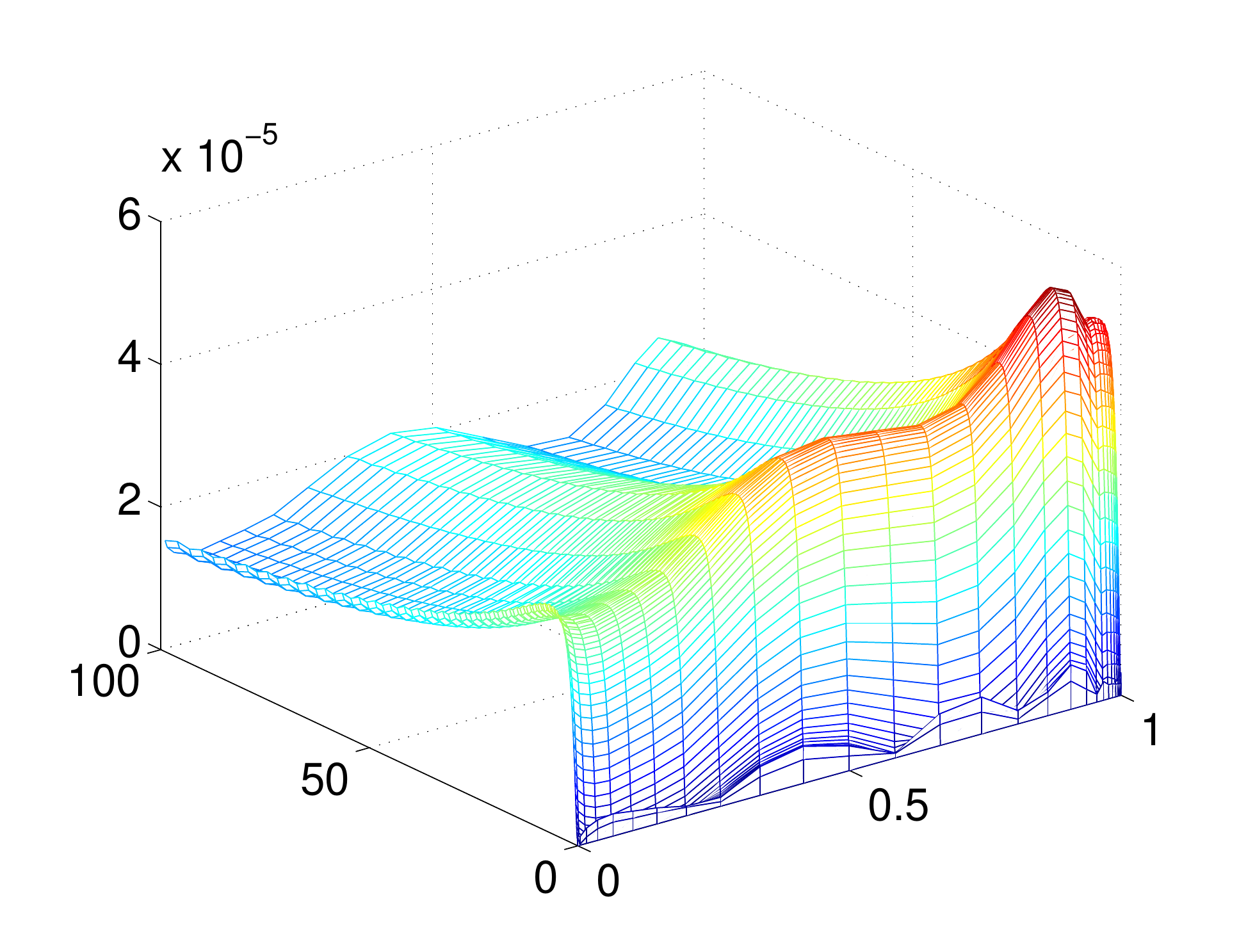}
    \put(-60,10){$x$}
    \put(-175,15){$t$}
    \put(-220,90){$\delta w$}
    \put(-220,160){$\textbf{a)}$}
    \hspace{2mm}
    \includegraphics [scale=0.40]{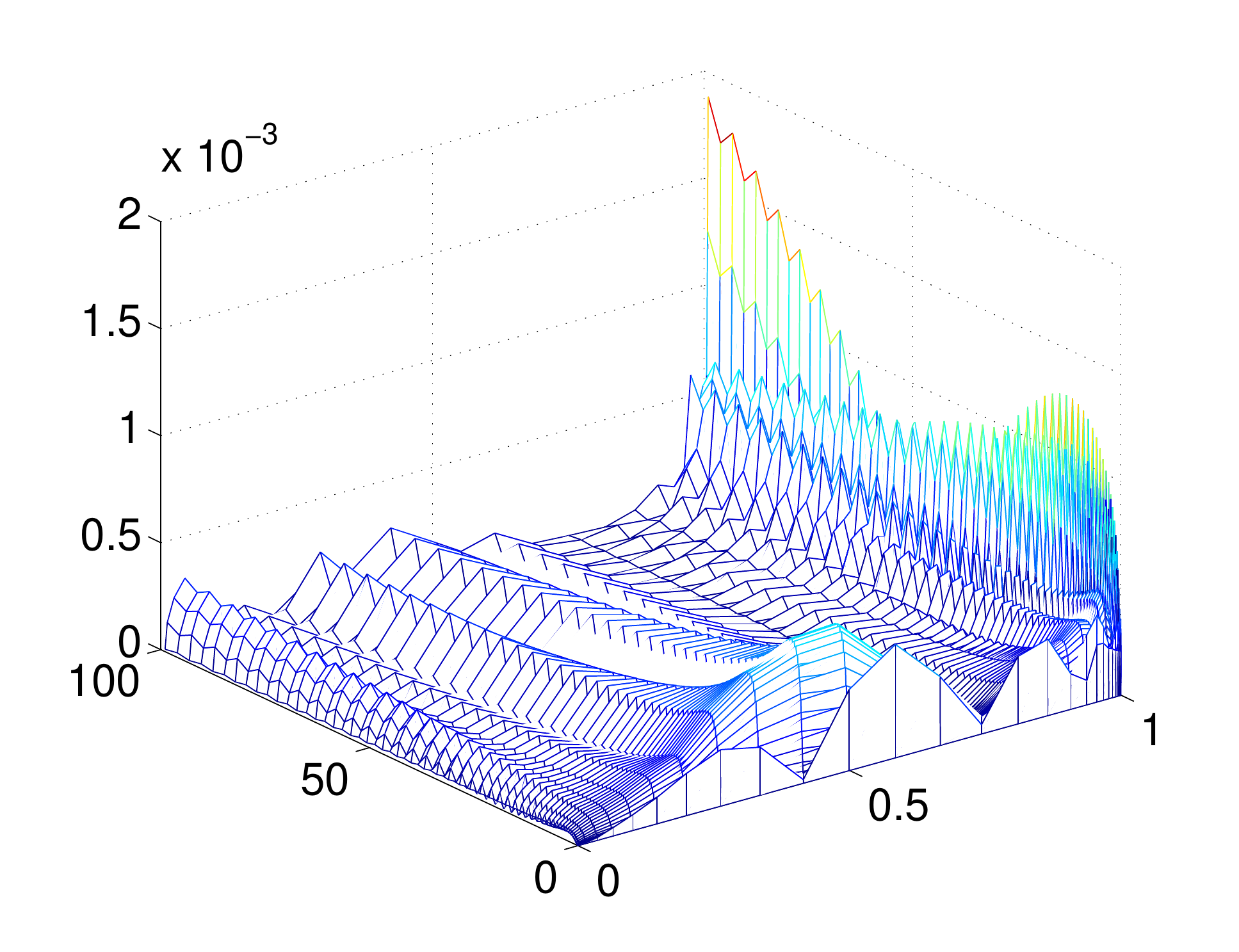}
    \put(-60,10){$x$}
    \put(-175,15){$t$}
    \put(-220,90){$\delta \phi$}
    \put(-220,160){$\textbf{b)}$}

    \caption{\emph{KGD model fluid driven regime.} The relative error of the solution for $N=30$ (spatial mesh), $M=100$ (temporal mesh): a) the error of crack opening, $\delta w$, b) the error of reduced particle velocity, $\delta \phi$. }

\label{bledy_KGD_f_2}
\end{figure}

\begin{figure}[h!]

    \includegraphics [scale=0.40]{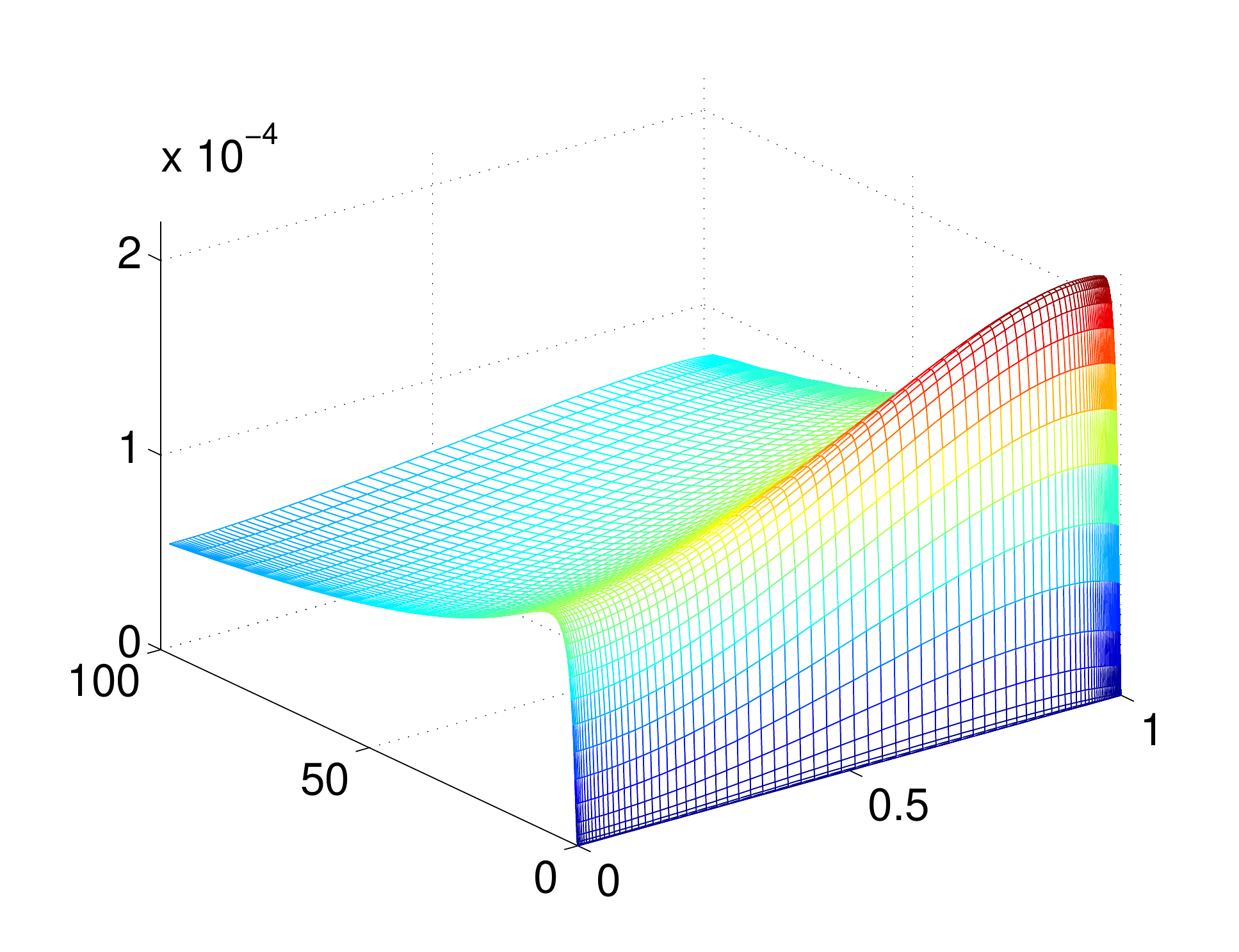}
    \put(-60,10){$x$}
    \put(-175,15){$t$}
    \put(-220,90){$\delta w$}
    \put(-220,160){$\textbf{a)}$}
    \hspace{2mm}
    \includegraphics [scale=0.40]{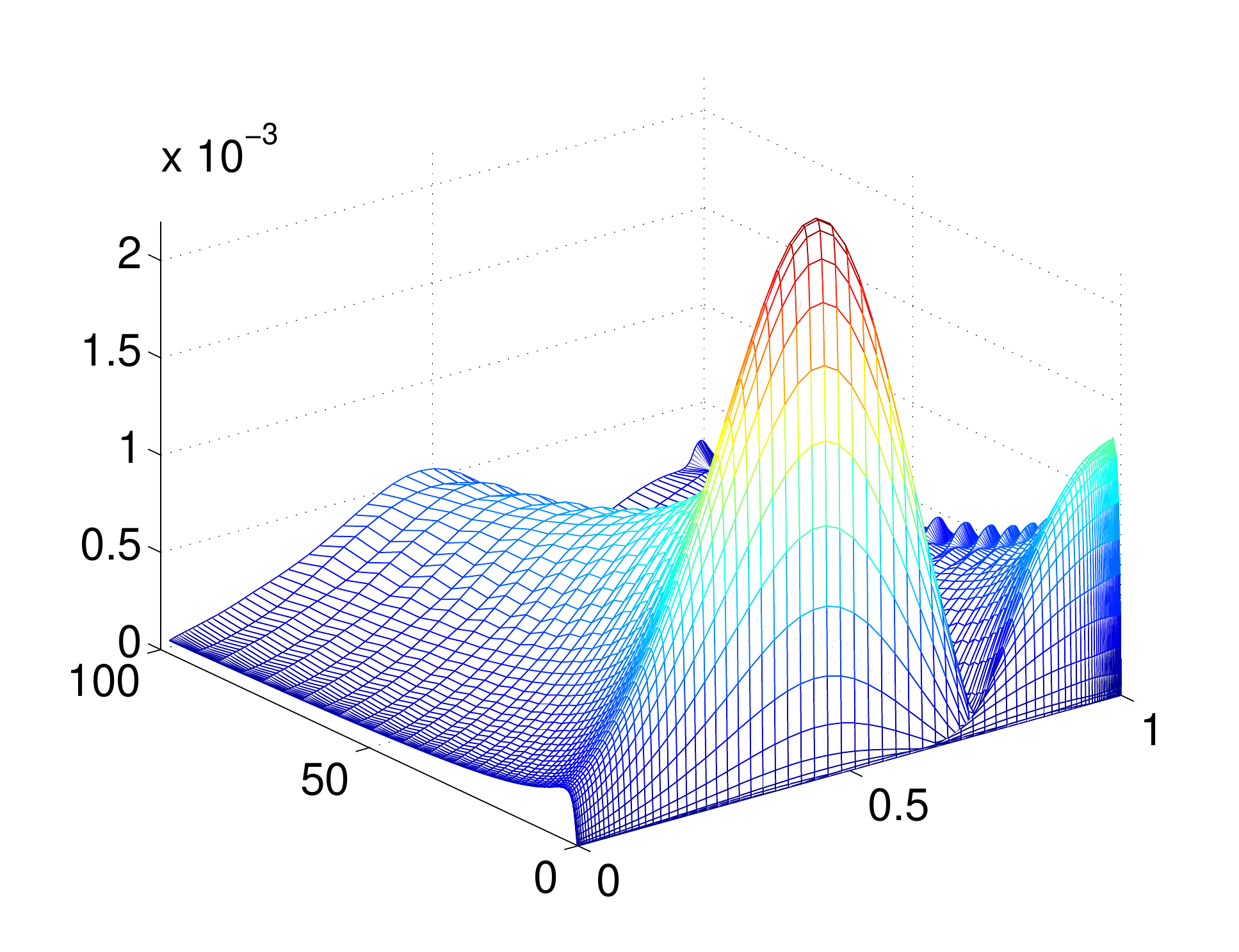}
    \put(-60,10){$x$}
    \put(-175,15){$t$}
    \put(-220,90){$\delta \phi$}
    \put(-220,160){$\textbf{b)}$}

    \caption{\emph{KGD model fluid driven regime.} The relative error of the solution for $N=100$ (spatial mesh), $M=50$ (temporal mesh): a) the error of crack opening, $\delta w$, b) the error of reduced particle velocity, $\delta \phi$. }

\label{bledy_KGD_f_3}
\end{figure}

\begin{figure}[h!]

    \includegraphics [scale=0.40]{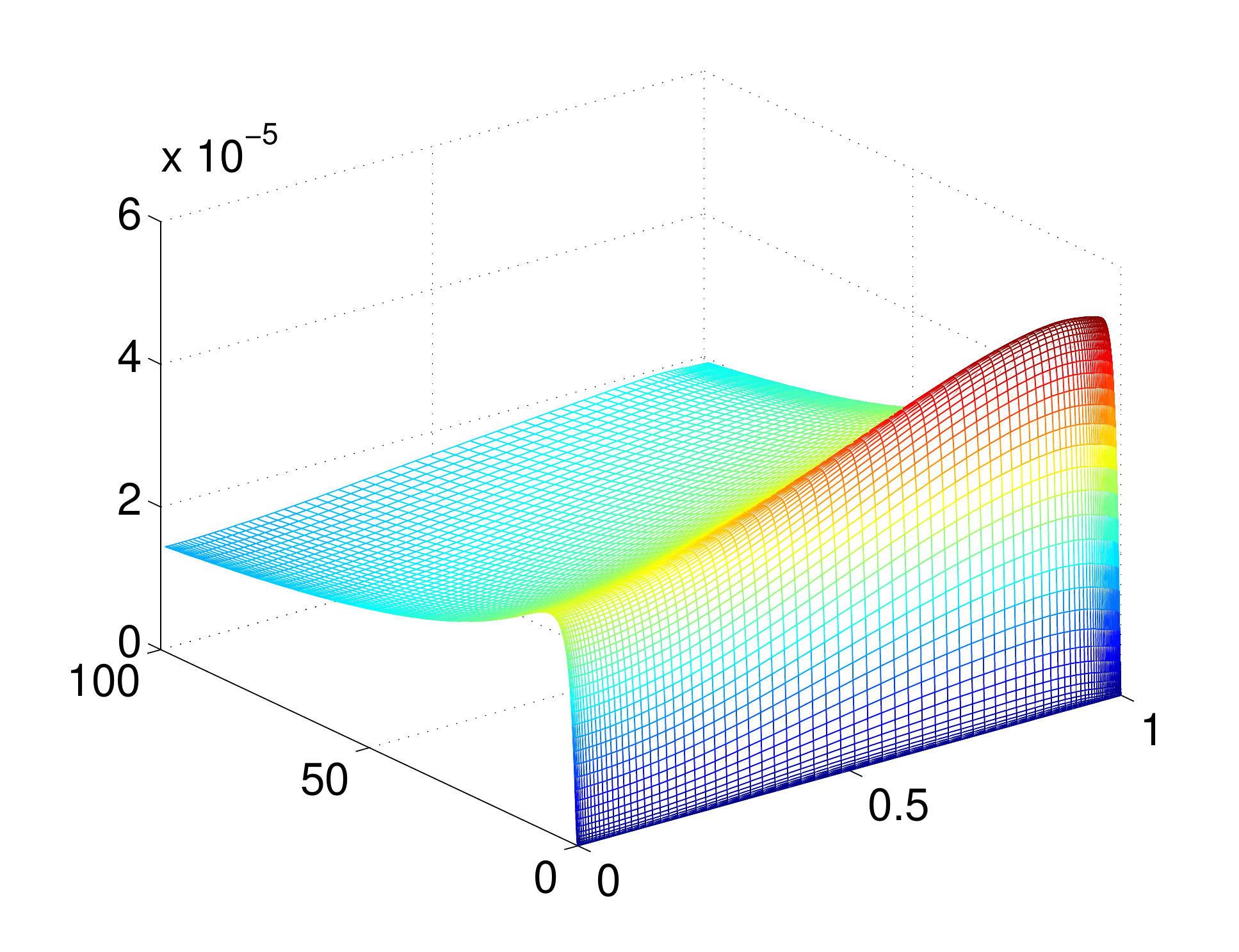}
    \put(-60,10){$x$}
    \put(-175,15){$t$}
    \put(-220,90){$\delta w$}
    \put(-220,160){$\textbf{a)}$}
    \hspace{2mm}
    \includegraphics [scale=0.40]{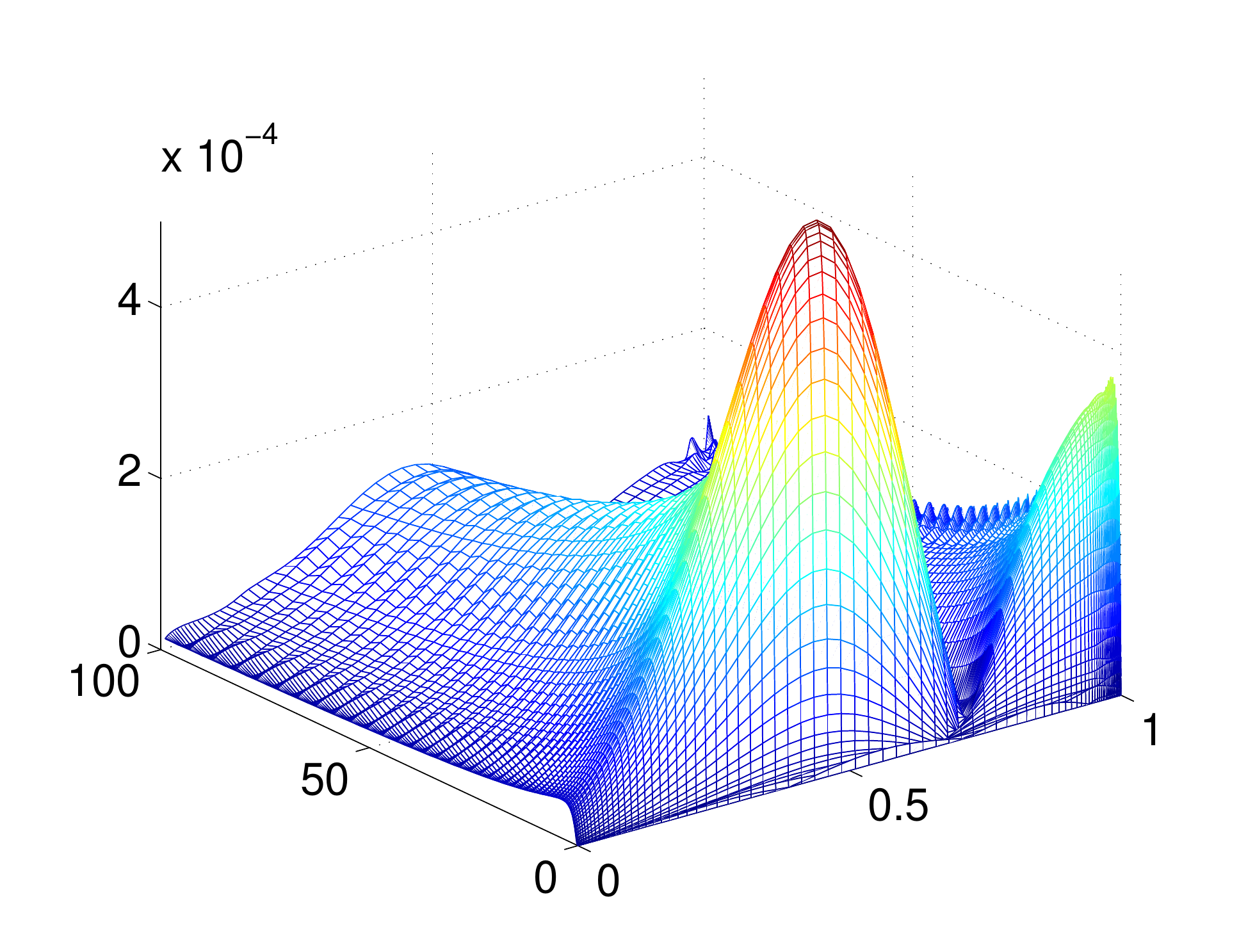}
    \put(-60,10){$x$}
    \put(-175,15){$t$}
    \put(-220,90){$\delta \phi$}
    \put(-220,160){$\textbf{b)}$}

    \caption{\emph{KGD model fluid driven regime.} The relative error of the solution for $N=100$ (spatial mesh), $M=100$ (temporal mesh): a) the error of crack opening, $\delta w$, b) the error of reduced particle velocity, $\delta \phi$. }

\label{bledy_KGD_f_4}
\end{figure}

\begin{figure}[h!]
    \center
    \includegraphics [scale=0.40]{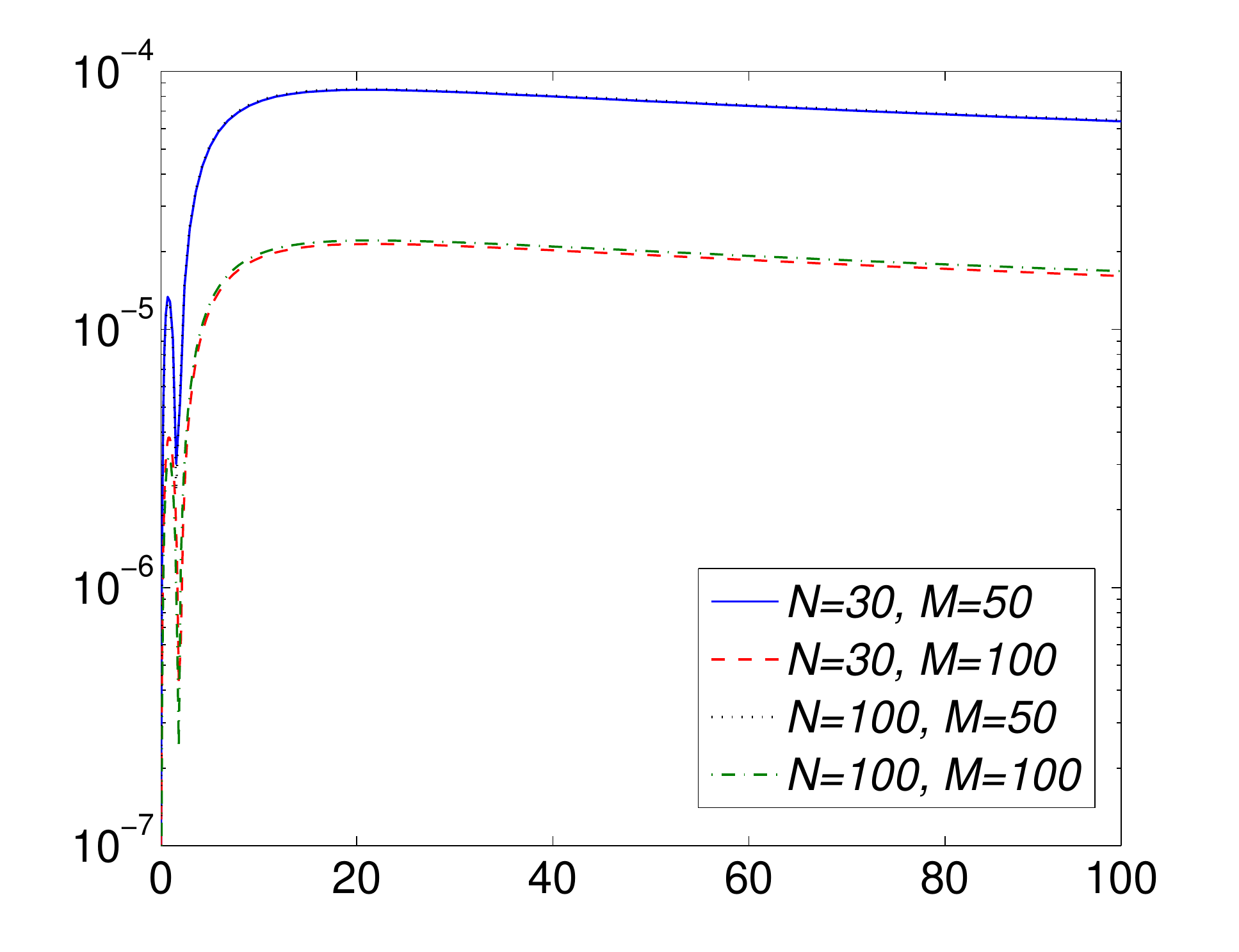}
    \put(-105,0){$t$}
    \put(-230,90){$ \delta L$}
    \caption{\emph{KGD model fluid driven regime.} The crack length error, $\delta L$.}
\label{bledy_KGD_f_5}
\end{figure}

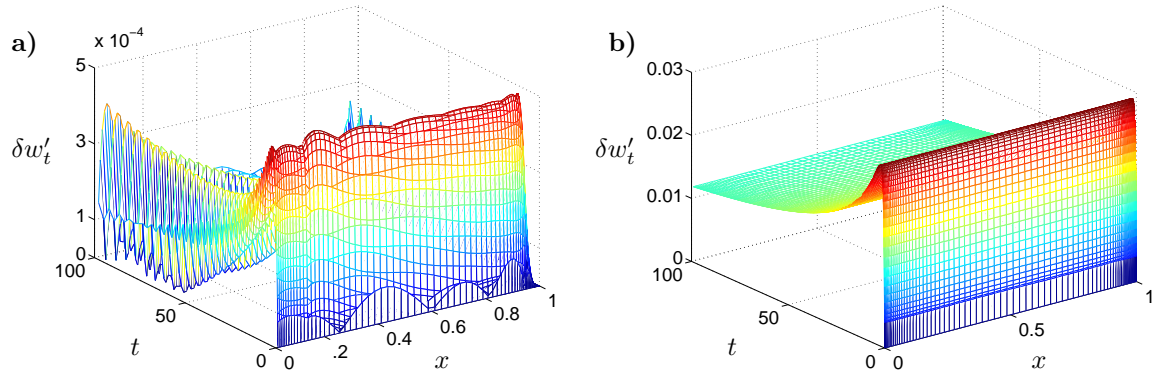
\begin{figure}[h!]
\begin{center}
    \includegraphics [scale=0.35]{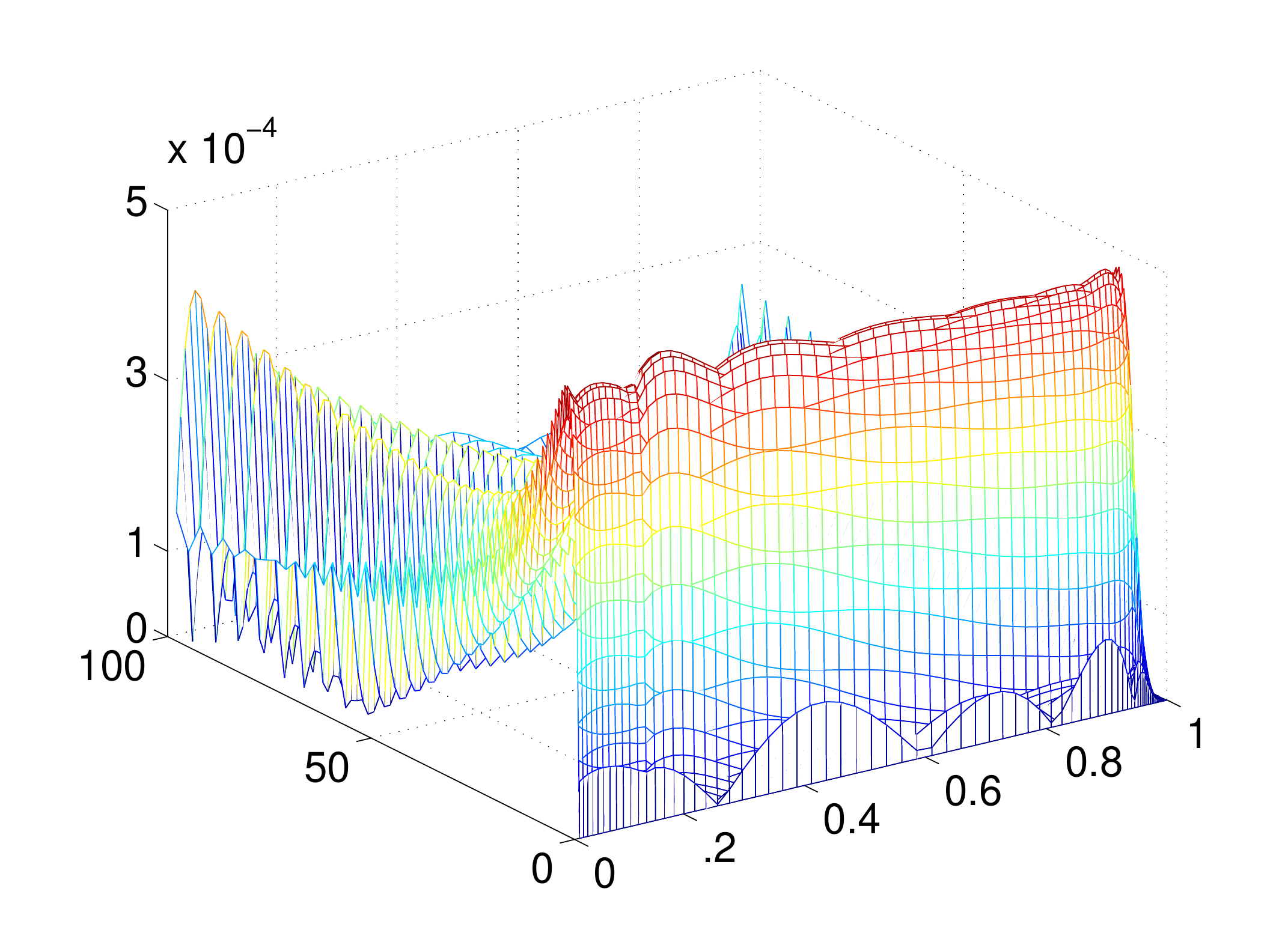}
    \put(-60,10){$x$}
    \put(-175,15){$t$}
    \put(-220,90){$\delta w'_t$}
    \put(-220,130){$\textbf{a)}$}
    \hspace{2mm}
    \includegraphics [scale=0.35]{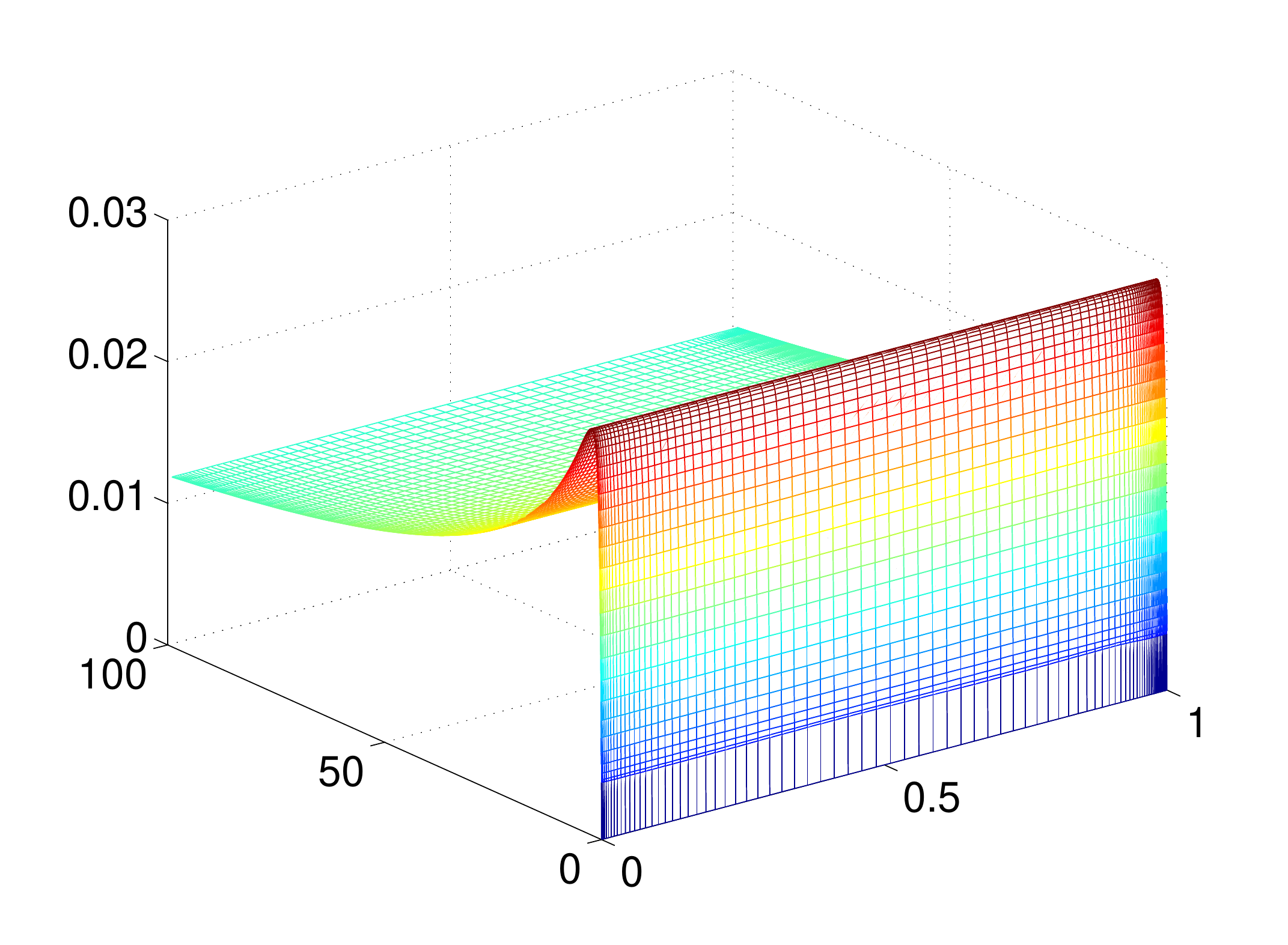}
    \put(-60,10){$x$}
    \put(-175,15){$t$}
    \put(-225,90){$\delta w'_t$}
    \put(-220,130){$\textbf{b)}$}
\end{center}
    \caption{\emph{KGD model - fluid driven regime.} The relative error of the solutions temporal derivative for $N=100$ (spatial mesh), $M=100$ (temporal mesh): a) improved temporal approximation b) ordinary finite difference. }

\label{dwt_KGD_fluid}
\end{figure}

In the following we present an analysis of the algorithm performance for the fluid driven KGD model, in a way analogous to that implemented in the previous chapter. To this end, we utilize the benchmark example already employed in subsection 2.3.2 for the self-similar variant of the problem. The time dependent term taken here, to construct the transient solution, is the power law type (see \eqref{psi_pow}), with  the parameter $\gamma=1/3$. The analyzed variants of mesh densities are defined by $N=30$, $N=100$ and $M=50$, $M=100$. This time we do not use the the lower value of $N$ ($N=20$), as the numerical computation of the inverse elasticity operator \eqref{inv_KGD_n_1} in such a case becomes more sensitive to the mesh density near the crack tip. As we do not want to include in this paper an additional analysis of the influence of this parameter (mesh density in the near-tip region) on the computations we decided to take $N=30$, providing 'fair' comparison for both spatial meshes. However even for $N=20$ the results are still very good, and can be of the same order as for $N=30$ when appropriately adjusting the mesh density near the tip.
To access the accuracy, we analyze the same parameters as previously for the PKN model. The results are displayed in Fig.~\ref{bledy_KGD_f_1} -- Fig.~\ref{dwt_KGD_fluid}.

When analyzing the error of the crack aperture, one can see that a mere 30 points of spatial meshing produces sufficient potential to improve the accuracy by taking more time steps. Indeed, for a fixed number of time steps,  $M=50$, the level of the relative error for the crack opening, $\delta w$, is the same for both $N=30$ and $N=100$ (only some improvement in smoothness of the error distribution can be observed - compare Fig.\ref{bledy_KGD_f_1} and Fig.\ref{bledy_KGD_f_3}). On the other hand, by using more time steps ($M=100$ instead of $M=50$), the error of the crack opening, $w$, can be reduced by an order of magnitude.

The situation is quite different, however, for the reduced velocity, $\phi$. In this case, $N=30$ points of spatial meshing is not enough to provide optimal results especially near the crack tip. By taking $N=100$, one can appreciably improve the accuracy of the reduced velocity (up to one order of magnitude) and prevent the escalation of the error near the fracture tip. For any particular number of the spatial mesh points, $N$, there exists an optimal number (within assumed time stepping strategy) of the time steps, $M$, at which the maximal achievable accuracy is obtained (saturation level). Further increases in $M$ do not produce better results unless the spatial mesh is refined.

Finally, the error in the crack length, $L$, is almost the same for a fixed number of the time steps $M$ regardless of the spatial mesh under consideration
(respective curves in Fig.~\ref{bledy_KGD_f_2} are hardly distinguishable). The explanation of this fact lies in the quality of computation for the parameter ${\cal{L}}(w)$ defining the crack speed: the better accuracy the of $w$ (especially near the crack tip), the better the accuracy of $L$. It is notable that, although this trend is the same as for the PKN model, its realization is obtained by different means: by increasing the number of the time steps $M$ in the PKN case, and by decreasing the step size for the spatial discretization (increasing $N$) for the KGD formulation.

Similarly to that done for the PKN model, we provide here a brief discussion on the application of the time derivative formula \eqref{dwdt}, comparing it with the standard two-points FD approach. In Fig.~\ref{dwt_KGD_fluid}, we display the distributions of the relative error of $w'_t$  for $N=M=100$.  Again, the superiority of approximation \eqref{dwdt} is clear. The values of $\delta w'_t$ are of the same level as for the PKN model, however the error distribution for the improved temporal approximation becomes non-uniform. When changing the mesh densities the same tendencies were observed as in the PKN case. Thus, the conclusions drawn beforehand also hold true here.

\subsubsection{The algorithm performance for the toughness driven KGD model}

\begin{figure}[h!]
    \includegraphics [scale=0.38]{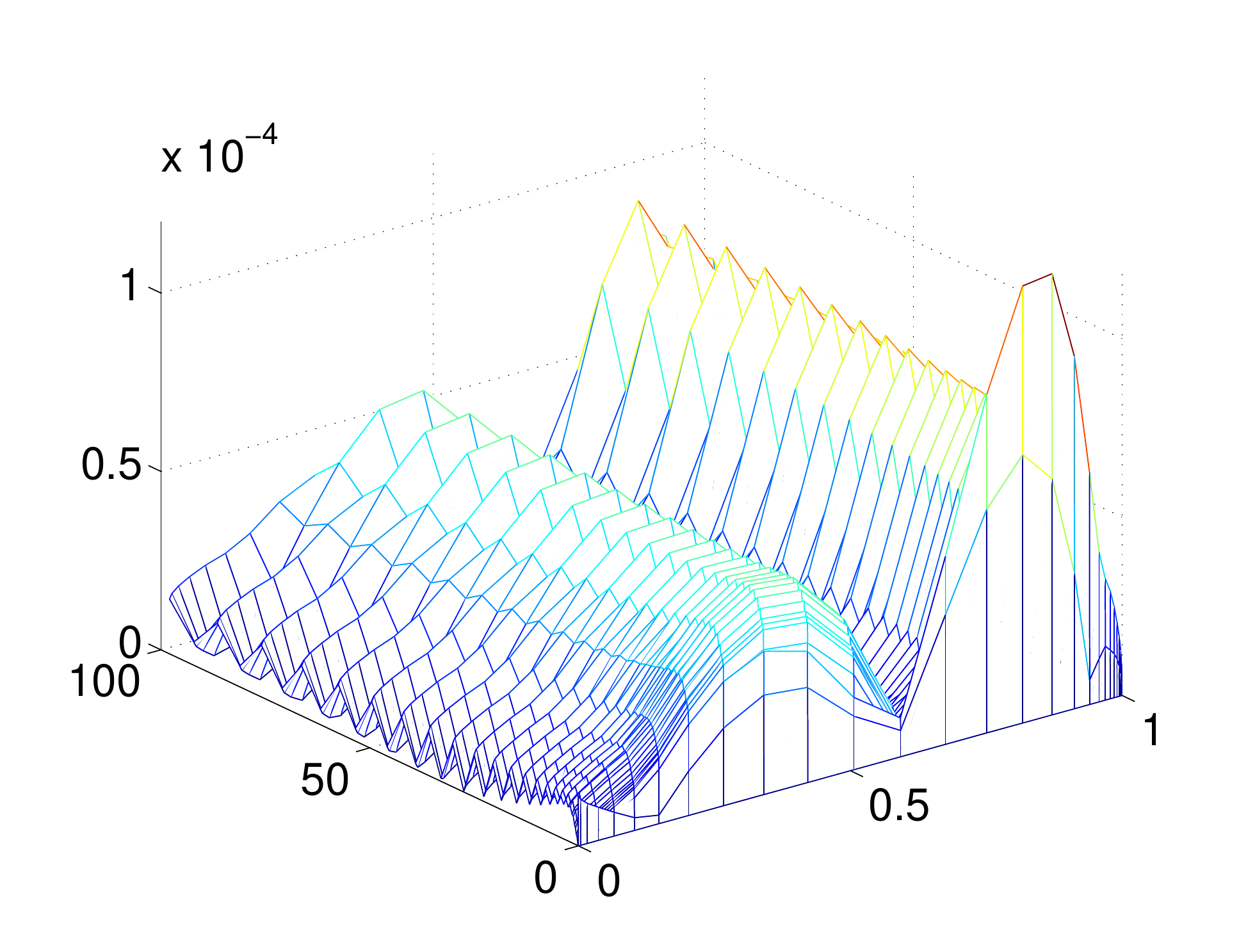}
    \put(-60,10){$x$}
    \put(-175,15){$t$}
    \put(-220,90){$\delta w$}
    \put(-220,160){$\textbf{a)}$}
    \hspace{2mm}
    \includegraphics [scale=0.38]{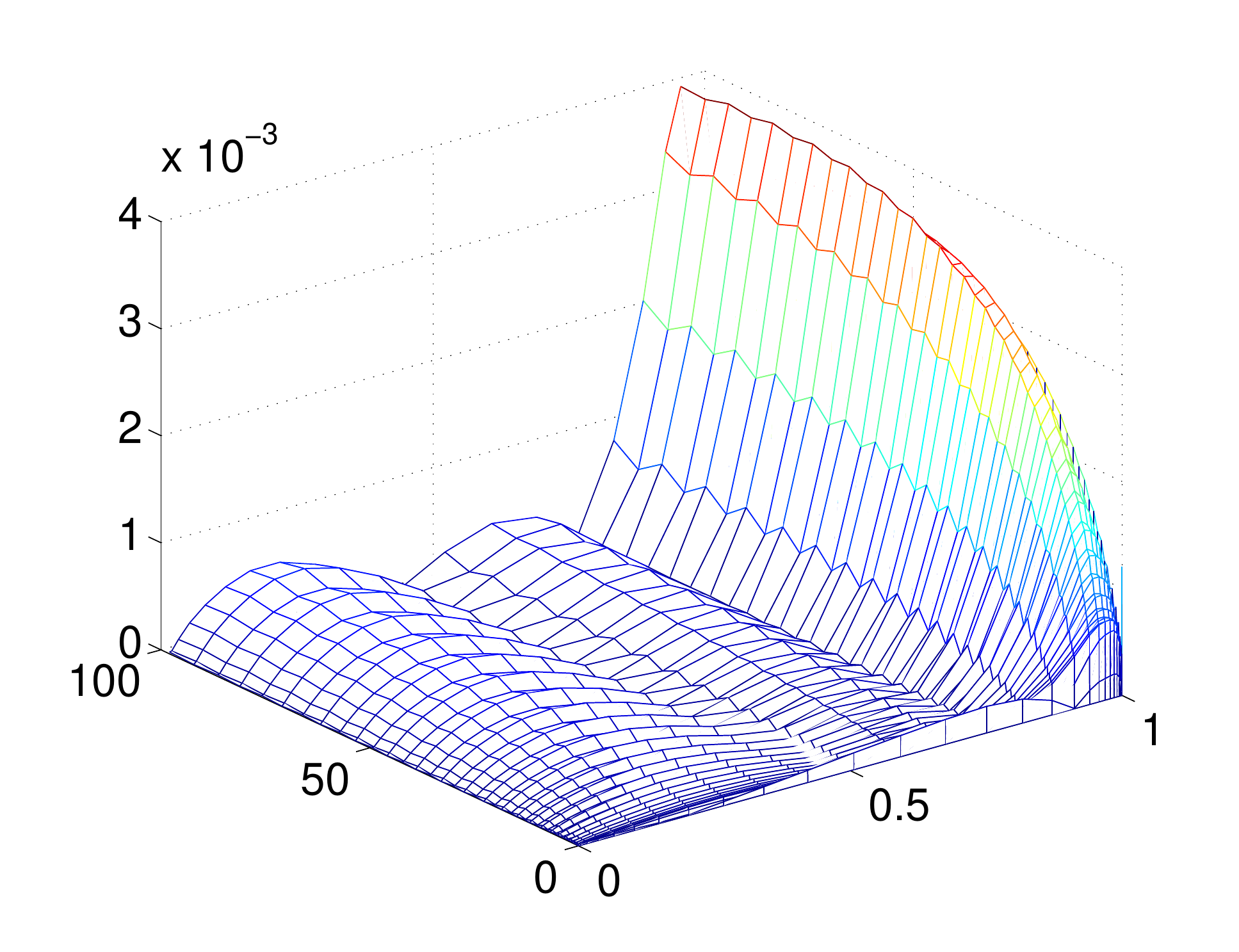}
    \put(-60,10){$x$}
    \put(-175,15){$t$}
    \put(-220,90){$\delta \phi$}
    \put(-220,160){$\textbf{b)}$}
    \caption{\emph{KGD model toughness driven regime.} The relative error of the solution for $N=30$ (spatial mesh), $M=50$ (temporal mesh): a) the error for the crack opening, $\delta w$, b) the error for the reduced particle velocity, $\delta \phi$. }
\label{bledy_KGD_t_1}
\end{figure}

\begin{figure}[h!]
    \includegraphics [scale=0.38]{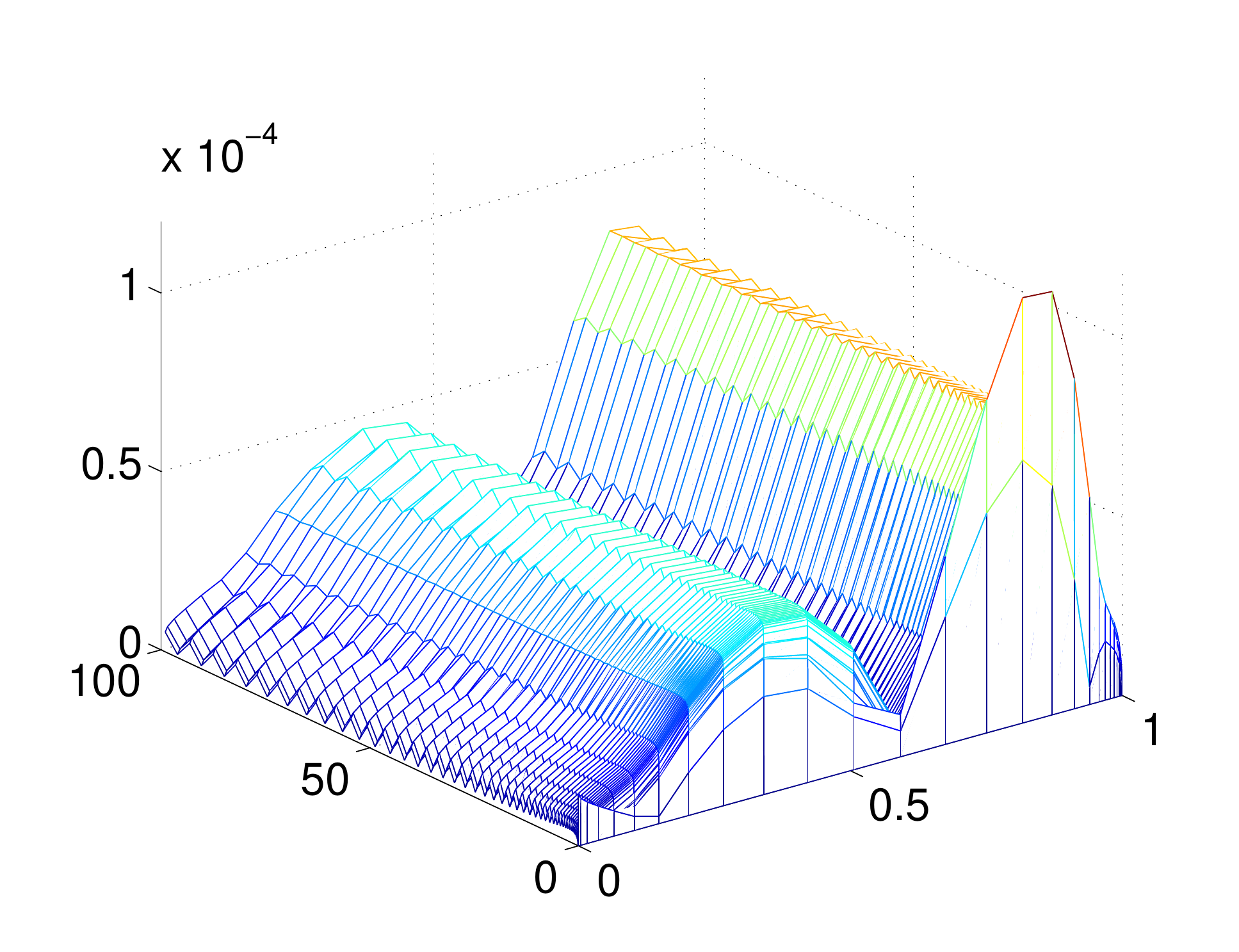}
    \put(-60,10){$x$}
    \put(-175,15){$t$}
    \put(-220,90){$\delta w$}
    \put(-220,160){$\textbf{a)}$}
    \hspace{2mm}
    \includegraphics [scale=0.38]{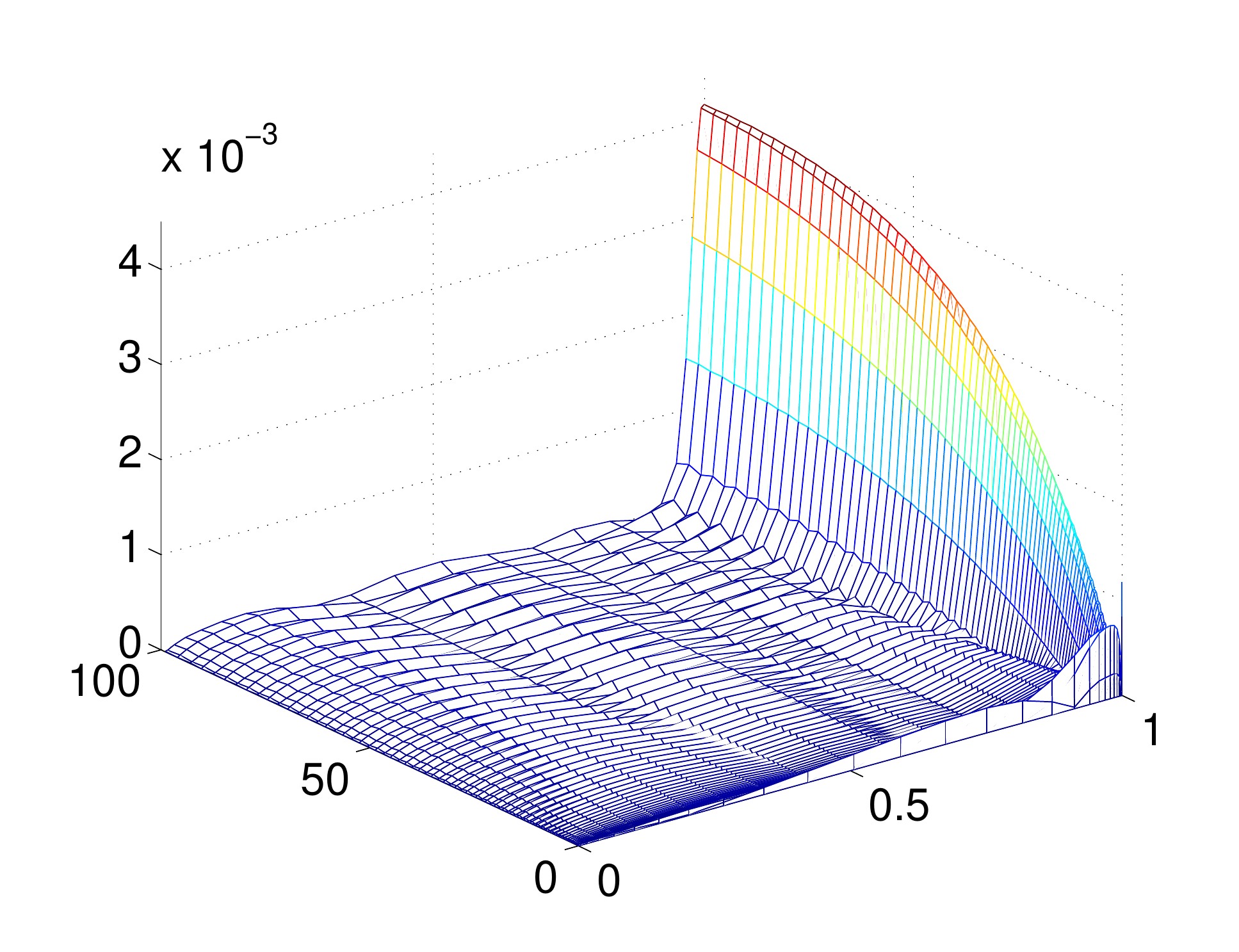}
    \put(-60,10){$x$}
    \put(-175,15){$t$}
    \put(-220,90){$\delta \phi$}
    \put(-220,160){$\textbf{b)}$}
    \caption{\emph{KGD model toughness driven regime.} The relative error of the solution for $N=30$ (spatial mesh), $M=50$ (temporal mesh): a) the error for the crack opening, $\delta w$, b) the error for the reduced particle velocity, $\delta \phi$. }
\label{bledy_KGD_t_2}
\end{figure}

\begin{figure}[h!]
    \includegraphics [scale=0.40]{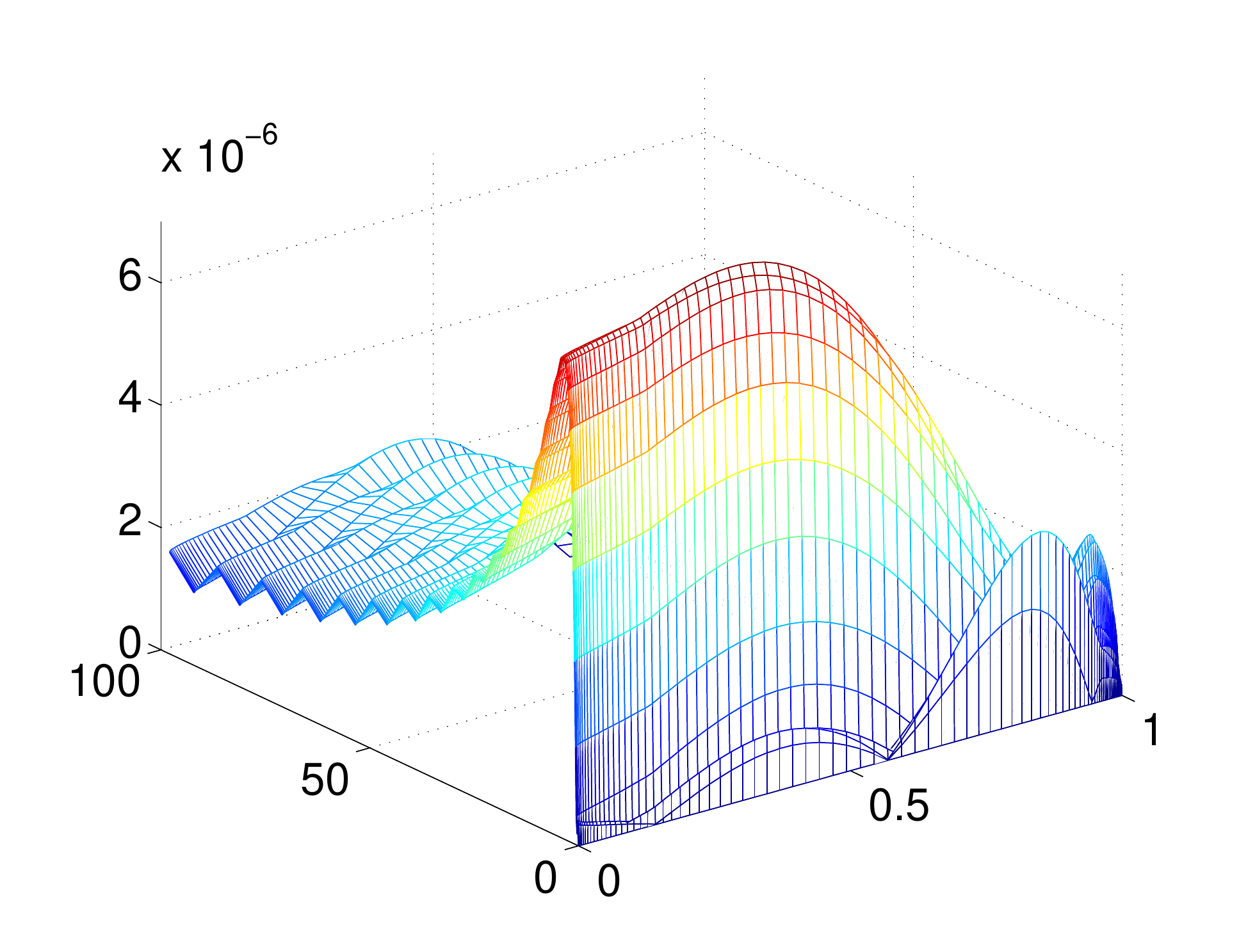}
    \put(-60,10){$x$}
    \put(-175,15){$t$}
    \put(-220,90){$\delta w$}
    \put(-220,160){$\textbf{a)}$}
    \hspace{2mm}
    \includegraphics [scale=0.40]{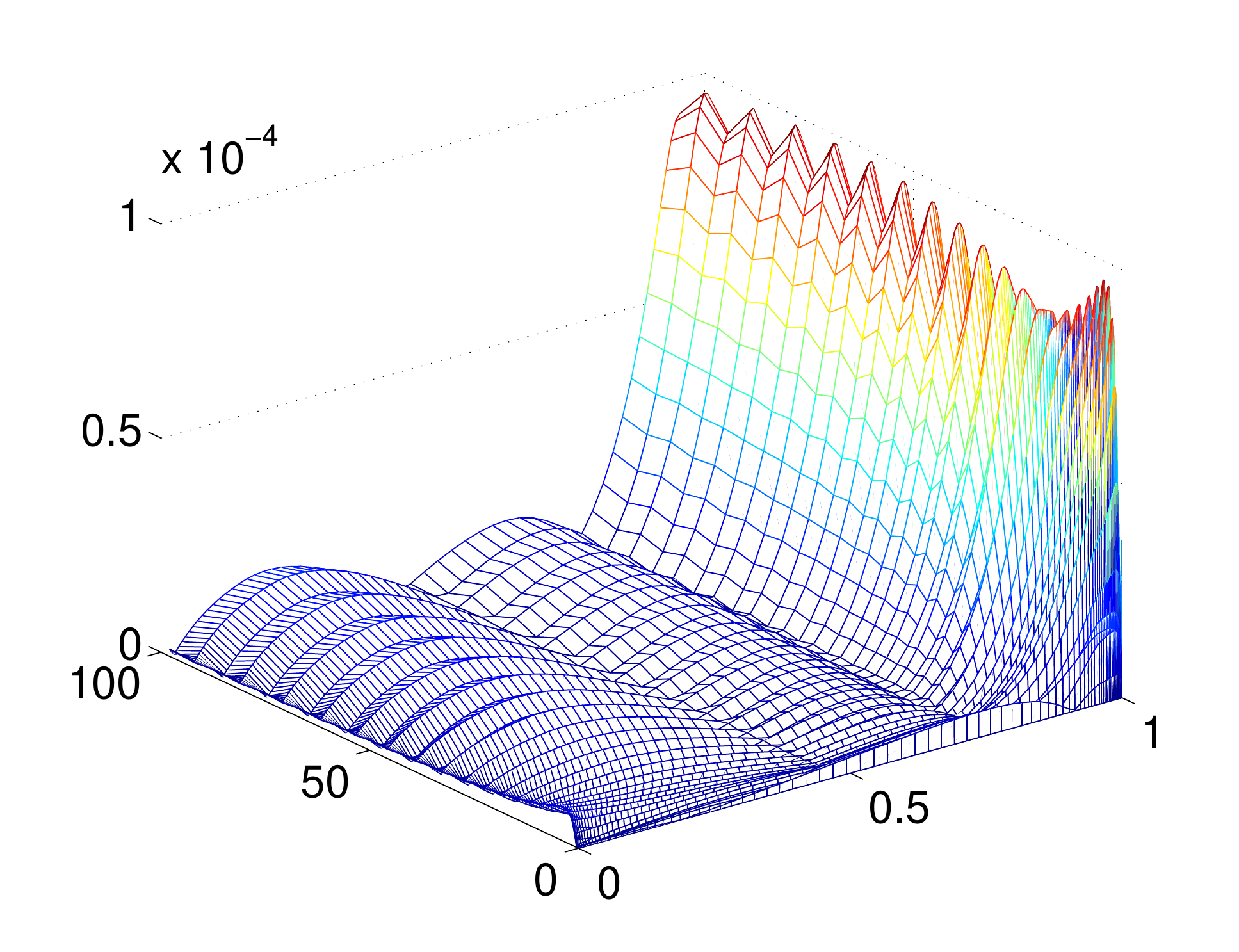}
    \put(-60,10){$x$}
    \put(-175,15){$t$}
    \put(-220,90){$\delta \phi$}
    \put(-220,160){$\textbf{b)}$}
    \caption{\emph{KGD model toughness driven regime.} The relative error of the solution for $N=100$ (spatial mesh), $M=60$ (temporal mesh): a) the error for the crack opening, $\delta w$, b) the error for the reduced particle velocity, $\delta \phi$. }
\label{bledy_KGD_t_3}
\end{figure}

\begin{figure}[h!]
    \includegraphics [scale=0.40]{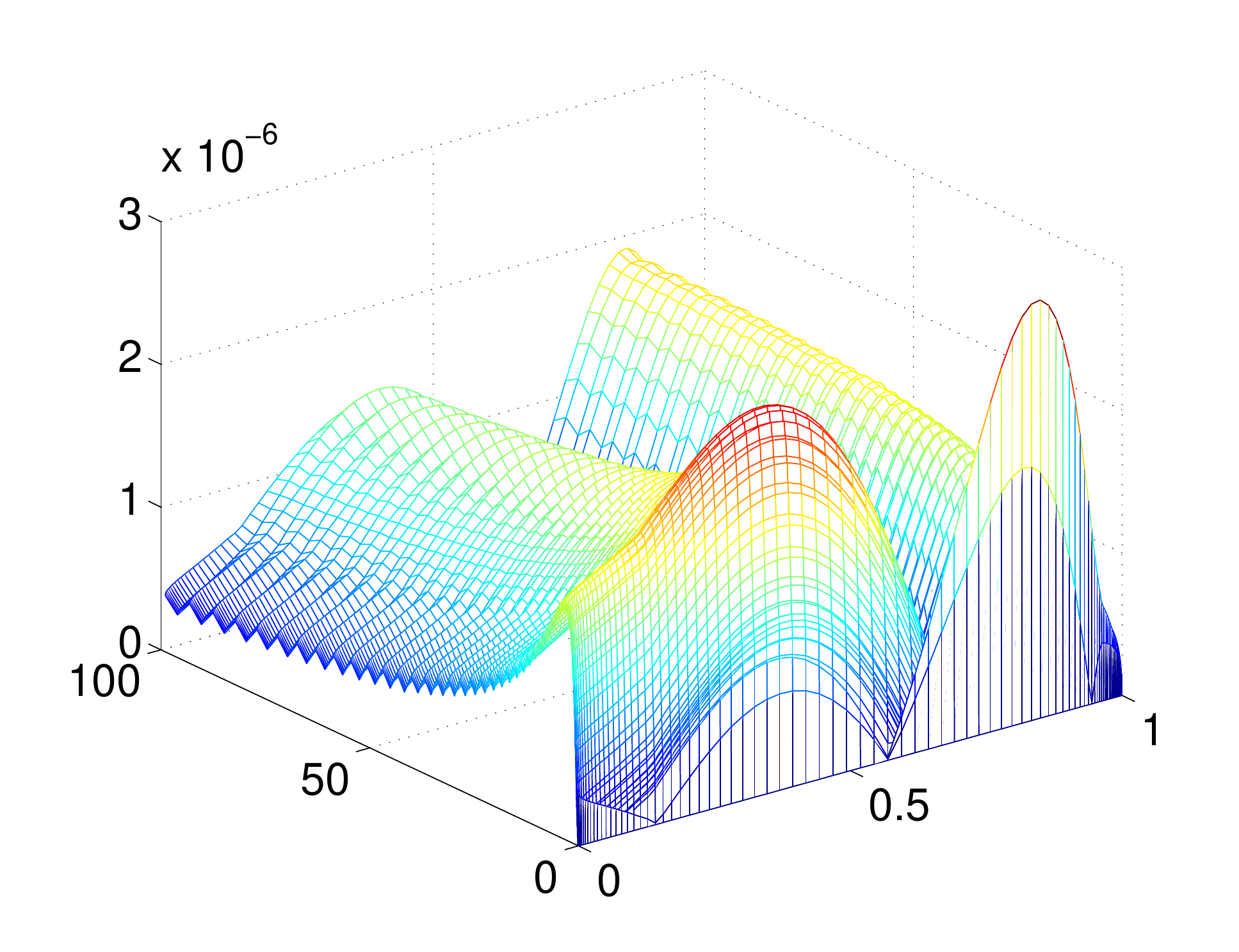}
    \put(-60,10){$x$}
    \put(-175,15){$t$}
    \put(-220,90){$\delta w$}
    \put(-220,160){$\textbf{a)}$}
    \hspace{2mm}
    \includegraphics [scale=0.40]{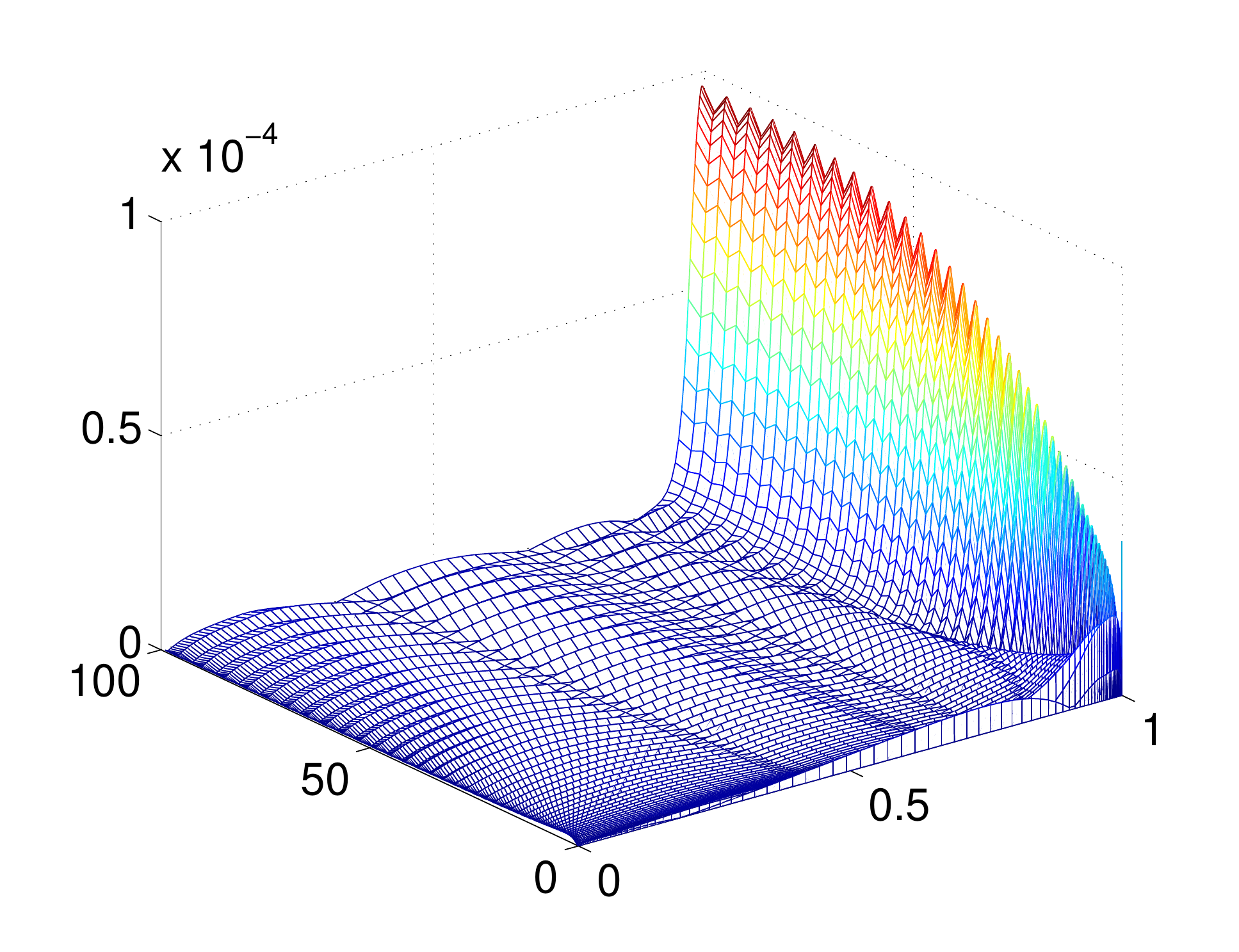}
    \put(-60,10){$x$}
    \put(-175,15){$t$}
    \put(-220,90){$\delta \phi$}
    \put(-220,160){$\textbf{b)}$}
    \caption{\emph{KGD model toughness driven regime.} The relative error of the solution for $N=100$ (spatial mesh), $M=100$ (temporal mesh): a) the error for the crack opening, $\delta w$, b) the error for the reduced particle velocity, $\delta \phi$. }
\label{bledy_KGD_t_4}
\end{figure}

\begin{figure}[h!]
    \center
    \includegraphics [scale=0.40]{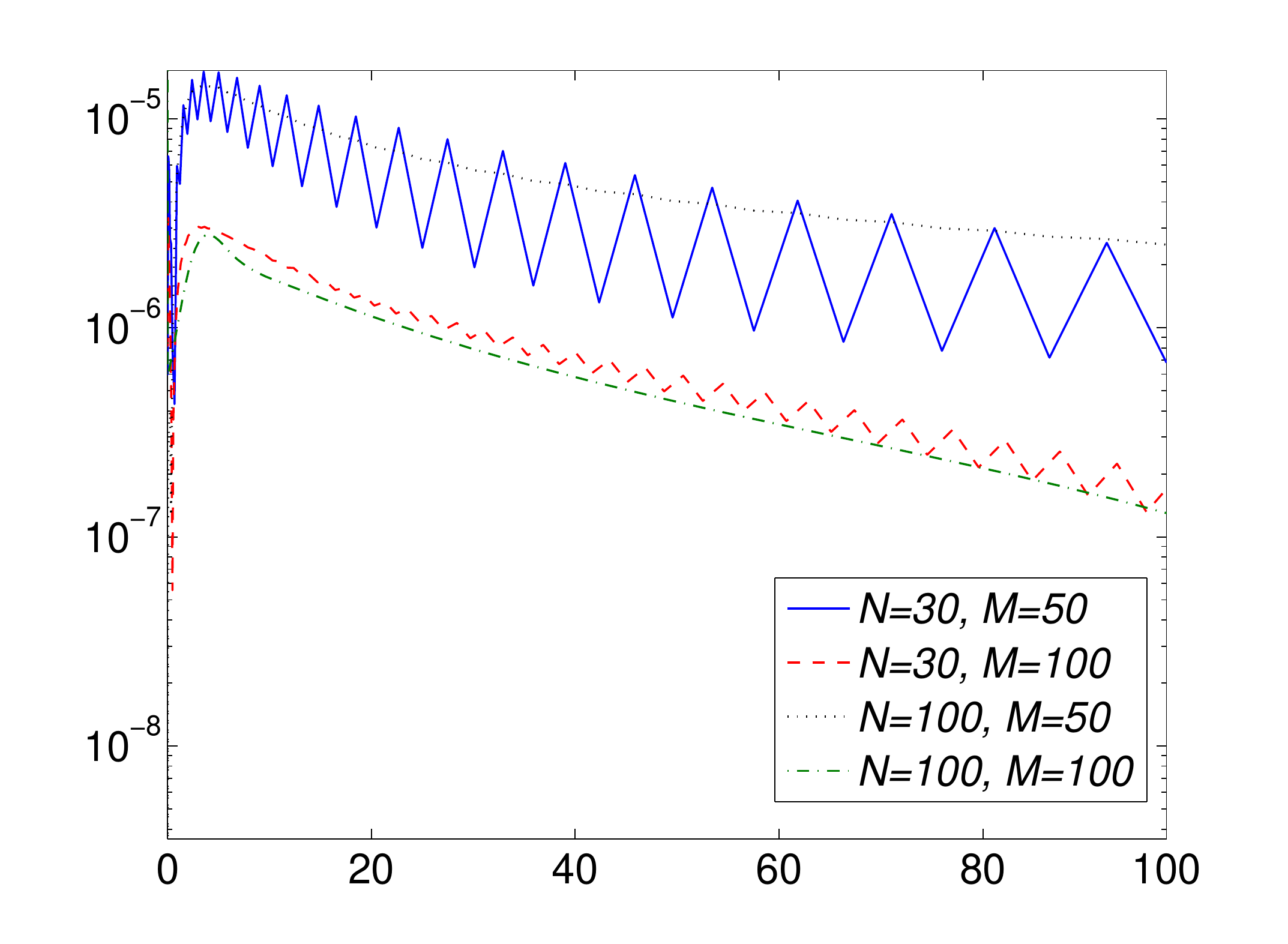}
    \put(-110,0){$t$}
    \put(-250,90){$ \delta L$}
    \caption{\emph{KGD model toughness driven regime.} The crack length error, $\delta L$. }
\label{bledy_KGD_t_5}
\end{figure}

In the last part of this subsection we investigate the algorithm performance for a transient regime of the toughness driven KGD model. The benchmark used in this case is constructed
 in the same way as that for the fluid driven regime. The time dependent term is the same as previously taken for the fluid driven variant. In the analysis of the accuracy of computations, the combinations of spatial and temporal meshing remain the same as in the case of the fluid drive regime. The results are displayed in Fig.~\ref{bledy_KGD_t_1} -- Fig.~\ref{bledy_KGD_t_5}.

The error distributions for $N=30$ show that, for this number of nodal points, the solution cannot be further improved by taking more time steps (the limiting factor here is the spatial meshing). When analyzing $N=100$ it turns out that, not only does the error of the solution decrease, but there also exists the potential for increasing the accuracy by taking more densely packed time steps. The graph for $\delta L$, Fig.~\ref{bledy_KGD_t_5}, exhibits rather a surprising result. First, we can see the fluctuations in the crack length error with time for $N=30$.
Respective curves for $N=100$ are already smooth, however, the level of the error, $\delta L$, does not decrease. As mentioned previously, the quality of the fracture length computation depends on the accuracy of the parameter ${\cal{L}}(w)$ defining the crack speed. In the toughness driven regime, this value utilizes the multipliers of the first two leading terms of the asymptotic expansion of $w$ (instead of the only one leading term in the PKN or the fluid driven KGD models) -- see \eqref{LC_KGD_toughness}.
In this way, the second term which by its nature is approximated with lower accuracy than the leading one,
appreciably affects the error of $L(t)$, limiting its potential for improvement with increasing $N$.

\begin{figure}[h!]
\begin{center}
    \includegraphics [scale=0.35]{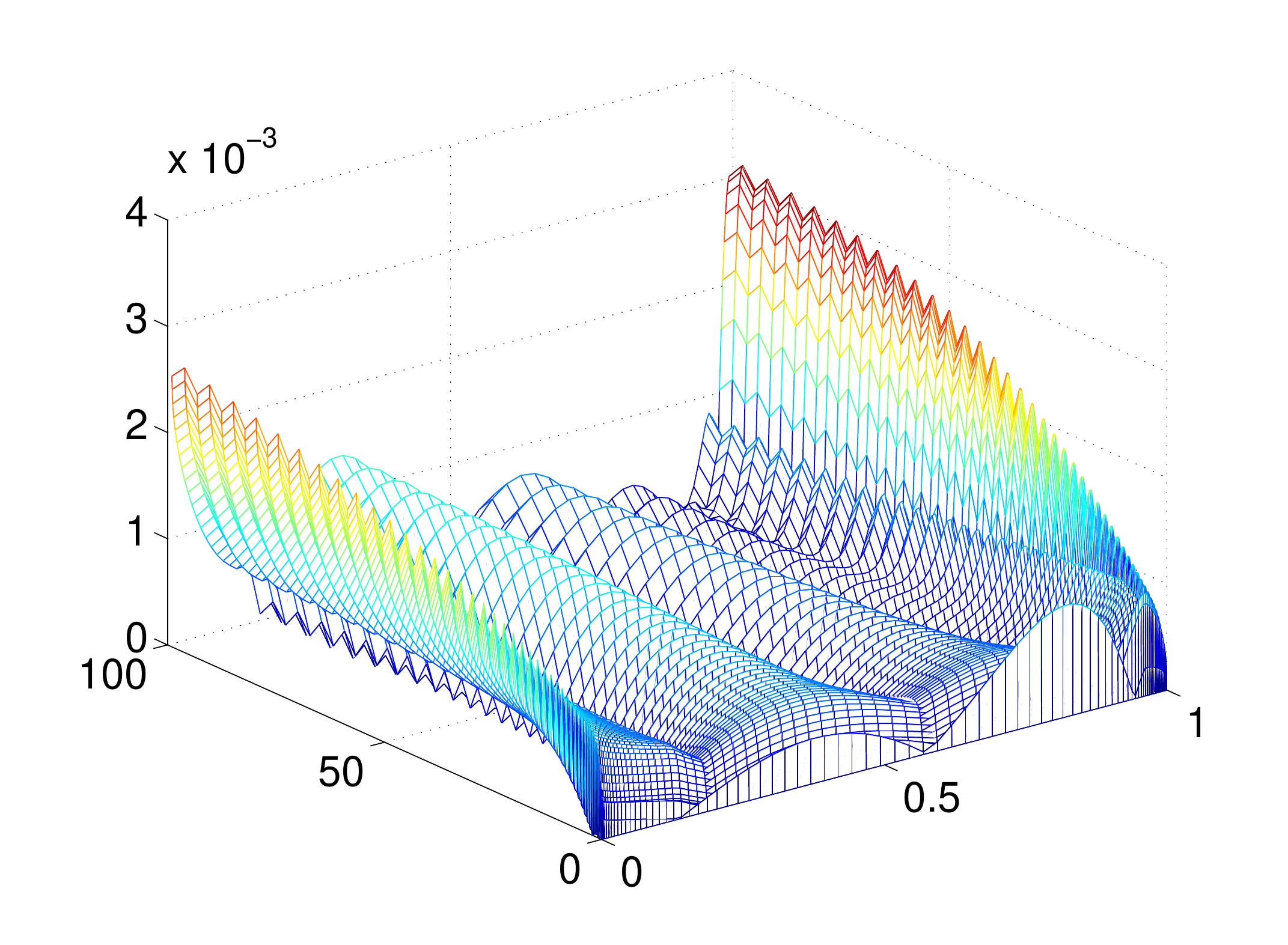}
    \put(-60,10){$x$}
    \put(-175,15){$t$}
    \put(-220,90){$\delta w'_t$}
    \put(-220,130){$\textbf{a)}$}
    \hspace{2mm}
    \includegraphics [scale=0.35]{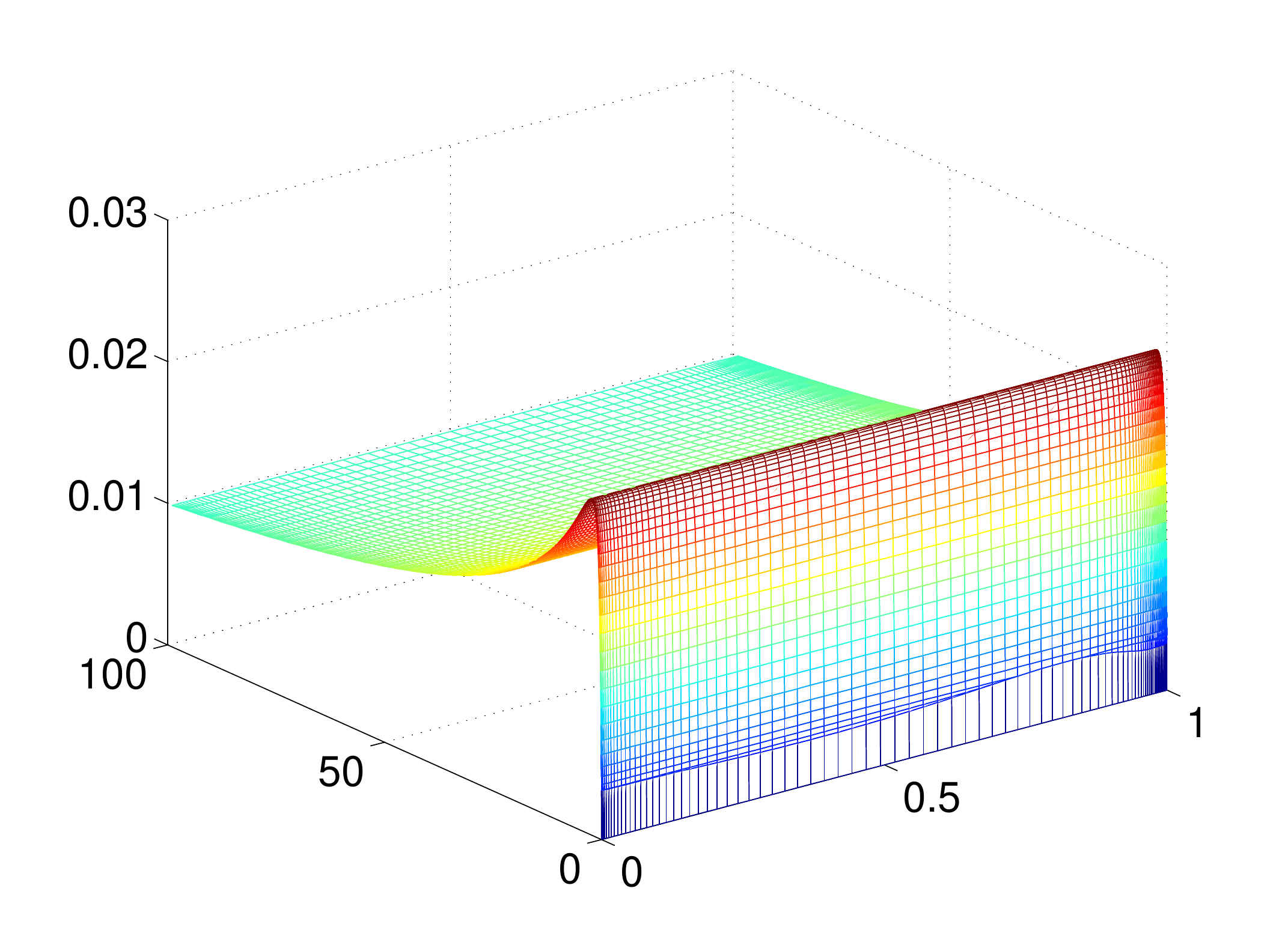}
    \put(-60,10){$x$}
    \put(-175,15){$t$}
    \put(-225,90){$\delta w'_t$}
    \put(-220,130){$\textbf{b)}$}
\end{center}
    \caption{\emph{KGD model - toughness driven regime.} The relative error of the solutions temporal derivative for $N=100$ (spatial mesh), $M=100$ (temporal mesh): a) improved temporal approximation b) two-point finite difference. }

\label{dwt_KGD_tough}
\end{figure}

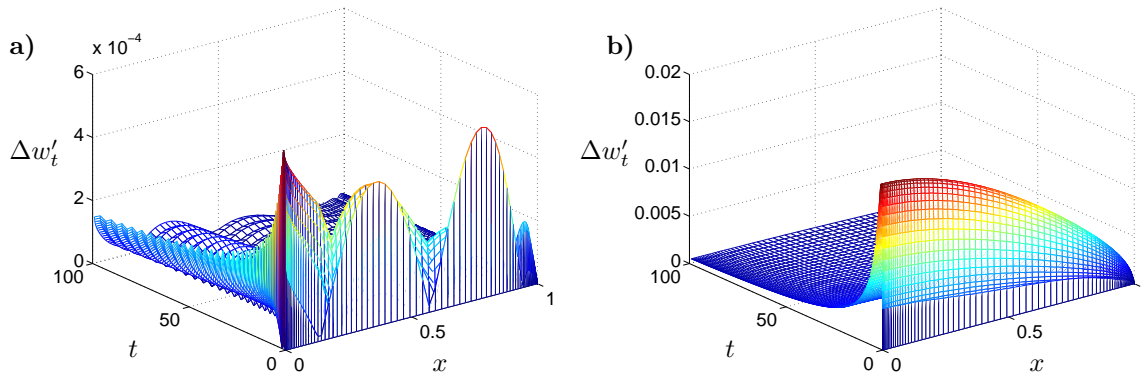
\begin{figure}[h!]
\begin{center}
    \includegraphics [scale=0.35]{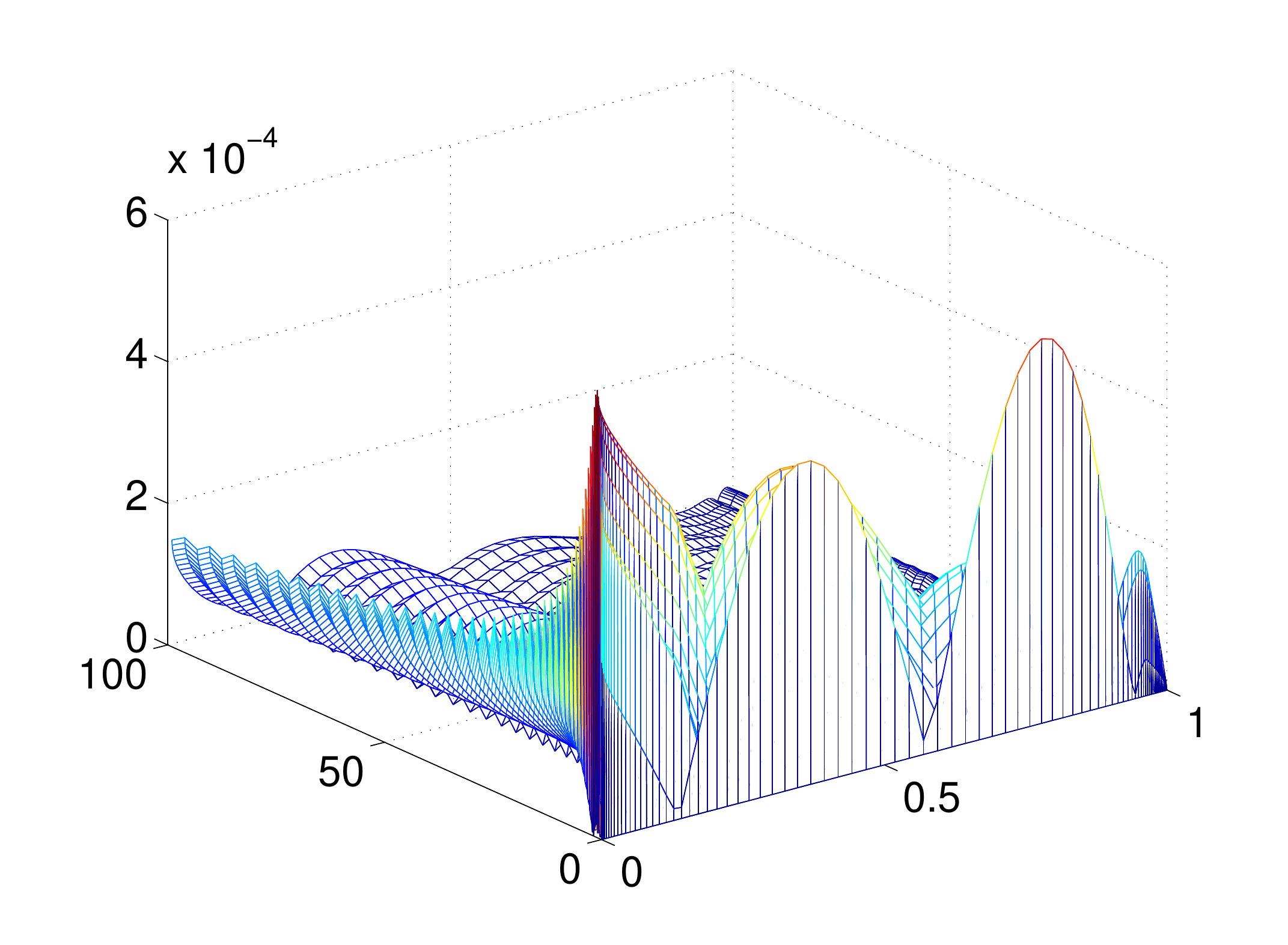}
    \put(-60,10){$x$}
    \put(-175,15){$t$}
    \put(-220,90){$\Delta w'_t$}
    \put(-220,130){$\textbf{a)}$}
    \hspace{2mm}
    \includegraphics [scale=0.35]{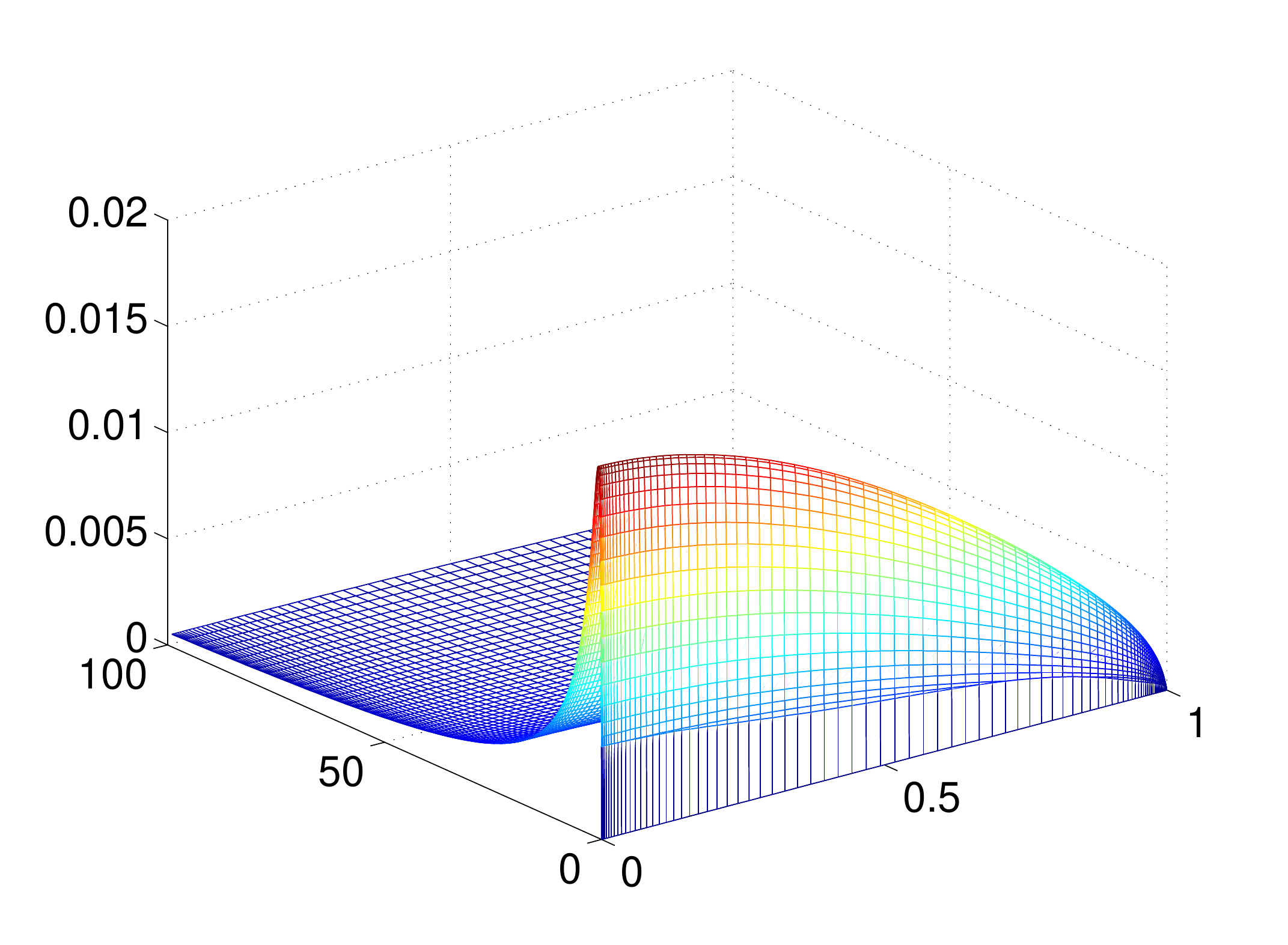}
    \put(-60,10){$x$}
    \put(-175,15){$t$}
    \put(-230,90){$\Delta w'_t$}
    \put(-220,130){$\textbf{b)}$}
\end{center}
    \caption{\emph{KGD model - toughness driven regime.} The absolute error of the solutions temporal derivative for $N=100$ (spatial mesh), $M=100$ (temporal mesh): a) improved temporal approximation b) two-point finite difference.}

\label{dwt_KGD_tough_abs}
\end{figure}

Finally, let us complete the discussion by presenting the results for the temporal derivative of the crack opening. The relative error of $w'_t$ for both methods of approximation (formula \eqref{dwdt} and the two-point FD scheme, respectively) are shown in Fig.~\ref{dwt_KGD_tough} ($N=M=100$). As can be seen, the relative error for the improved temporal approximation is of the order $10^{-3}$ -- one order of magnitude worse than that previously revealed for the PKN and fluid driven KGD models.  However it is still much lower than that of the classic FD approach (see Fig.~\ref{dwt_KGD_tough}b)). One can also observe in Fig.~\ref{dwt_KGD_tough}a) a pronounced growth of $\delta w'_t$ at both ends of the spatial interval, which magnifies with time. This trend is caused, to a large extent, by the behaviour of $w'_t$ itself, which yields a time asymptote proportional to $(a+t)^{\gamma-1}$ (compare \eqref{psi_pow}). Indeed, in Fig.~\ref{dwt_KGD_tough_abs} we show the absolute values of the error of the temporal derivative $\Delta w'_t$, and for formula \eqref{dwdt} one obtains its stable (or even decreasing with time) levels at both ends of the spatial interval. The general trends in the quality of $w'_t$ approximation for the coarser meshes were the same as previously obtained for the PKN and fluid driven KGD models.

\newpage

\section{Discussion and Conclusions}

In this paper the classic problem of hydraulic fracture, considered in its general form, has been revisited and reformulated in terms of a new pair of the dependent variables, with both having a clear physical sense:
the crack opening, $w$, and \emph{\textbf{the reduced particle velocity}}, $\phi$. The latter is directly related to the average (over the fracture cross section) speed of fluid flow. It was shown that the new formulation is mathematically complete and well defined.
In particular, \emph{\textbf{unified equation, (\ref{v_0_univ}), directly relating the crack propagation speed to the asymptotics of the crack opening}} were
evaluated, using the speed equation, for various elasticity operators and fracture propagation regimes. Self-similar formulations of the problem for different hydraulic fracture models have been given.
In this setting, a universal algorithm for the computational simulation of hydraulic fracture has been developed.
It enables one to account for various elasticity operators, fluid flow and fracture propagation regimes within the framework of a unified scheme.

The proposed algorithm has a modular architecture and consists of two basic modules:
 \begin{itemize}
  \item
the first one, a {\it universal} one, computes the reduced particle velocity,
and is the same regardless of the variant of the problem under consideration,
  \item
the second one, evaluating the crack opening, should be adjusted depending on which elasticity operator
is used. In order to account for a specific crack propagation (fluid/toughness driven) regime, one only needs to adjust
the values of the respective parameters which describe the asymptotics of the crack opening and the reduced velocity near the crack tip. Thus, preliminary knowledge of asymptotic behaviour of the solution is crucial.
 \end{itemize}
The crack propagation speed is computed using respective explicit formula, (\ref{v_0_univ}), derived from the speed equation with utilization of the solvability condition (the fluid balance equation) providing \emph{\textbf{the explicit formula, (\ref{L_int_PKN}), linking the crack length to the tip asymptotics of the crack opening}}.

Various numerical techniques
are utilized in the proposed method with the reduced particle velocity as the main component.
The key points of the algorithm are:
(i) proper handling of the independent variables (appropriate spatial and temporal meshing),
(ii) relevant regularization techniques, in particular the so-called $\varepsilon$-regularization and operator regularization of the governing equations, taking the solvability condition into account when necessary,
(iii) explicit formula for the crack propagation speed for the fracture front tracing,
(iv) rigorous utilization of the solution tip asymptotics,
(v) improved approximation of the temporal derivative of the crack opening.

Extensive analysis of performance of the algorithm, for both its self-similar and the general time-dependent formulations, has been done and comparison made with analytical benchmark solutions developed for various models of hydraulic fracture.  Some of them were adopted from the authors previous papers, while others (KGD models) are discussed in Appendix B. Also, reference solutions available in the literature have been used.The following conclusions can be drawn:
\begin{itemize}
\item[(i)] the algorithm is numerically stable regardless of the hydraulic fracture model used;
\item[(ii)] the accuracy of produced results is appreciably better than that of other solutions available in the literature;
\item[(iii)] in most cases the computational cost is very low. A properly distributed spatial mesh composed of several dozen of points  provides accuracy better than 0.1$\%$. Only the case of very small toughness may necessitate finer meshing.
\end{itemize}

To summarize, \emph{\textbf{the particle velocity based universal algorithm}} developed in this paper is capable of tackling various hydraulic fracturing models under different crack propagation regimes. Its flexibility is a result of its adaptive character and modular code architecture. The method of tracing the fracture front, based on the speed equation, is stable and accurate. The key issue related to its realization is application of the explicit formulae for the crack propagation
speed and the solution tip asymptotics, together with resulting formulae for the crack length. It was shown that the new algorithm is more accurate than any other available in the literature.

 Additionally, taking advantage of the algorithm's accuracy,
semi-analytic formulae for solving the KGD model in the fluid driven regime have been evaluated, with a solution which yields all necessary components: the crack opening, $w$, the particle velocity, $v$, the fluid pressure, $p$ (the fluid flux should be computed as $q=wv$).

Although we restrict ourselves in this paper only to the Newtonian fluids, the algorithm may be easily adapted to other rheological models. This approach can also be extended to 2D fractures.

\vspace{3mm}

{\bf Acknowledgements}. The authors acknowledge support from FP7 Marie Curie IAPP projects PIAP-GA-2009-251475 and PIAP-GA-2011-286110.

\section*{Appendix A: Self-similar solutions}

\subsection*{A1: General representation}

Let us assume the following separation of variables for the crack opening and the net pressure:
\begin{equation}
\label{w_sep}
w(t,x)=\psi(t) \hat w(x),\quad p(t,x)=\frac{\psi(t)}{L^{m-1}(t)}\hat p(x).
\end{equation}
where $\psi(t)$ is a smooth continuous function of time and will be specified later. As a consequence, the qualitative asymptotic behaviour of the respective spatial functions in \eqref{w_sep} remains the same as their time dependent counterparts (e.g $\hat w(x)$ complies with \eqref{w_roz}).
The elasticity equations (\ref{A_norm_PKN_0}), (\ref{A_norm_PKN}) and (\ref{inv_KGD_n_1}) are transformed to:
\begin{equation}
\label{p_ss}
\hat w(x)={\cal A}^{-1}\hat p(x),
\end{equation}
\begin{equation}
\label{p_ss0}
\hat w(x)=\hat p(x),
\end{equation}
\begin{equation}
\label{Am2_eq}
\hat w=-\frac{4}{\pi }\int_0^1  \frac{d \hat p }{ds}K(s,x)ds+\frac{4}{\sqrt{\pi}}\hat K_I \sqrt{1-x^2},
\end{equation}
for the PKN and KGD models, respectively. For a constant and non-zero value of the dimensionless toughness, (\ref{inv_KGD_n_1}) can be transformed to its time independent counterpart \eqref{Am2_eq} only if:
\begin{equation}
\label{tough_ss_codn}
\frac{\psi(t)}{\sqrt{L(t)}}=\textbf{const}=c.
\end{equation}
Then the self-similar stress intensity factor, $\hat K_I$, is expressed as:
\begin{equation}
\label{K_Iss}
\hat K_I=\frac{2c}{ \sqrt{\pi}}\int_0^1 \frac{\hat p(\eta)}{\sqrt{1-\eta^2}}d\eta.
\end{equation}
However, if one assumes that the normalized material toughness, $\tilde K_I$, changes with time as:
\begin{equation}
\label{K_Iss_cond}
\tilde K_I(t)=\hat K_I\frac{\psi(t)}{\sqrt{L(t)}},
\end{equation}
then \eqref{Am2_eq} and \eqref{K_Iss} are satisfied automatically (the latter for $c=1$).

Equation \eqref{Am2_eq} is the inversion of the self-similar form of the  operator \eqref{A_norm_KGD}:
\begin{equation}
\label{Am1_ss}
\hat p(x)=-\frac{1}{2\pi}\int_0^1\frac{d \hat w}{d \eta}\frac{\eta d\eta}{\eta^2-x^2}.
\end{equation}

The fluid flow rate, particle velocity and reduced particle velocity functions can be expressed as:
\begin{equation}
\label{q_vphi_sep}
q(t,x)=\frac{\psi^4(t)}{L^m}\hat q(x), \quad v(t,x)=\frac{\psi^3(t)}{L^m(t)}\hat v(x),\quad \phi(t,x)=\frac{\psi^3(t)}{L^m(t)}\hat \phi(x),
\end{equation}
\begin{equation}
\label{q_phi_hat}
\hat q(x)=-\hat w^3 \frac{d\hat p}{d x}, \quad \hat v(x)=\hat \phi(x)+x \hat v_0,
\end{equation}
where
\begin{equation}
\label{v_hat}
\hat v=-\hat w^2 \frac{d \hat p}{d x},\quad \mbox{or}\quad \frac{d \hat p}{d x}=-\frac{1}{\hat w^2}\left(\hat \phi(x)+x \hat v_0\right).
\end{equation}

\noindent
Based on these assumptions, equation \eqref{Phi_cont} can be transformed into:
\begin{equation}
\label{cont_sep}
\frac{d \psi}{dt}\hat w+\frac{\psi^4}{L^{m+1}}\frac{d}{dx}\left(\hat w \hat \phi\right)+\frac{L'}{L}\psi \hat w +q_l=0.
\end{equation}
Functions $\psi(t)$, $L(t)$ and $q_l(t,x)$ should be properly specified in order to eliminate the time variable from the above equation. Then, the problem reduces to the time-independent form.

\subsubsection*{A2: Self-similar solution of exponential type}
Let the function $\psi(t)$ have the form:
\begin{equation}
\label{psi_exp}
\psi(t)=e^{\gamma t},
\end{equation}
where $\gamma >0$ is an arbitrary constant. Then, substituting \eqref{q_vphi_sep}$_2$ into \eqref{L_int_PKN}, and taking the initial crack length as:
\[
L(0)=\left[\frac{(m+1)C_{\cal A} {\cal L} (\hat w)}{3\gamma}\right]^\frac{1}{m+1},
\]
one can derive the following relation for $L(t)$:
\begin{equation}
\label{L_exp}
L(t)=\left[\frac{(m+1)C_{\cal A}{\cal L}(\hat w)}{3\gamma}\right]^\frac{1}{m+1} e^\frac{3\gamma t}{m+1}.
\end{equation}
Then, if one assumes that the leak-off function complies with the representation:
\begin{equation}
\label{leak_off_exp}
q_l(t,x)=e^{\gamma t}\hat q_l(x),
\end{equation}
the governing equation  \eqref{cont_sep}  after simple transformations can be reduced to:
\begin{equation}
\label{cont_exp}
\frac{1}{C_{\cal A} {\cal L}(\hat w)} \frac{d}{dx}(\hat w \hat \phi)=-\frac{m+4}{3}\hat w-\frac{m+1}{3\gamma}\hat q_l.
\end{equation}
In this way we obtain an ordinary differential equation equipped with the following boundary conditions:
\begin{equation}
\label{exp_ss_BC}
\hat w(1)=0,\quad \hat \phi(1)=0,\quad \hat w(0) \hat \phi(0)=\hat q_0.
\end{equation}
Additionally, for the KGD model the following symmetry condition holds:
\begin{equation}
\label{sym_ss}
\frac{d \hat w}{dx}\big | _{x=0}=0.
\end{equation}
The equivalent of the fluid balance equation \eqref{global_balance_n} is:
\begin{equation}
\label{balance_exp}
\frac{1}{C_{\cal A}{\cal L}(\hat w)}\hat q_0-\frac{m+4}{3}\int_0^1 \hat w dx-\frac{m+1}{3\gamma}\int_0^1 \hat q_l dx=0.
\end{equation}

\subsubsection*{A3: Self-similar solution of the power law type}
In this variant of the self-similar solution let us take:
\begin{equation}
\label{psi_pow}
\psi(t)=(a+t)^\gamma,
\end{equation}
where $a \geq 0$ and $\gamma >0$ are some constants.
In the same way as previously, one can obtain a relation for the crack length in the form:
\begin{equation}
\label{L_pow}
L(t)=\left[\frac{(m+1)C_{\cal A}{\cal L}(\hat w)}{3\gamma+1}\right]^\frac{1}{m+1}(a+t)^\frac{3\gamma+1}{m+1},
\end{equation}
provided that:
\[
L(0)=\left[\frac{(m+1)C_{\cal A}{\cal L} (\hat w)}{3\gamma +1}\right]^\frac{1}{m+1}a^\frac{3\gamma +1}{m+1}.
\]
Additionally if the leak-off function can be expressed as:
\begin{equation}
\label{ql_pow}
q_l(t,x)=(a+t)^{\gamma-1}\hat q_l(x),
\end{equation}
equation \eqref{cont_sep} converts to:
\begin{equation}
\label{cont_pow}
\frac{1}{C_{\cal A}{\cal L}(\hat w)}\frac{d}{dx}(\hat w \hat \phi)=-\left[\frac{(m+1)\gamma}{3\gamma+1}+1\right]\hat w-\frac{m+1}{3\gamma+1}\hat q_l.
\end{equation}
Boundary conditions for the above differential equation remain the same as \eqref{exp_ss_BC} -\eqref{sym_ss}.
Finally, the global fluid balance equation can be transformed to:
\begin{equation}
\label{balance_pow}
\frac{1}{C_{\cal A} {\cal L}(\hat w)}\hat q_0-\left[\frac{(m+1)\gamma}{3\gamma+1}+1\right]\int_0^1 \hat w dx-\frac{m+1}{3\gamma+1}\int_0^1 \hat q_l dx=0.
\end{equation}

\section*{Appendix B: Construction of the benchmarks solutions}

In the following we introduce a set of analytical benchmark solutions for the considered problem. The ideas behind their construction are the same regardless of the hydraulic fracture model in use,
and have already been employed in \cite{M_W_L,solver_calkowy,Kusmierczyk} for PKN model.
The basic idea is to use an analytical solution to the self-similar problem defined in
Appendix A, and extended it into the time-dependent form using relations \eqref{w_sep}, \eqref{q_vphi_sep} -- \eqref{q_phi_hat} and \eqref{L_exp} or \eqref{L_pow}.

Let us concentrate now on finding some examples of analytical solutions to the self-similar equation:
\begin{equation}
\label{ODE_gen}
\frac{1}{{\cal L}(\hat w)} \frac{d}{dx}(\hat w \hat \phi)=-\chi \hat w-\varkappa \hat q_l,
\end{equation}
which is a generalization of \eqref{cont_exp} and \eqref{cont_pow}. The corresponding constants $\chi$ and $\varkappa$ in \eqref{cont_exp} can be
determined by direct comparison with equations \eqref{cont_exp} and \eqref{cont_pow}, and are given in Table~2.

The boundary conditions \eqref{exp_ss_BC} are to be satisfied, together with their respective form of the balance equation (\eqref{balance_exp} or \eqref{balance_pow}). Depending on the hydraulic fracture model one of the operators
\eqref{p_ss} shall be in use.

The general idea behind constructing a benchmark solution is quite straightforward. At first, assume that the crack opening function can be expressed as a weighted sum of properly chosen base functions:
\begin{equation}
\label{w_rep}
\hat w(x)=\sum_{i=0}^N \lambda_i \hat w_i(x).
\end{equation}
The functions $\hat w_i(x)$ should be selected in a way which enables one to: i)  comply with the respective asymptotic representation \eqref{w_roz},
ii) analytically compute the pressure operators \eqref{p_ss}, iii) satisfy the boundary conditions \eqref{exp_ss_BC} (and \eqref{sym_ss} for KGD).
Provided that ii) is fulfilled, the pressure function can be calculated in a closed form from \eqref{p_ss} to give:
\begin{equation}
\label{p_rep}
\hat p(x)=\sum_{i=0}^N \lambda_i \hat p_i(x),
\end{equation}
where each of the functions $\hat p_i(x)$ corresponds to the respective function $\hat w_i(x)$.
Then, according to \eqref{v_hat}$_1$ the particle velocity is defined:
\begin{equation}
\label{v_rep}
\hat v(x)=-\left[\sum_{i=0}^N \lambda_i \hat w_i(x)\right]^2\sum_{i=0}^N \lambda_i \frac{d}{dx} \hat p_i(x),
\end{equation}
and consequently we have its leading term as:
\begin{equation}
\label{v0_rep}
\hat v_0=\hat v(1)=C_{\cal A}{\cal L}(\hat w).
\end{equation}
The reduced velocity is determined using \eqref{v_rep} -- \eqref{v0_rep} according to \eqref{q_phi_hat}$_2$:
\begin{equation}
\label{fi_rep}
\hat \phi(x)=-\left[\sum_{i=0}^N \lambda_i \hat w_i(x)\right]^2\sum_{i=0}^N \lambda_i \frac{d}{dx} \hat p_i(x)-xC_{\cal A}{\cal L}(\hat w).
\end{equation}

Next, by substituting \eqref{w_rep} and \eqref{fi_rep}
into equation \eqref{ODE_gen}, we can define the benchmark leak-off function $\hat q_l(x)$.
Finally the value of $\hat q_0$ is determined by substituting \eqref{w_rep} and \eqref{fi_rep} into the boundary condition \eqref{exp_ss_BC}$_3$:
\begin{equation}
\label{q0_rep}
\hat q_0=-\left[\sum_{i=0}^N \lambda_i \hat w_i(0)\right]^3\sum_{i=0}^N \lambda_i \frac{d\hat p_i}{dx} \big|_{x=0}.
\end{equation}
 In this way, the analytic benchmark solution is fully defined by \eqref{w_rep},
\eqref{fi_rep}, corresponding leak-off function and corresponding influx value \eqref{q0_rep}. Clearly, the fluid balance equation (\eqref{balance_exp} or \eqref{balance_pow}) is satisfied automatically.

\subsection*{B1: PKN model}
For the PKN model let us adopt the following $N+1$ test functions ($i=0,1,...,N$):
\begin{equation}
\label{h_PKN}
\hat w_i(x)=(1-x)^{i+1/3}, \quad i\le N-1, \quad \hat w_N(x)=e^x(1-x)^{N+1/3}.
\end{equation}
The base function $\hat w_N(x)$ was taken in the specified form in order to introduce an additional non-linear effect to the benchmark, without
violating the asymptotic behaviour.

By applying representation \eqref{h_PKN} in \eqref{w_rep} and elasticity operator \eqref{p_ss}$_1$ one obtains the formula for the pressure function, $\hat p(x)$,
which after differentiation yields:
\begin{equation}
\label{pp_PKN}
\frac{d}{dx}\hat p(x)=-\left[\sum_{i=0}^{N-1}\lambda_i (i+1/3)(1-x)^{i-2/3}+\lambda_N(N+x-2/3)e^x(1-x)^{N-2/3}\right].
\end{equation}
Then by formulae \eqref{v_rep}-\eqref{q0_rep} one can construct the benchmark solution, taking ${\cal L}(\hat w)=\lambda_0^3$.

In general, the leak-off behaviour near the crack tip can be controlled by the powers of the base functions $\hat w_i$, for $i>1$ (e.g. if one wants to mimic the Carter law, each power starting from the second one should be 1/6 greater than the previous - compare \cite{Kusmierczyk}).  The representation used in this paper gives:
\begin{equation}
\label{q_l_banch}
\hat q_l(x)=O\left((1-x)^{\alpha_0}\right),\quad x\to1,
\end{equation}
which is the same as the asymptotics of the crack opening itself. However, even in such a case, proper manipulation of the multipliers can provide better behaviour of $q_l$.

\subsection*{B2: KGD model: fluid driven regime}

For this model we assume that the self-similar crack aperture is defined by the following base functions:
\begin{equation}
\label{h_KGD fluid}
\hat w_i(x)=(1-x^2)^{\alpha_i}C_{2(i+1)-2}^{\alpha_i-1/2}(x), \quad i\le N-1, \quad \hat w_N(x)=\sqrt{1-x^2}-\frac{2}{3}(1-x^2)^{3/2}-x^2\ln{\frac{1+\sqrt{1-x^2}}{x}},
\end{equation}
where $C_{2(i+1)-2}^{\alpha_i-1/2}(x)$ is the ultraspherical or Gegenbauer polynomial. The term $w_N(x)$ was introduced to obtain a non-zero pressure gradient for $x=0$.
In our case we take in computations $N=2$, where $\alpha_0=2/3$, $\alpha_1=5/3$. Note that:
\begin{equation}
\label{w_N fluid}
\hat w_N(x)=O\left((1-x^2)^{5/2}\right), \quad x\to 1.
\end{equation}

A representation similar to \eqref{h_KGD fluid} (except for the last term) was used in \cite{Adachi_Detournay} to define the base functions for the series approximation of solution. This general representation was utilized in order to solve the problem for a variety of shear-thinning fluids. In a similar way \eqref{h_KGD fluid} with the proper values of $\alpha_i$ can be employed to construct benchmark solutions for non-Newtonian fluids.

By applying \eqref{h_KGD fluid} in \eqref{Am1_ss} and subsequent differentiation one obtains an analytical formula for the pressure gradient:
\begin{equation}
\label{pp_PKNa}
\frac{d}{dx}\hat p(x)=-\left[\sum_{i=0}^{N-1}\lambda_i F_i(\alpha_i,x)+\lambda_N (x-\pi/4)\right],
\end{equation}
where:
\begin{equation}
\label{F_0_def}
F_0(\alpha_0,x)=\pi^{-1}\alpha_0(2\alpha_0-1)B(1/2,\alpha_0)x\cdot_2\!F_1(3/2-\alpha_0,2;3/2;x^2)
\end{equation}
\begin{equation}
\label{F_i_def}
F_i(\alpha_i,x)=\frac{(1-2\alpha_i)(2i-1)x}{4\pi(i-1+\alpha_i)}B\left(1/2-i,\alpha_i+i\right)
\end{equation}
\[
\left\{[3-5i+2i(i+\alpha_i)]_2F_1(5/2-i-\alpha_i,i;3/2,x^2)-\frac{2\alpha_i}{3}i(2i-5+2\alpha_i)
x^2 \cdot_2F_1(7/2-i-\alpha_i,i+1;5/2;x^2)  \right\},
\]
\[i>1.\]
$B$ is the beta function, and $_2F_1$ is the Gauss hypergeometric function.

Then from  \eqref{v_rep} -- \eqref{q0_rep} we have the complete benchmark solution, where ${\cal L}(\hat w)=\lambda_0^3$.

\subsection*{B3: KGD model: toughness driven regime}

In the case of the toughness driven KGD model we consider the benchmark where the crack opening is represented by a sum of four functions:
\begin{equation}
\label{h_KGD tough}
\hat w_i(x)=(1-x^2)^\frac{i}{2},\quad i=1,2, \quad \hat w_3(x)=(1-x^2)^\frac{3}{2}\ln{(1-x^2)}, \quad \hat w_4(x)=2\sqrt{1-x^2}+x^2\ln{\frac{1-\sqrt{1-x^2}}{1+\sqrt{1-x^2}}}.
\end{equation}
The above representation is consistent with \eqref{w_roz} and Table~\ref{T1}. The special term $\hat w_N$ gives a non-zero pressure gradient at $x=0$ and:
\begin{equation}
\label{w_N tough}
\hat w_N(x)=O\left((1-x^2)^{3/2}\right), \quad x\to 1.
\end{equation}
Obviously it can be replaced by $\hat w_n$ from \eqref{h_KGD fluid} if convenient. Also further terms of type $\hat w_i$ from \eqref{h_KGD fluid} can be used here, provided that the powers $\alpha_i$ are greater than those from the leading terms of the asymptotics.

Then by operator \eqref{Am1_ss}$_2$ one obtains respective pressure components $\hat p_i$ from \eqref{p_rep}, where $\hat p_0(x)=\pi/2$. Thus, the non-zero components of the pressure derivative are:
\[
\hat p'_1(x)=0,\quad \hat p_2'(x)=\frac{1}{\pi x}\left[-\frac{1}{1-x^2}+_2\!F_1(-1/2,1;1/2;x^2)\right],
\]
\begin{equation}
\label{p_i_tough}
\end{equation}
\[\hat p'_3(x)=-\frac{x}{2}(5-6\ln2)-\frac{3}{2} \arcsin(x)\left(\sqrt{1-x^2}-\frac{x^2}{\sqrt{1-x^2}}\right),\quad \hat p_4'(x)= -\frac{\pi}{2}\]
Finally, by applying \eqref{h_KGD tough} and \eqref{p_i_tough} in  \eqref{v_rep} -- \eqref{q0_rep} one constructs the benchmark solution, where ${\cal L}(\hat w)=\lambda_0^2\lambda_1$.
Moreover, decreasing/increasing this value, one tests the algorithms on the small/large toughness regimes.
Finally, in the case of KGD problems (both regimes), the leak-off function, $q_l$, constructed above for the benchmarks, generally speaking, will behave in the same manner as in (\ref{q_l_banch}).

\end{document}